\documentclass[aps,prl, 10 pt, twocolumn,showpacs,superscriptaddress, preprintnumbers, nofootinbib]{revtex4-2}

\usepackage[pdftex,colorlinks]{hyperref}
\usepackage[dvipsnames]{xcolor}
\definecolor{phthaloblue}{rgb}{0.0, 0.06, 0.54}
\hypersetup{
    colorlinks=true,
    linkcolor=NavyBlue,
    citecolor=NavyBlue,
    filecolor=NavyBlue,
    urlcolor=NavyBlue,
    }
\usepackage[pdftex]{graphicx}
\usepackage{bm} 
\usepackage{cancel}
\usepackage{here}
\usepackage{comment,braket}
\usepackage{epsf}
\usepackage{amsmath}
\usepackage{graphics} 
\usepackage{subfigure}
\usepackage{amsfonts}
\usepackage{amssymb}
\usepackage{latexsym}
\usepackage{color}
\usepackage{natbib}
\usepackage[normalem]{ulem}
\usepackage{orcidlink}
\usepackage{cleveref}
\usepackage{physics}
\usepackage{upgreek}
\usepackage{makecell}

\crefname{section}{Sec.}{Sec.}
\crefformat{section}{Sec.~#2#1#3}
\crefmultiformat{section}{Sec.~#2#1#3}{-#2#1#3}{, #2#1#3}{-#2#1#3}
\allowdisplaybreaks

\usepackage{booktabs}

\usepackage{array} 

\usepackage{orcidlink}
\usepackage{fontawesome5}

\usepackage{dcolumn}

\newcommand{\lettersection}[1]{\noindent{\bf \emph{#1.}---}}
\newcommand{\ipc}[1]{\textcolor{Maroon}{#1}}

\begin{document}


\title{On the cosmic window for neutrino self-interactions}

\author{Ivan Pérez-Castro\,\orcidlink{0009-0003-1060-6858}}
\email[~]{ivan.perez.c@cinvestav.mx}
\affiliation{Departamento de F\'{\i}sica, Centro de Investigaci\'on y de Estudios Avanzados del Instituto Politécnico Nacional \\ Apartado Postal 14-740, 07000, Ciudad de M\'exico, M\'exico.}

\author{Jorge Venzor\,\orcidlink{0000-0003-4510-2089}}
\email[~]{jorge.venzor@cinvestav.mx}
\affiliation{Departamento de F\'{\i}sica, Centro de Investigaci\'on y de Estudios Avanzados del Instituto Politécnico Nacional \\ Apartado Postal 14-740, 07000, Ciudad de M\'exico, M\'exico.}

\author{Josue De-Santiago\,\orcidlink{0000-0002-1163-3730}}
\email[~]{josue.desantiago@cinvestav.mx}
\affiliation{Departamento de F\'{\i}sica, Centro de Investigaci\'on y de Estudios Avanzados del Instituto Politécnico Nacional \\ Apartado Postal 14-740, 07000, Ciudad de M\'exico, M\'exico.}
\affiliation{Secretar\'ia de Ciencia, Humanidades, Tecnolog\'ia e Innovaci\'on.
Av. Insurgentes Sur 1582, 03940, Ciudad de M\'exico, M\'exico}

\author{Pablo Roig\,\orcidlink{0000-0002-6612-7157}}
\email[~]{pablo.roig@cinvestav.mx}
\affiliation{Departamento de F\'{\i}sica, Centro de Investigaci\'on y de Estudios Avanzados del Instituto Politécnico Nacional \\ Apartado Postal 14-740, 07000, Ciudad de M\'exico, M\'exico.}

\author{Abdel Pérez-Lorenzana\,\orcidlink{0000-0001-9442-3538}}
\email[~]{abdel.perez@cinvestav.mx}
\affiliation{Departamento de F\'{\i}sica, Centro de Investigaci\'on y de Estudios Avanzados del Instituto Politécnico Nacional \\ Apartado Postal 14-740, 07000, Ciudad de M\'exico, M\'exico.}


\begin{abstract}
Neutrino self-interactions ($\nu$SI) beyond the Standard Model have arisen as a promising probe of new physics in cosmology.
In this Letter, we show that $\nu$SI emerging in the low-energy limit of renormalizable theories implies resonant mediator behavior.
We identify two regions for $\nu$SI: (i) neutrinos become tightly coupled
with background evolution; for the first time, we establish how this evolution modifies the Boltzmann hierarchy equations (BHEs), and (ii) resonant $\nu$SI dominates the dynamics with negligible background evolution, relevant for
mediator masses $\sim 10^{-6}-1$ MeV. 
In region (i), before matter-radiation equality, neutrino free-streaming is determined by the effective four-fermion interaction, known as the heavy-mediator limit (HML); we explore the impact of background evolution on this scenario, and from our modified BHEs, we show that self-interacting dark radiation scenarios require larger couplings than previously estimated to mimic $\nu$SI.
Region (ii) shows that the simplest realization of $\nu$SI mediated by a (pseudo)scalar particle leads to an extra radiation contribution of $\Delta N_{\rm eff}^{\rm CMB} \approx 0$, instead of the commonly quoted $\simeq 0.12$, at the CMB epoch. Overall, we highlight that our findings should be considered in any plausible scenario involving $\nu$SI.
\end{abstract}

\maketitle

\lettersection{Introduction}
Beyond the Standard Model (BSM) of particle physics, non-standard $\nu$SI
constitute an intriguing possibility in neutrino phenomenology~\cite{Berryman2023neutrino}, whereby neutrino–neutrino scattering may significantly impact astrophysical signals~\cite{Ng2014cosmic, Ioka2014icecube, Bustamante2020bounds, Creque2021, Esteban2021probing, Carpio2023simulating, Chang2022towards, Cerdeno2023constraints, Hooper2023signals, Wu2024shedding, Dhuria2024probing, De2024mu, Wang2025probing, Francener2025probing, He2025implications, Natwariya2026constraining, Machado:2025ltu, Bharadwaj2026diffuse, Das2026sn1987a} as well as cosmological probes. 
In the latter, the effective four-neutrino interaction, $\sim G_{\rm eff} (\overline{\nu}\nu)(\overline{\nu}\nu)$, has been widely explored, with the parameter
$G_{\rm eff} \gg G_F$~\cite{CyrRacine2014, Archidiacono2014, Oldengott2015, Lancaster2017, Oldengott2017, Kreisch2020, Barenboim2019inflation, Park2019, Das2021, Choudhury2021, Brinckmann2020self, Mazumdar2022flavour, Kumar_2022, Kreisch:2022zxp, Das:2023npl, Camarena2023confronting, He2024self, Camarena2025strong,  Poudou:2025qcx, Calabrese2025atacama, He:2025jwp, Noriega:2026uhw, Taule2022global, RoyChoudhury2022, Bostan_2024, Pal_2025, Libanore2025, Montefalcone:2025ibh, Whitford:2025dmq, Pal2026neutrino, Parashari:2026dxo, Mogensen2026self, Barenboim:2026jef, Mondol:2026irr, Uggerhoj:2026jet}, where
$G_F$ is the Fermi constant. 
A key feature of this model is that neutrinos remain self-collisional after their thermal decoupling from the primordial plasma at $T_{\nu, {\rm dec}} \sim 2 \, {\rm MeV}$~\cite{De2016relic, Akita2020precision, Froustey2020neutrino, Bennett_2021}, hence delaying the onset of neutrino free-streaming.
Consequently, some data analyses have hinted at indirect evidence of BSM physics, with the $\nu$SI parameter bounds alternating between a preference for strong $\nu$SI ($G_{\rm eff} \sim 10^{-3}- 10^{-1} \, {\rm MeV}^{-2}$) and for a region akin to $\Lambda$CDM~\cite{CyrRacine2014, Archidiacono2014, Oldengott2015, Lancaster2017, Oldengott2017, Barenboim2019inflation, Kreisch2020, Park2019, Das2021, Choudhury2021, Brinckmann2020self, Mazumdar2022flavour, Kumar_2022, Kreisch:2022zxp, Das:2023npl, Camarena2023confronting, He2024self, Camarena2025strong, Poudou:2025qcx, Calabrese2025atacama, He:2025jwp, Noriega:2026uhw}.

The effective interaction can arise at low energies from neutrino interactions involving sub-MeV mediators.
Some realizations involve  Majoron-like models~\cite{Chikashige1981there, Gelmini1981left, Berryman2018, Dutta:2023fdt},
vector boson mediators~\cite{Berbig2020, Ghosh2024hubble}, as well as Dirac neutrino scenarios~\cite{He2020}. Indeed, the $G_{\rm eff}$ values observed in cosmology are large enough that they should have already left smoking-gun signatures in existing neutrino experiments~\cite{Gerda2015results, Blum2018neutrinoless, Krnjaic2020probing, Blinov2019, Deppisch2020, Lyu2021self, Demirci2024, Dev2025new, Foroughi2026precision} and in astrophysical observations~\cite{Heurtier2017supernova, Shalgar2021, Akita2022, Fiorillo2022, Doring2024testing, Akita2024limits}. 
Thus, some mechanisms were formulated with the purpose of reproducing cosmological phenomenology and simultaneously avoiding external constraints~\cite{Foroughi2025enabling, Das2025impostor}.
In particular, the authors of Ref.~\cite{Das2025impostor} proposed a framework 
in which strongly self-interacting dark radiation mimics $\nu$SI while depleting the neutrino abundance and thereby relaxing their cosmological mass bound~\cite{Farzan2016neutrinos, Escudero2023seesaw}, which is in potential tension with neutrino oscillation data~\cite{Gariazzo2023quantifying, Naredo2024critical, Elbers2025constraints, Di2025cosmoverse, SPT_2026_CMBSPA}.

In this Letter, we present the first self-consistent 
treatment of $\nu$SI for both 
simple and extended models below $T_{\nu, {\rm dec}}$.
To do so, we reassess two prevalent assumptions in the literature.
First, background effects at the perturbation level are commonly neglected under the premise that $\nu$SI preserve background distribution functions; however, new interactions may invalidate this.
Second, other regimes such as resonant $\nu$SI, which treats the mediator as an unstable particle over cosmic timescales, are largely overlooked (with the exception of Refs.~\cite{venzor2023resonant, Noriega2025resonant} for mediator masses $M \lesssim 10^2 \, {\rm eV}$), although resonances should be prevalent at high temperatures, or even at low temperatures in models with light mediators (e.g.,~\cite{Chikashige1981there, Gelmini1981left, Berryman2018, Dutta:2023fdt, Berbig2020, Ghosh2024hubble}). Furthermore, 
similar results arise from thermalization via decays \cite{Escudero2020CMBsearch} within the neutrino–mediator 
fluid approximation after mediator production via inverse neutrino decays, raising 
questions about this ``equivalence''.

Motivated by these observations, we construct BHEs with a non-vanishing zeroth-order collision term, allowing us to capture background effects and assess their implications for the allowed parameter space and phenomenology of cosmic $\nu$SI. 

\lettersection{Cosmology with neutrino self-interactions}
The modern treatment for investigating the impact of $\nu$SI relies on the so-called \emph{thermal approximation}~\cite{Kreisch2020, Perez2026towards}. In this framework, the neutrino multipole moments, denoted ${\cal F}_\ell$, are introduced to construct the BHEs at first order 
in cosmological perturbation theory, assuming that the collision term vanishes at the background level, $C_{\nu_i}^{(0)}=0$, once the cosmic neutrino background temperature $T_\nu$ drops below $T_{\nu, {\rm dec}}$.

To derive BHEs with $C_{\nu_i}^{(0)} \neq 0$, we consider the cosmological time evolution of the background neutrino distribution $\bar{f}_{\nu_i}^{(0)}$, $d\bar{f}_{\nu_i}^{(0)}/dt = C_{\nu_i}^{(0)}$. We parametrize the solution as $\bar{f}_{\nu_i}^{(0)} = \xi_{i,r} f_{\nu_i}^{(0)}$, where $f_{\nu_i}^{(0)}$ retains its thermal form of each $i$ mass eigenstate, while $\xi_{i,r}$ encodes momentum- and temperature-dependent deviations from $f_{\nu_i}^{(0)}$.
In the conformal Newtonian (CN) gauge, we propose the modified BHEs in terms of neutrino multipoles ${\cal F}_{i,\ell}$:
    \begin{equation}
    \begin{split}
        \frac{\partial {\cal F}_{i,\ell}}{\partial \eta}
        &+ k \bigg[
        \frac{\ell +1}{2\ell +1} {\cal F}_{i,\ell +1}
        - \frac{\ell}{2\ell +1} {\cal F}_{i,\ell -1}
        \bigg]
        -4 R_i f_{g,\ell} \\
        &- a {\cal I}_{{\rm CN},i}^{(0)} \tilde{\psi} \delta_{0 \ell}
        = - a\sum_j \Gamma_{ij, \ell} {\cal F}_{j,\ell}
        + a{\cal I}_{{\rm eff},i}^{(0)} {\cal F}_{i,\ell} \, ,
    \end{split}
    \label{BHE-0-CN-S}
    \end{equation}
where $a$ is the scale factor, $\eta$ the conformal time and $k$ the wavenumber of the perturbations. The standard metric perturbation
variables
$\tilde{\phi}$ and $\tilde{\psi}$ determine
    \begin{equation}
        f_{g,\ell} = \frac{\partial \tilde{\phi}}{\partial \eta} \delta_{0\ell} + \frac{k}{3} \tilde{\psi} \delta_{1\ell} \, , \label{f_g-massless}
    \end{equation}
while the terms arising from a non-null $C_{\nu_i}^{(0)}$ are
    \begin{align} 
        {\cal I}^{(0)}_{{\rm eff},i} &= \frac{1}{2 \pi^2 \bar{\rho}_{\nu_i}} \int dp p^3 C_{\nu_i}^{(0)} \, , \label{Gamma-0}\\
        {\cal I}^{(0)}_{{\rm CN},i} &= \frac{2} {\pi^2 \bar{\rho}_{\nu_i}} \int dp p^3 \left( \frac{\partial \ln \bar{f}_{\nu_i}^{(0)}}{\partial \ln p} - \Delta_i \right)^{-1} C_{\nu_i}^{(0)} \, , \label{Gamma-CN} \\
        R_i &= \frac{1} { 2\pi^2\bar{\rho}_{\nu_i}} \int dp p^3 \frac{\partial \bar{f}_{\nu_i}^{(0)}}{\partial \ln p} \left( \frac{\partial \ln \bar{f}_{\nu_i}^{(0)}}{\partial \ln p} - \Delta_i \right)^{-1} , \label{R} \\
        \Delta_i &= \frac{\partial \ln \xi_{i,r}}{\partial \ln p} + \frac{\partial \ln \xi_{i,r}}{\partial \ln T_\nu} \, , \label{Delta}
    \end{align}
where $p$ denotes the magnitude of the local momentum of a reference neutrino, and the energy density per degree of freedom $\bar{\rho}_{\nu_i}$ evolves due to the non-trivial background evolution, as in neutrino decay scenarios~\cite{Barenboim2021invisible, Chen2022weaker, Terzaghi2025impact, Abellan2026neutrino}.
The interaction rates $\Gamma_{ij,\ell}$ arise from the first-order collision term.\footnote{The modified BHEs~\eqref{BHE-0-CN-S} can be constructed in the synchronous gauge by replacing the term $f_{g, \ell}$ as in Ref.~\cite{Perez2026towards}, while the term~\eqref{Gamma-CN} does not appear as long as $\tilde{\psi}$ is only a potential in the CN gauge.} Consult Sec.~S1 of the Supplemental Material (SuM) for further derivations.

In the
HML usually: ${\cal F}_{i,\ell} \approx {\cal F}_\ell$, and $\sum_j \Gamma_{ij,\ell} \approx \Gamma_\ell = G_{\rm eff}^2 T_\nu^5 \alpha_\ell/4$,
where the coefficients $\alpha_\ell$ are determined numerically (see Refs.~\cite{Kreisch2020, Perez2026towards}).
None of the cosmological analyses studying the effects of $\nu$SI on neutrino perturbations has considered the
contribution of $C_{\nu_i}^{(0)}$ to the BHEs, 
working with a simplified version of Eqs.~\eqref{BHE-0-CN-S}.
The same assumptions are also made in analyses involving ultralight mediators, where $\Gamma_\ell \sim g_{\rm eff}^4 T_\nu$~\cite{Forastieri:2015paa, Forastieri:2019cuf, Venzor2022}, as well as in resonant $\nu$SI scenarios~\cite{venzor2023resonant, Noriega2025resonant}. We will discuss when we need to implement these modified BHEs.

\lettersection{Renormalizable completions for neutrino self-interactions}
The effective four-neutrino interaction breaks down at high energies, as it is non-renormalizable and must be replaced by interactions involving new degree(s) of freedom. 
A simple realization that reproduces the HML is a real scalar mediator $\varphi$ that couples to neutrinos, with mass $m_\varphi$. The corresponding Lagrangian in the mass basis for Majorana neutrinos is
    \begin{equation}
        {\cal L} \supset - \frac{1}{2} m_\varphi^2 \varphi^2 + \frac{1}{2} \left(g_{ij} \overline{\nu_{i L}^c} \nu_{j L} \varphi + \rm{H.c.} \right) \, , \label{L-scalar}
    \end{equation}
where $\nu_{i L}$ are left-handed neutrinos, with $\nu_{i L}^c$ their charge conjugates, and $g_{ij}$ denotes the coupling between neutrino mass eigenstates $i$ and $j$.\footnote{The couplings $g^{\alpha \beta}$ in the flavor basis are related to the mass basis via the PMNS matrix $U_{\alpha i}$; in the Majorana case, $g_{ij} = g^{\alpha\beta} U_{\alpha i} U_{\beta j}$. Here, $\alpha$ denotes the flavors (e.g., $\alpha=e,\mu,\tau$).}

The full set of dimension-four operators
follows by generalizing the bilinear term $\overline{\nu_{i L}^c} \nu_{j L} \to \overline{\nu_{i L}^c} \Gamma^a \nu_{j L}$. The independent Dirac structures are given by:
    \begin{equation}
        \Gamma^a = \big\{ I, i\gamma_5, \gamma^\mu, \gamma^\mu \gamma_5, \sigma^{\mu\nu} := \frac{i}{2} \left[ \gamma^\mu, \gamma^\nu \right] \big\} \, . \label{In-operators}
    \end{equation}
The first case corresponds to a scalar interaction, while the remaining structures correspond, respectively, to pseudoscalar, vector, axial-vector, and tensor interactions. In each case, the mediator field must be replaced by the appropriate field type (e.g., scalar, vector, etc.) consistent with the Lorentz structure of the interaction. Note that scalar and pseudoscalar (and likewise vector and axial-vector) interactions are indistinguishable in cosmic $\nu$SI when $T_\nu \gg m_\nu$.

The operators like Eq.~\eqref{In-operators} appear in several extensions of the Standard Model at low energy~\cite{Berbig2020, He2020, Beacom2004,Ghosh2024hubble, Chikashige1981there, Beacom2004, De2020dodelson, Berryman2018, De2024minimal}, and are often referred to as \emph{secret neutrino interactions} (SNI)~\cite{Huang2018}.
In such scenarios, the mediator propagator features a simple pole of the form $\sim ({\rm q}^2 - M^2)^{-1}$, where ${\rm q}$ is the transferred momentum and $M$ is the mediator mass. When $M \gg \sqrt{|{\rm q}^2|}$, it reproduces an effective field theory like Fermi’s theory, the so-called HML, and the mediators can be produced on-shell when ${\rm q}^2 = M^2$.

\begin{figure} [t!]
\center
\includegraphics[width=1\linewidth]{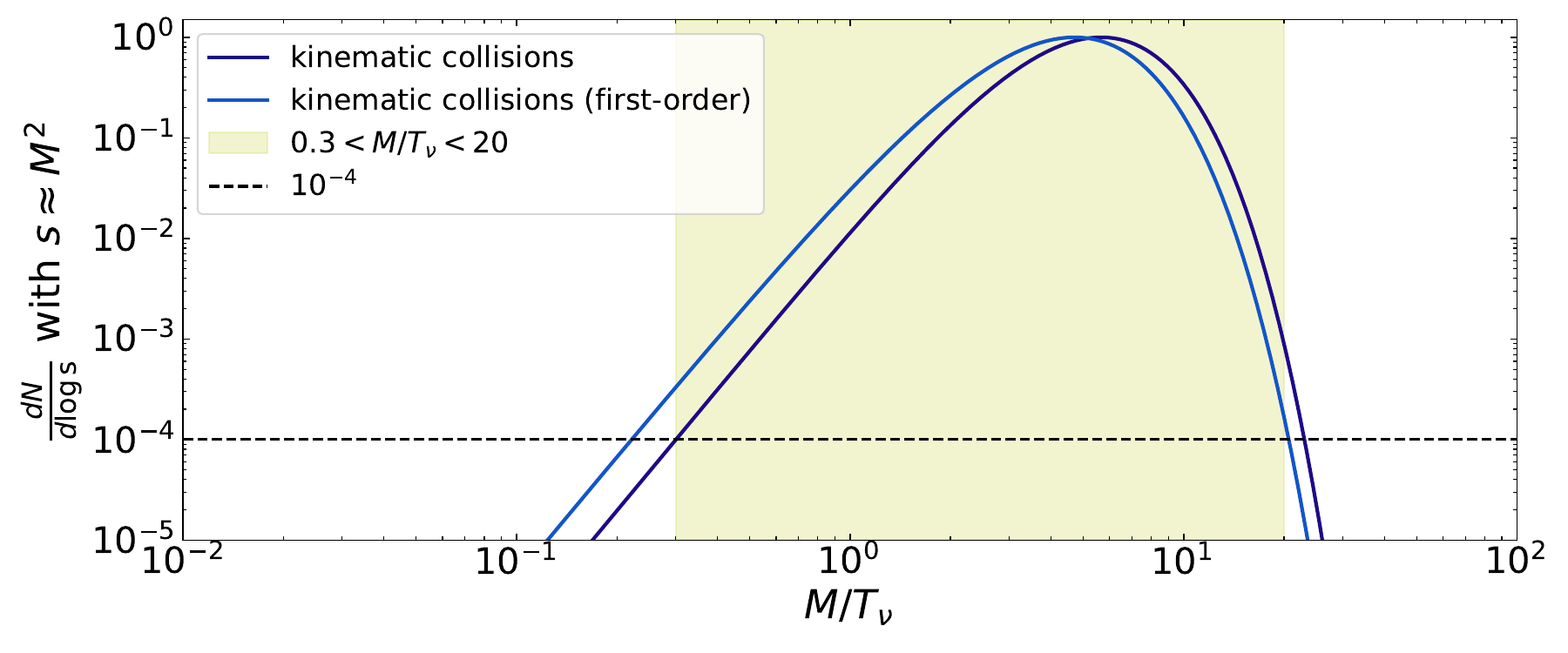}
\caption{\label{fig:Events} Distribution per logarithmic interval at $s\approx M^2$,  shown as a function of $M/T_\nu$ and normalized to unity. We show two parametrizations: one based on the background-level interaction rate in the model-independent framework, labeled \emph{kinematic collisions}, and the other based on the multipole-independent contribution of $\nu$SI to the first-order collision term ($M=m_\varphi$).}
\end{figure}

The typical scale used in cosmology to produce the resonance is $T_\nu \sim M$. However, for cosmic neutrinos, this approximation does not hold 
because when choosing a reference neutrino, all scattering configurations must be taken into account. 
We find that the kinematically allowed on-shell production in the $s$-channel contribution is more frequent for $T_\nu \sim (1/20, 10/3) \, M$; see Fig.~\ref{fig:Events} and consult the details in Sec.~S3.1 of the SuM. Its relevance also depends on the neutrino--mediator coupling, as we will discuss in what follows.

\lettersection{Resonant neutrino self-interactions}
Rather than computing an effective interaction rate $\Gamma_\ell$, the effects of $\nu$SI can be incorporated into the BHEs via the relaxation-time approximation (RTA), where $\Gamma_\ell$ is replaced by $\Gamma_{bc \to df} = n_c \langle \sigma v \rangle_{bc \to df}$, which arises in the \emph{pseudo-chemical approximation} to evolve the particle number density~\cite{Bernstein1985cosmological}, wherein $n_c$ is the number density of the particles that scatter with a reference neutrino, and the thermally averaged cross-section $\langle \sigma v \rangle_{bc \to df}$ is computed in the center-of-mass frame, neglecting quantum-statistical effects such as Pauli blocking. Moreover, since these interaction rates 
arise at the background level, comparing them with the Hubble expansion rate $H$ provides an estimate of whether the process significantly affects neutrino evolution ($\Gamma \gtrsim H$) or is subdominant ($\Gamma < H$). 

In the following, we compute interaction rates for the processes allowed by~\eqref{L-scalar}, assuming $g_{ij} = g \delta_{ij}$ with $i=1,2,3$.\footnote{Non-diagonal interaction leads to different evolution of the multipoles ${\cal F}_{i,\ell}$ for each $i$.} We improve the estimation, as in Ref.~\cite{venzor2023resonant}, by using $f_\nu^{(0)}(p) = (e^{p/T_\nu} + 1)^{-1}$ (see Sec.~S2 of the SuM for detailed derivations).

For resonant $\nu$SI, we adopt the narrow-width approximation (NWA),
    \begin{equation}
        \Gamma_{\nu\nu \to \nu\nu} = g^2\frac{ \mu_\varphi^2 T_\nu}{12\pi \zeta(3)} F(\mu_\varphi^2) \quad \text{with} \quad \mu_\varphi= \frac{m_\varphi}{T_\nu}\, . \label{Resonat-NSI}
    \end{equation}
Here, we employed the complex-mass scheme, including the total decay width $\Gamma_\varphi = 3g^2m_\varphi/16\pi$ to properly account for the resonant pole\footnote{The complex-mass scheme avoids ambiguities in the calculation of cross sections and differs from Breit–Wigner approaches that sometimes omit a consistent treatment of all allowed channels and couplings~\cite{Creque2021, venzor2023resonant}. We verified that the NWA agrees with the exact result for $g \lesssim 10^{-1}$.} (see, e.g., Refs.~\cite{Denner1999predictions, Willenbrock2024mass}).
Meanwhile, the pair-annihilation $\nu\nu \to \varphi \varphi$ has the interaction rate:
    \begin{align} 
        \Gamma_{\nu \nu \to \varphi \varphi} &= \frac{g^4 T_\nu}{96 \pi^3 \zeta(3)} \int_{4\mu_\varphi^2}^\infty V(\tilde{s}, \mu_\varphi) F(\tilde{s}) \, d\tilde{s} \,, \label{Rate-pair-annihilation}\\
        V(\tilde{s}, \mu_\varphi) &= \left[\frac{\tilde{s}^2 - 4\mu_\varphi^2 \tilde{s} + 6\mu_\varphi^4}{\tilde{s}(\tilde{s} - 2\mu_\varphi^2)} \right] \ln \frac{(1+\beta_\varphi)}{(1-\beta_\varphi)} - \beta_\varphi \,,
    \end{align}
where $\beta_\varphi = \sqrt{1-4 \mu_\varphi^2/\tilde{s}}$ and $\tilde{s}=s/T_\nu^2$.

\begin{figure} [t!]
\includegraphics[width=1\linewidth]{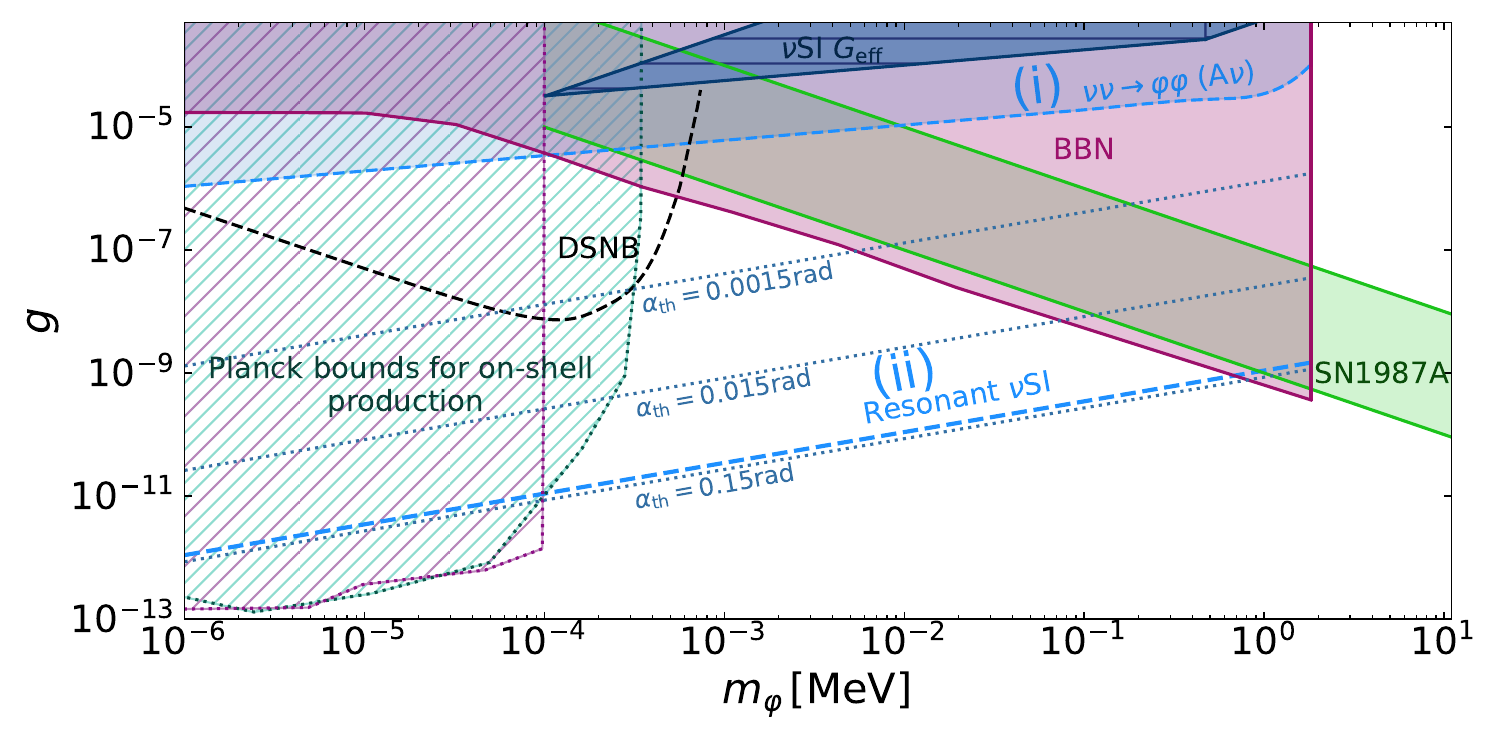}
\caption{\label{fig:PS} Parameter space in the $g\text{--}m_\varphi$ plane with regions: (i) pair-annihilation (A$\nu$), the HML for $\log_{10}(G_{\rm eff} {\rm MeV}^{2}) \in [-6.5,-1.0]$ ($\nu$SI $G_{\rm eff}$)~\cite{Kreisch2020, Poudou:2025qcx, He:2025jwp, Perez2026towards}, BBN bounds~\cite{Huang2018}, Supernova 1987A bounds~\cite{Fiorillo2022}, signatures of testable $\nu$SI in the DSNB~\cite{Wang2025probing}. We also show Planck bounds from thermalization via decays~\cite{Escudero2020CMBsearch}, or ``equivalently'' resonant $\nu$SI~\cite{venzor2023resonant} (dotted lines), together with the less-explored parameter space (ii) for resonant $\nu$SI (with $g$ defined between our lower limit in Eq.~\eqref{g-low} and the region (i)) compared with $g$ determined by the angular suppression $\alpha_{\rm th}$ of on-shell production.}
\end{figure}

In Fig.~\ref{fig:PS}, 
we highlight the region
$\Gamma_{\nu\nu\to\varphi\varphi}\gtrsim H$ after neutrino--photon decoupling in the $g$--$m_\varphi$ plane, indicating where the interaction can modify the background evolution. Once the scalar population becomes non-negligible, its interactions with neutrinos, $\nu\varphi\leftrightarrow\nu\varphi$, must also be included. The figure also shows the region where $\alpha_{\rm th}$
characterizes the angular suppression of scalar production before thermalization via $2 \leftrightarrow 1$ processes; we discuss its definition and implications below.

For couplings satisfying $\Gamma_{\nu\nu\to\varphi\varphi} < H$,
the dominant phenomenon is typically described by $2 \leftrightarrow 1$ processes. Their contributions
can be separated into three stages \cite{Escudero2020CMBsearch, Li2023neff}: (I) \emph{Population}, in the freeze-in regime, scalars are produced via $\nu\nu \to \varphi$; (II) \emph{Thermalization}, when $\varphi \to \nu\nu$ equilibrates with $\nu\nu \to \varphi$;
and (III) \emph{Depletion}, when $\nu\nu \to \varphi$ becomes inefficient and since their lifetime is short compared to cosmological timescales, all $\varphi$ particles decay into neutrinos. Based on this evolution, one might include
decays at the background level along with the $2 \leftrightarrow 2$ processes~\cite{Huang2018}; however,
we notice that, near the resonance, doing both counts the same physical process twice.

\begin{figure} [t!]
\includegraphics[width=1\linewidth]{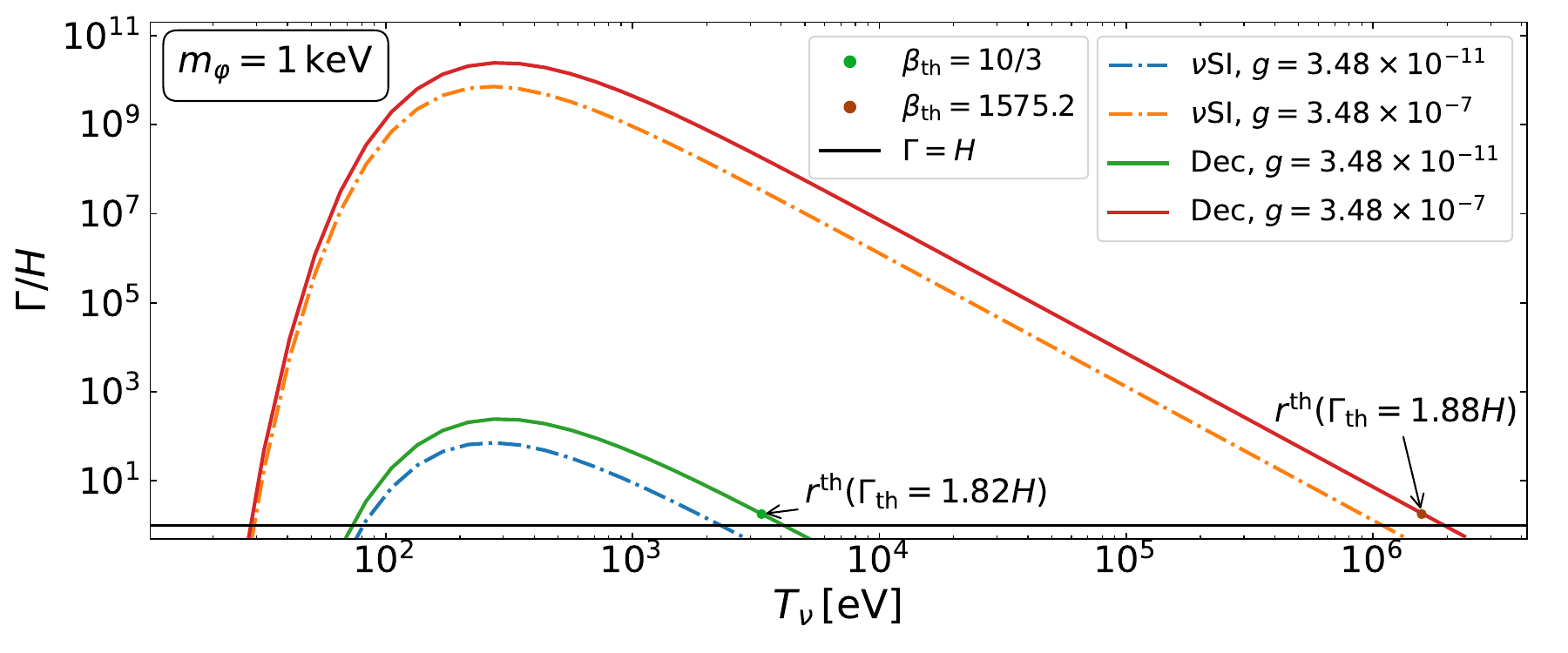}
\caption{Interaction rates used in the RTA: resonant $\nu$SI in the NWA and $\nu\text{--}\varphi$ interaction (Dec) in the MB approximation for $\varphi \leftrightarrow \nu\nu$~\cite{Escudero2020CMBsearch}. For the latter case, it also describes $\nu$–$\varphi$ thermalization in the instantaneous limit.}
\label{fig:Rates}
\end{figure}

Our first clue is that the interaction rate~\eqref{Resonat-NSI} is similar to that for decays (see Fig.~\ref{fig:Rates}). This is not surprising, since the squared $s$-channel amplitude for $\nu\nu \to \nu\nu$ factorizes into a product of decay amplitudes,
    \begin{equation}
        |{\cal M}_{\nu \nu \to \nu\nu}|^2  \bigg|_{\sqrt{s} \sim  m_\varphi} = \frac{|{\cal M}_{\nu \nu \to \varphi}|^2 \, |{\cal M}_{\varphi \to \nu\nu}|^2}{(s - m_\varphi^2)^2 + m_\varphi^2 \Gamma_{\varphi}^2} \, ,
    \end{equation}
which, in the NWA, reduces to a decay amplitude. 
In other words, the resonant $2\to2$ process can be understood as $\nu\nu \to \varphi$ followed by $\varphi \to \nu \nu$.
Therefore, including both 
processes results in double counting.
Thus, a natural question arises: Should we consider thermalization via decays or resonant $\nu$SI?

First, we will study the $2 \leftrightarrow 1$ processes using the Maxwell–Boltzmann (MB) statistics to simplify the analysis, and show that this scenario is not self-consistent
if thermalization occurs at high temperature ($T_\nu \gg m_\varphi/3$).

In a comoving volume, the effective total number of $\varphi$ and $\nu$ (per chirality) particles is conserved when the interaction $\varphi \leftrightarrow \nu\nu$ dominates over other neutrino interactions~\cite{Li2023neff}.
This means $(\bar{n}_\varphi + N_\nu \bar{n}_\nu) a^3(t) = {\rm const.}$, where $\bar{n}_\varphi$ is the number density of the $\varphi$ particle, $N_\nu =3$ for three mass states, and $\bar{n}_\nu$ is the neutrino number density per degree of freedom. Then, we define the \emph{relative fraction of populated $\varphi$ particles} as
    \begin{equation}
        r := \frac{\bar{n}_\varphi}{(\bar{n}_\varphi + N_\nu \bar{n}_\nu)} \,.
        \label{fr_def}
    \end{equation}
If the scalar production begins at $T_{\nu,0} = \beta_0 m_\varphi$ and $a(t) \propto T_\nu^{-1}$ with $T_{\nu} = \beta m_\varphi$ during stage I, from the conservation of the effective total number, we can estimate the number density of the produced scalars as $\bar{n}_\varphi(\beta) = N_\nu r(\beta) n_{\rm MB}$, where $n_{\rm MB}$ is the MB number density and $r(\beta)$ acts as an attenuation term.

If instantaneous $\nu$--$\varphi$ thermalization starts at $T_{\nu,{\rm th}} = \beta_{\rm th} m_\varphi$ ($\beta_0 \gg \beta_{\rm th}$), it has the following implications (see technical details in Sec.~S3.2 of the SuM):
\begin{enumerate}
    \item The collision term at zeroth order vanishes; this leads to $r^{\rm th} \approx 0.209$ during stage II.
    \item For fixed $\beta_{\rm th}$ ($\gtrsim 10/3$), the thermalization fraction $r^{\rm th}$ can always be reached by a high enough coupling $g$, namely
    \begin{equation}
        g = \left[\frac{32\pi g_H}{M_{\rm pl}} \left(\frac{r^{\rm th}}{I(\beta_{\rm th}, \beta_0)} \right) m_\varphi \right]^{1/2} \, , \label{g}
    \end{equation}
where $g_H = (4\pi^3 g_{*}(T_\nu)/5)^{1/2}/3$ depends on the effective number of relativistic degrees of freedom $g_{*}(T_\nu)$, $M_{\rm pl}$ denotes the Planck mass, and
    \begin{equation}
        I(\beta_{\rm th}, \beta_0) = \int_{\beta_{\rm th}}^{\beta_0} {\rm z}^{-5} K_1(1/{\rm z}) \, d{\rm z} \, .
    \end{equation}
    At the start of stage II, we need to make sure that the interaction is efficient enough. For the fixed coupling~\eqref{g}, we obtain that $\Gamma_{\rm th}\gtrsim 1.8 H$ where $\Gamma_{\rm th} = \Gamma_\varphi \mu_\varphi^2 K_1(\mu_\varphi)/2 |_{T_\nu = T_{\nu, {\rm th}}}$ (see Fig.~\ref{fig:Rates}).
    \item During the radiation-dominated era, for $\beta_{\rm th} \gtrsim 10/3$, the boosted lifetime of the produced scalar satisfies, on average, $\tau_\varphi^\prime \gtrsim 3(1-r^{\rm th}) \, t_{\rm th}$, where $t_{\rm th} = M_{\rm pl}[2g_H \beta_{\rm th}^2 m_\varphi^2]^{-1}$ is the cosmic time at which the $\nu\text{--}\varphi$ fluid thermalizes.
\end{enumerate}
Point 3 implies that decays $\varphi \to \nu\nu$ near $T_{\nu, {\rm th}}$ cannot balance the inverse neutrino decays because the production is most efficient as long as $T_\nu \to m_\varphi$, which contradicts the fact that $r^{\rm th}$ freezes during stage II.

Moreover, scalar production via $\nu\nu \to \varphi$ is, on average, temperature-suppressed because the incident angle of neutrinos is restricted to $\alpha_p \lesssim \alpha_{\rm th} = [3 (1-r^{\rm th})\beta_{\rm th}]^{-1}$ (which fixes the coupling through Eq.~\eqref{g}, see Fig.~\ref{fig:PS}). Therefore, we have overestimated scalar production in the freeze-in regime, where we also neglect $\varphi$ decays to evolve~\eqref{fr_def}, leading to spurious thermalization.

Approaches that evolve chemical potentials~\cite{Abenza2020precision, Escudero2020CMBsearch} 
establish a $\nu\text{--}\varphi$ equilibrium  while $\varphi$ is relativistic; this sets the
initial conditions that yield $\Delta N_{\rm eff}^{\rm CMB} = 0.12$~\cite{Abenza2020precision, Kelly2021intimate, Li2023neff} (see Sec.~S3.3 of the SuM). In the first-order collision term, however, on-shell production is kinematically suppressed at
$T_\nu \gg m_\varphi/3$, precisely where the estimate~\eqref{Resonat-NSI} of the interaction rate fails. So the scalars are produced efficiently near $T_\nu \sim m_\varphi/3$, without developing a sizable population
(see Sec.~S3 of the SuM and~\cite{Perez2026Addressing}). This leads to $\Delta N_{\rm eff}^{\rm CMB} \approx 0$; the argument extends to other mediator types.
 
For stage II, an effectively massless coupled neutrino-(pseudo)scalar fluid is typically assumed when deriving the BHEs~\cite {Escudero2020CMBsearch, Sandner2023precision},
which
cannot accurately 
describe massive particles. Therefore, we take $\beta_{\rm th} \gtrsim 10/3$ as the threshold beyond which resonant $\nu$SI characterize the system throughout cosmic evolution. In this case,
    \begin{equation}
        g \gtrsim 1.1 \times 10^{-9} \sqrt{\frac{m_\varphi}{\rm MeV}}\, . \label{g-low}
    \end{equation} 
In Fig.~\ref{fig:PS}, within region (i), neutrinos become tightly coupled in the resonant regime, and their contribution must be considered together with non-resonant processes; before matter-radiation equality, the HML with $G_{\rm eff}= g^2/m_\varphi^2$ delays the neutrino free-streaming whenever\footnote{This bound comes from $\Gamma_\ell = G_{\rm eff}^2 T_\nu^5 \alpha_\ell/4 \gtrsim H$, together with the validity of the HML at $T_\nu < m_\varphi/100$~\cite{Perez2026towards}. In Fig.~\ref{fig:PS}, we have only considered one contribution with $\alpha_{\ell\geq 2} \approx 0.19$ (see Sec. S4.2 of the SuM).}
    \begin{equation}
        g \gtrsim \left(\frac{4 g_H}{(12.2) \alpha_\ell} \right)^{1/4} 10^{-3.75} \left( \frac{m_\varphi}{{\rm MeV}} \right)^{1/4} \, .
    \end{equation}
For region (ii), between $\Gamma_{\nu\nu \to \varphi\varphi}<H$ and $g$ above the lower bound in Eq.~\eqref{g-low}, resonant $\nu$SI is the only relevant process. This region overlaps with BBN bounds~\cite{Huang2018}, which come from thermalization via
decays in the relativistic regime; if on-shell production is suppressed there, as argued above, they would be relaxed. 

\lettersection{Outlook for cosmic neutrino self-interactions}
From Eqs.~\eqref{BHE-0-CN-S}, we discuss the two regions of cosmic $\nu$SI.

\begin{figure}
\includegraphics[width=1\linewidth]{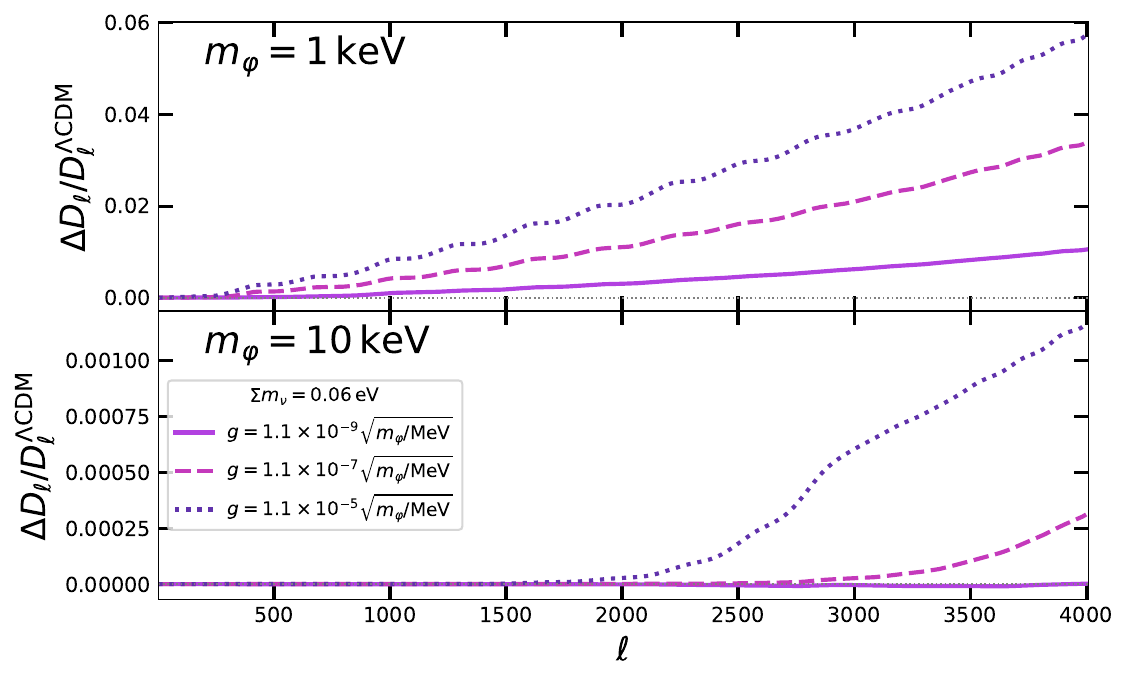}
\addtocounter{footnote}{1}
\caption{\label{fig:CMB} Resonant $\nu$SI effects on the CMB TT power spectrum for a scalar mediator of mass $m_\varphi$ within the RTA. As the coupling $g$ increases, neutrino free-streaming sets in at later times (lower redshifts), 
progressively affecting lower CMB multipoles. Spectra computed with the public code~\cite{venzor_2026_res_class}.
}
\end{figure}

\noindent{\textit{Cosmological impact of resonant $\nu$SI---}}Deriving the BHEs as Eqs.~\eqref{BHE-0-CN-S} neglects neutrino masses. By contrast, within the RTA with $C_{\nu_i}^{(0)} \approx 0$, one can work with the Legendre multipole moments $\Psi_{i,\ell}$ of the perturbation to $f_{\nu_i}^{(0)}$.\footnote{This approach has been used to study $\nu$SI between massive neutrinos~\cite{Choudhury2021, Venzor2022, Taule2022global, venzor2023resonant, Noriega2025resonant} (see also Fig.~\ref{fig:CMB}). Here, the Legendre multipole moments $\Psi_{i,\ell}$ are defined in Fourier space through the expansion of $\tilde{\Theta}_{\nu_i} = {\cal F}[\Theta_{\nu_i}]$, where $f_{\nu_i} = f_{\nu_i}^{(0)}[1 + \Theta_{\nu_i}]$.} 
Beyond approximate regimes, the non-trivial temperature dependence of the first-order collision terms~\cite{Perez2026towards, Perez2026Addressing} motivates implementing $\nu$SI in the BHEs through an \emph{effective relaxation-time approximation} (eRTA), $\Gamma \Psi_{i,\ell}\to \sum_j \Gamma_{ij,\ell}^{(r)} \Psi_{j,\ell}$, which leads to the Boltzmann hierarchy,
    \begin{equation}
    \begin{aligned}
        \dot{\Psi}_{i,\ell}
    &= \frac{q_i k}{(2\ell + 1)\epsilon_i}
    \left(
    \ell \Psi_{i,\ell-1}- (\ell+1)\Psi_{i,\ell+1}\right) \\
    &\quad - \frac{d \ln f_{\nu_i}^{(0)}}{d \ln q_i} f_{g,\ell}^i
    - a\sum_j \Gamma_{ij,\ell}^{(r)}\Psi_{j,\ell}\, .
    \end{aligned}
    \end{equation}
Here, dots denote derivatives with respect to the conformal time, and the index $i$ labels the neutrino mass eigenstates with comoving energies $\epsilon_i = (a^2 m_i^2 + q_i^2)^{1/2}$. Following the notation introduced in Ref.~\cite{Perez2026towards}, $f_{g,\ell} \to f_{g,\ell}^i$ ($\epsilon \to \epsilon_i$) is gauge-dependent, and the terms $\Gamma_{ij,\ell}^{(r)}$, which are kinematically suppressed at $T_\nu \gg m_\varphi/3$, 
are derived within the thermal approximation in Ref.~\cite{Perez2026Addressing}.

Using the RTA, we explore resonant $\nu$SI beyond the regime $m_\varphi \lesssim 10^2 \, {\rm eV}$~\cite{venzor2023resonant, Noriega2025resonant}. For $m_\varphi \lesssim 1 \, {\rm eV}$, neutrino masses must be included in the interaction rate, and for $g > 1.1 \times 10^{-5} \,(m_\varphi/{\rm MeV})^{1/2}$ the standard \texttt{CLASS} solver~\cite{Blas2011cosmic, Lesgourgues2011cosmic} requires modifications due to the tightly coupled regime induced by resonant $\nu$SI. We find that the largest enhancement of the CMB TT spectrum occurs for $m_\varphi \lesssim 1 \, {\rm keV}$ (see Fig.~\ref{fig:CMB}), whereas for $m_\varphi > 1 \, {\rm keV}$, the effect is expected to be compatible with $\Lambda$CDM, similarly to the HML case with $\log_{10}(G_{\rm eff} {\rm MeV}^{2}) \ll -2.5$~\cite{Lancaster2017, Poudou:2025qcx}.

In extended models, when the mediator of mass $M$ couples predominantly to a fermion $X$ with coupling $g_X$ rather than to neutrinos, its total decay width is $\Gamma_M \propto g_X^2 M$. In the NWA, the interaction rate for $\nu$SI  scales as $g^4/g_X^2$, accompanied by resonant neutrino conversion into $X$, followed by $X\text{--}\nu$ thermalization. Our analysis suggests that neutrino conversion does not significantly affect the CMB spectrum for $M > 1 \,{\rm keV}$, relevant to the scenarios in Refs.~\cite{Farzan2016neutrinos, Das2025impostor}.

Equation~\eqref{L-scalar}, along with its generalizations, arises in a variety of BSM scenarios, including neutrino dark matter portals~\cite{Berlin2018thermal, Berlin2019thermal, Kelly2021intimate}, dark sectors~\cite{Aloni2023dark, Benso2025large, Das2025impostor}, models with new symmetries~\cite{Chacko2004cmb, Escudero2019cosmology, Dror2020discovering, Kelly2020origin, Arias2021neutrino, Araki2021resolving, Fernandez2021inverse, Esseili2024cosmological, Asai2024contribution, Li2024cosmological, Escudero2023seesaw, De2024minimal}, other models mentioned above, and references in~\cite{Berryman2023neutrino}. This opens a broad parameter space where mediators with $M \sim 10^{-6}-1 \, {\rm MeV}$ can be probed through resonant $\nu$SI, even when the mediator is weakly resonant~\cite{Foroughi2025enabling}.

\noindent{\textit{The heavy mediator paradigm---}}
The HML has been widely explored (see, e.g., Refs.~\cite{CyrRacine2014, Archidiacono2014, Oldengott2015, Lancaster2017, Oldengott2017, Park2019, Kreisch2020, Das2021, Choudhury2021, Brinckmann2020self, Mazumdar2022flavour, Kumar_2022, Kreisch:2022zxp, Das:2023npl, Camarena2023confronting, He2024self, Camarena2025strong,  Poudou:2025qcx, Calabrese2025atacama, He:2025jwp, Noriega:2026uhw}),
while neglecting other processes allowed by the SNI Lagrangian~\eqref{L-scalar}, see Fig.~\ref{fig:PS}. Including these processes
modifies the neutrino distribution function at high redshift, leading to a nonzero $\Delta$ ($\equiv \Delta_i$) in the thermal approximation (see Eq.~\eqref{Delta}): $\Delta > 0$ corresponds to the loss of neutrino energy into the production of scalars; a thermalization phase follows; once this phase ends, energy is transferred back to the neutrino sector ($\Delta < 0$), rendering the thermal approximation ill-defined unless $\Delta$ is sufficiently small. As scalar production ceases at $T_\nu \sim m_\varphi/3$, $\varphi$ decays reheat the neutrino sector. This rapidly drives $\Delta \to 0$, while slightly modifying the $\alpha_\ell$ coefficients due to the rapid reheating process (see details in Sec.~S4 of the SuM).

Motivated by cosmological data hinting at strong $\nu$SI, Ref.~\cite{Das2025impostor} proposed a dark sector consisting of a scalar $\phi$ and $n_\chi$ Majorana fermions to reproduce the HML. In this scenario, the additional processes modify the background. Once a population of $\chi$ fermions is produced, their BHEs must be constructed; Ref.~\cite{Das2025impostor} adopts $\alpha_{\ell \geq 2} \approx 0.424$. 
However, even in the MB-like approximation, we obtain much smaller values, 
$\alpha_{\ell \geq 2} \sim (0.0354, 0.0387)$ ($\sim (3.297, 3.601) \times 10^{-3}$) for $n_\chi=1$ ($=40$), where $T_\chi \approx 2T_\gamma/3 \, (\approx 4T_\gamma/11)$. This is because the resulting distribution function of the $\chi$ particles yields a lower energy density than the exact thermal form while preserving $N_{\rm eff} (\approx 3.04)$. As $n_\chi$ increases, $\alpha_\ell$ decreases further, and the relation between $T_\chi$ and redshift is modified. To compensate for this effect, a larger $G_{{\rm eff},\chi} = \lambda_{\phi\chi}^2/m_\phi^2$ is required to mimic $\nu$SI, increasing the coupling $\lambda_{\phi\chi}$ between the $\phi$ and $\chi$ (see Sec.~S4.3 of the SuM).

\lettersection{Summary and conclusions}
In this \emph{Letter}, we have proposed modified BHEs~\eqref{BHE-0-CN-S} for non-trivial background evolution to study the non-standard thermal history of active neutrinos, allowing us to reveal previously unknown features of cosmological $\nu$SI. We have shown that the low-energy limit of renormalizable theories that allow non-standard $\nu$SI imply resonant mediator
behavior due to on-shell mediator production, $s= M^2$, with $M \sim 10^{-6} - 1 \, {\rm MeV}$, rather than the phenomena described by the known frameworks based on $2\leftrightarrow1$ processes, motivating a reexamination of the BBN bounds on SNI. 
We identified two regions for $\nu$SI: (i) $\nu$SI have resonant and non-resonant contributions with background evolution induced by neutrino--mediator interactions; in particular, we argued that rapid energy restoration in the neutrino sector, which is driven by $\varphi \to \nu\nu$ in SNI scenarios, only mildly affects the HML, and (ii) the background remains unchanged, while resonant $\nu$SI dominates the dynamics with $\Delta N_{\rm eff}^{\rm CMB} \approx 0$. This region enables exploration of a wide range of couplings, mediator masses, and mediator types, with cosmological observations in the era of precision cosmology.
Our modified BHEs can be naturally extended to study self-interacting dark radiation scenarios~\cite{Das2025impostor}; we showed that larger couplings than previously estimated are required for such interactions to mimic $\nu$SI.

\lettersection{Acknowledgments}
I. P. C. thanks Secihti for the scholarship as ``Ayudante de Investigador SNII III''. J. V.  acknowledges support from a Secihti postdoctoral fellowship.
P. R. acknowledges Conahcyt-Secihti (Mexico) funding, through the project CBF2023-2024-3226.

\bibliography{main}

@article{Berryman2023neutrino,
  title="{Neutrino self-interactions: A white paper}",
  author={Berryman, Jeffrey M and others},
  journal={Phys. Dark Univ.},
  volume={42},
  pages={101267},
  year={2023},
  publisher={Elsevier},
  doi={10.1016/j.dark.2023.101267},
  url={https://doi.org/10.1016/j.dark.2023.101267},
  eprint={2203.01955},
  archivePrefix = "arXiv",
  primaryClass = "hep-ph"
}

@article{Ng2014cosmic,
  title={Cosmic neutrino cascades from secret neutrino interactions},
  author={Ng, Kenny CY and Beacom, John F},
  journal={Phys. Rev. D},
  volume={90},
  number={6},
  pages={065035},
  year={2014},
  publisher={APS},
  doi= {10.1103/PhysRevD.90.065035},
  url={https://link.aps.org/doi/10.1103/PhysRevD.90.065035},
  eprint = {1404.2288},
  archivePrefix = "arXiv",
  primaryClass = "astro-ph.HE"
}

@article{Ioka2014icecube,
  title="{IceCube PeV--EeV neutrinos and secret interactions of neutrinos}",
  author={Ioka, Kunihito and Murase, Kohta},
  journal={PTEP},
  volume={2014},
  number={6},
  pages={061E01},
  year={2014},
  publisher={Oxford University Press},
  doi={10.1093/ptep/ptu090},
  url={https://doi.org/10.1093/ptep/ptu090},
  eprint = {1404.2279},
  archivePrefix = "arXiv",
  primaryClass = "astro-ph.HE"
}

@article{Bustamante2020bounds,
  title={Bounds on secret neutrino interactions from high-energy astrophysical neutrinos},
  author={Bustamante, Mauricio and Rosenstr{\o}m, Charlotte and Shalgar, Shashank and Tamborra, Irene},
  journal={Phys. Rev. D},
  volume={101},
  number={12},
  pages={123024},
  year={2020},
  publisher={APS},
  doi={10.1103/PhysRevD.101.123024},
  url={https://link.aps.org/10.1103/PhysRevD.101.123024},
  eprint={2001.04994},
  archivePrefix = "arXiv",
  primaryClass = "astro-ph.HE"
}

@article{Creque2021,
  title = {Resonant neutrino self-interactions},
  author = {Creque-Sarbinowski, Cyril and Hyde, Jeffrey and Kamionkowski, Marc},
  journal = {Phys. Rev. D},
  volume = {103},
  issue = {2},
  pages = {023527},
  numpages = {11},
  year = {2021},
  month = {Jan},
  publisher = {American Physical Society},
  doi = {10.1103/PhysRevD.103.023527},
  url = {https://link.aps.org/doi/10.1103/PhysRevD.103.023527},
  eprint={2005.05332},
  archivePrefix = "arXiv",
  primaryClass = "hep-ph"
}

@article{Esteban2021probing,
  title = {Probing secret interactions of astrophysical neutrinos in the high-statistics era},
  author = {Esteban, Ivan and Pandey, Sujata and Brdar, Vedran and Beacom, John F.},
  journal = {Phys. Rev. D},
  volume = {104},
  issue = {12},
  pages = {123014},
  numpages = {17},
  year = {2021},
  month = {Dec},
  publisher = {American Physical Society},
  doi = {10.1103/PhysRevD.104.123014},
  url = {https://link.aps.org/doi/10.1103/PhysRevD.104.123014},
  eprint = "2107.13568",
  archivePrefix = "arXiv",
  primaryClass = "hep-ph"
}

@article{Carpio2023simulating,
  title={Simulating neutrino echoes induced by secret neutrino interactions},
  author={Carpio, Jose Alonso and Murase, Kohta},
  journal={JCAP},
  volume={2023},
  number={02},
  pages={042},
  year={2023},
  publisher={IOP Publishing},
  doi={10.1088/1475-7516/2023/02/042},
  url={https://iopscience.iop.org/article/10.1088/1475-7516/2023/02/042},
  eprint = "2204.09029",
  archivePrefix = "arXiv",
  primaryClass = "hep-ph"
}

@article{Chang2022towards,
  title = "{Toward Powerful Probes of Neutrino Self-Interactions in Supernovae}",
  author = {Chang, Po-Wen and Esteban, Ivan and Beacom, John F. and Thompson, Todd A. and Hirata, Christopher M.},
  journal = {Phys. Rev. Lett.},
  volume = {131},
  issue = {7},
  pages = {071002},
  numpages = {8},
  year = {2023},
  month = {Aug},
  publisher = {American Physical Society},
  doi = {10.1103/PhysRevLett.131.071002},
  url = {https://link.aps.org/doi/10.1103/PhysRevLett.131.071002},
  eprint = {2206.12426},
  archivePrefix = "arXiv",
  primaryClass = "hep-ph"
}

@article{Cerdeno2023constraints,
  title = {Constraints from the duration of supernova neutrino burst on on-shell light gauge boson production by neutrinos},
  author = {Cerde\~no, David and Cerme\~no, Marina and Farzan, Yasaman},
  journal = {Phys. Rev. D},
  volume = {107},
  issue = {12},
  pages = {123012},
  numpages = {15},
  year = {2023},
  month = {Jun},
  publisher = {American Physical Society},
  doi = {10.1103/PhysRevD.107.123012},
  url = {https://link.aps.org/doi/10.1103/PhysRevD.107.123012},
  eprint = {2301.00661},
  archivePrefix = "arXiv",
  primaryClass = "hep-ph"
}

@article{Hooper2023signals,
  title="{Signals of a new gauge boson from IceCube and the muon $g-2$}",
  author={Hooper, Dan and Juan, Joaquim Iguaz and Serpico, Pasquale D},
  journal={Phys. Rev. D},
  volume={108},
  number={2},
  pages={023007},
  year={2023},
  publisher={APS},
  doi={10.1103/PhysRevD.108.023007 },
  url={https://doi.org/10.1103/PhysRevD.108.023007},
  eprint = "2302.03571",
  archivePrefix = "arXiv",
  primaryClass = "astro-ph.HE"
}

@article{Wu2024shedding,
  title={Shedding light on neutrino self-interactions with solar antineutrino searches},
  author={Wu, Quan-feng and Xu, Xun-Jie},
  journal={JCAP},
  volume={2024},
  number={02},
  pages={037},
  year={2024},
  publisher={IOP Publishing},
  doi={10.1088/1475-7516/2024/02/037},
  url={https://iopscience.iop.org/article/10.1088/1475-7516/2024/02/037},
  eprint = "2308.15849",
  archivePrefix = "arXiv",
  primaryClass = "hep-ph"
}

@article{Dhuria2024probing,
  title="{Probing self-interacting ultrahigh-energy neutrinos with the cosmic 21-cm signal}",
  author={Dhuria, Mansi and Gupta Teli, Bishnu},
  journal={Phys. Rev. D},
  volume={110},
  number={12},
  pages={123033},
  year={2024},
  publisher={APS},
  doi={10.1103/PhysRevD.110.123033},
  url={https://doi.org/10.1103/PhysRevD.110.123033},
  eprint = "2406.19279",
  archivePrefix = "arXiv",
  primaryClass = "hep-ph"
}

@article{De2024mu,
  title="{$L_\mu-L_\tau$ solution to the IceCube ultrahigh-energy neutrino deficit in light of NA64}",
  author={de la Vega, Leon MG and Peinado, Eduardo and Wudka, Jos{\'e}},
  journal={Phys. Rev. D},
  volume={110},
  number={9},
  pages={095032},
  year={2024},
  publisher={APS},
  doi={10.1103/PhysRevD.110.095032},
  url={https://journals.aps.org/prd/abstract/10.1103/PhysRevD.110.095032},
  eprint = "2406.19968",
  archivePrefix = "arXiv",
  primaryClass = "hep-ph"
}

@article{Wang2025probing,
  title="{Probing a New Regime of Neutrino Self-Interactions with Astrophysical Neutrinos and the Relativistic Cosmic Neutrino Background}",
  author={Wang, Isaac R and Xu, Xun-Jie and Zhou, Bei},
  journal={Phys. Rev. Lett.},
  volume={135},
  number={18},
  pages={181002},
  year={2025},
  publisher={APS},
  doi = {doi.org/10.1103/9ddp-j1z9},
  url = {https://link.aps.org/doi/10.1103/9ddp-j1z9},
  eprint = "2501.07624",
  archivePrefix = "arXiv",
  primaryClass = "hep-ph"
}

@article{Francener2025probing,
  title="{Probing a low-mass $Z^\prime$ gauge boson at IceCube and prospects for IceCube-Gen2}",
  author={Francener, Reinaldo and Gon{\c{c}}alves, Victor P and Gratieri, Diego R},
  journal={Phys. Rev. D},
  volume={111},
  number={9},
  pages={095005},
  year={2025},
  publisher={APS},
  doi={10.1103/PhysRevD.111.095005},
  url={https://doi.org/10.1103/PhysRevD.111.095005},
  eprint = "2502.19338",
  archivePrefix = "arXiv",
  primaryClass = "hep-ph"
}

@article{He2025implications,
  title="{Implications of the KM3NeT Ultrahigh-energy Event on Neutrino Self-interactions}",
  author={He, Yuxuan and Liu, Jia and Wang, Xiao-Ping and Zhong, Yi-Ming},
  journal={Phys. Rev. D},
  volume={113},
  number={4},
  pages={043022},
  year={2026},
  publisher={APS},
  doi={10.1103/bb85-7kyf},
  url={https://doi.org/10.1103/bb85-7kyf},
  eprint = "2504.20163",
  archivePrefix = "arXiv",
  primaryClass = "hep-ph"
}

@article{Natwariya2026constraining,
  title={Constraining self-interacting ultrahigh-energy muon neutrinos by cosmic microwave background spectral distortion},
  author={Natwariya, Pravin Kumar and Si, Shibsankar and Nayak, Alekha C and Srivastava, Tripurari},
  journal={Phys. Rev. D},
  volume={113},
  number={10},
  pages={103017},
  year={2026},
  publisher={APS},
  doi={10.1103/jsf2-krmq},
  url={https://doi.org/10.1103/jsf2-krmq},
  eprint = "2506.24110",
  archivePrefix = "arXiv",
  primaryClass = "astro-ph.CO"
}

@article{Machado:2025ltu,
    author = "Machado, Pedro A. N. and Wang, Isaac R. and Xu, Xun-Jie and Zhou, Bei",
    title = "{Widen the resonance at ultra-high energies: novel probes of neutrino self-interactions in the high-mass regime}",
    reportNumber = "FERMILAB-PUB-25-0853-T",
    doi = "10.1088/1475-7516/2026/07/064",
    journal = "JCAP",
    volume = "2026",
    number={07},
    pages = "064",
    year = "2026",
    url={https://iopscience.iop.org/article/10.1088/1475-7516/2026/07/064},
    eprint = "2512.00165",
    archivePrefix = "arXiv",
    primaryClass = "hep-ph",
}

@article{Bharadwaj2026diffuse,
  title="{Diffuse Supernova Neutrinos with Secret Neutrino Interactions}",
  author={Bharadwaj, Praveen and Chattopadhyay, Utpal and Ghosh, Dilip Kumar and Sarker, Arnab},
  journal={JCAP},
  volume={2026},
  number={10},
  pages={023},
  year={2026},
  publisher={IOP Publishing},
  doi = {10.1088/1475-7516/2026/10/023},
  url = {https://doi.org/10.1088/1475-7516/2026/10/023},
  eprint = "2606.22898",
  archivePrefix = "arXiv",
  primaryClass = "hep-ph"
}

@article{Das2026sn1987a,
  title="{SN1987A Constraints of Light $Z^\prime$ with Non-Mixing Polarisations}",
  author={Das, Debottam and Ghosh, Purusottam and Puri, Rahul},
  journal={},
  year={2026},
  eprint = "2606.23769",
  archivePrefix = "arXiv",
  primaryClass = "hep-ph"
}

@article{Heurtier2017supernova,
  title={Supernova constraints on massive (pseudo) scalar coupling to neutrinos},
  author={Heurtier, Lucien and Zhang, Yongchao},
  journal={JCAP},
  volume={2017},
  number={02},
  pages={042--042},
  year={2017},
  doi={10.1088/1475-7516/2017/02/042},
  url={https://iopscience.iop.org/article/10.1088/1475-7516/2017/02/042},
  eprint = "1609.05882",
  archivePrefix = "arXiv",
  primaryClass = "hep-ph"
}

@article{Shalgar2021,
  title = {Core-collapse supernova stymie secret neutrino interactions},
  author = {Shalgar, Shashank and Tamborra, Irene and Bustamante, Mauricio},
  journal = {Phys. Rev. D},
  volume = {103},
  issue = {12},
  pages = {123008},
  numpages = {9},
  year = {2021},
  month = {Jun},
  publisher = {American Physical Society},
  doi = {10.1103/PhysRevD.103.123008},
  url = {https://link.aps.org/doi/10.1103/PhysRevD.103.123008},
  eprint = "1912.09115",
  archivePrefix = "arXiv",
  primaryClass="astro-ph.HE"
}

@Article{Akita2022,
  author={Akita, Kensuke
  and Im, Sang Hui
  and Masud, Mehedi},
  title={Probing non-standard neutrino interactions with a light boson from next galactic and diffuse supernova neutrinos},
  journal={JHEP},
  year={2022},
  month={Dec},
  day={09},
  volume={2022},
  number={12},
  pages={50},
  issn={1029-8479},
  doi={10.1007/JHEP12(2022)050},
  url={https://doi.org/10.1007/JHEP12(2022)050},
  eprint = {2206.06852},
  archivePrefix = "arXiv",
  primaryClass = "hep-ph"
}

@article{Fiorillo2022,
  title = "{Strong Supernova 1987A Constraints on Bosons Decaying to Neutrinos}",
  author = {Fiorillo, Damiano F. G. and Raffelt, Georg G. and Vitagliano, Edoardo},
  journal = {Phys. Rev. Lett.},
  volume = {131},
  issue = {2},
  pages = {021001},
  numpages = {7},
  year = {2023},
  month = {Jul},
  publisher = {American Physical Society},
  doi = {10.1103/PhysRevLett.131.021001},
  url = {https://link.aps.org/doi/10.1103/PhysRevLett.131.021001},
  eprint = {2209.11773},
  archivePrefix = "arXiv",
  primaryClass = "hep-ph"
}

@article{Doring2024testing,
  title={Testing secret interaction with astrophysical neutrino point sources},
  author={D{\"o}ring, Christian and Vogl, Stefan},
  journal={JCAP},
  volume={2024},
  number={07},
  pages={015},
  year={2024},
  publisher={IOP Publishing},
  doi={10.1088/1475-7516/2024/07/015},
  url={https://iopscience.iop.org/article/10.1088/1475-7516/2024/07/015},
  eprint = "2304.08533",
  archivePrefix = "arXiv",
  primaryClass = "hep-ph"
}

@article{Akita2024limits,
  title="{Limits on heavy neutral leptons, Z$^\prime$ bosons and majorons from high-energy supernova neutrinos}",
  author={Akita, Kensuke and Im, Sang Hui and Masud, Mehedi and Yun, Seokhoon},
  journal={JHEP},
  volume={2024},
  number={7},
  pages={1--38},
  year={2024},
  publisher={Springer},
  doi={10.1007/JHEP07(2024)057},
  url={https://link.springer.com/article/10.1007/JHEP07(2024)057},
  eprint = "2312.13627",
  archivePrefix = "arXiv",
  primaryClass = "hep-ph"
}

@article{CyrRacine2014,
  title = "{Limits on neutrino-neutrino scattering in the early Universe}",
  author = {Cyr-Racine, Francis-Yan and Sigurdson, Kris},
  journal = {Phys. Rev. D},
  volume = {90},
  issue = {12},
  pages = {123533},
  numpages = {8},
  year = {2014},
  month = {Dec},
  publisher = {American Physical Society},
  doi = {10.1103/PhysRevD.90.123533},
  url = {https://link.aps.org/doi/10.1103/PhysRevD.90.123533},
  eprint={1306.1536},
  archivePrefix = "arXiv",
  primaryClass = "astro-ph.CO"
}

@article{Archidiacono2014,
  year = 2014,
  month = {jul},
  publisher = {{IOP} Publishing},
  volume = {2014},
  number = {07},
  pages = {046--046},
  author = {Maria Archidiacono and Steen Hannestad},
  title = "{Updated constraints on non-standard neutrino interactions from Planck}",
  journal = {JCAP},
  doi = {10.1088/1475-7516/2014/07/046},
  url = {https://doi.org/10.1088\%2F1475-7516\%2F2014\%2F07\%2F046},
  eprint={1311.3873},
  archivePrefix = "arXiv",
  primaryClass = "astro-ph.CO"
}

@article{Oldengott2015,
  author = {Isabel M. Oldengott and Cornelius Rampf and Yvonne Y.Y. Wong},
  doi = {10.1088/1475-7516/2015/04/016},
  journal = {JCAP},
  month = {apr},
  number = {04},
  pages = {016},
  publisher = {Springer},
  title = {{Boltzmann hierarchy for interacting neutrinos I: formalism}},
  url = {https://dx.doi.org/10.1088/1475-7516/2015/04/016},
  volume = {2015},
  year = {2015},
  eprint = "1409.1577",
  archivePrefix = "arXiv",
  primaryClass = "astro-ph.CO"
}

@article{Lancaster2017,
  doi = {10.1088/1475-7516/2017/07/033},
  url = {https://doi.org/10.1088\%2F1475-7516\%2F2017\%2F07\%2F033},
  year = 2017,
  month = {jul},
  publisher = {{IOP} Publishing},
  volume = {2017},
  number = {07},
  pages = {033--033},
  author = {Lachlan Lancaster and Francis-Yan Cyr-Racine and Lloyd Knox and Zhen Pan},
  title = {A tale of two modes: neutrino free-streaming in the early universe},
  journal = {JCAP},
  eprint = {1704.06657},
  archivePrefix = "arXiv",
  primaryClass = "astro-ph.CO"
}

@article{Oldengott2017,
  author = {Isabel M. Oldengott and Thomas Tram and Cornelius Rampf and Yvonne Y.Y. Wong},
  doi = {10.1088/1475-7516/2017/11/027},
  journal = {JCAP},
  month = {nov},
  number = {11},
  pages = {027},
  publisher = {},
  title = {Interacting neutrinos in cosmology: exact description and constraints},
  url = {https://dx.doi.org/10.1088/1475-7516/2017/11/027},
  volume = {2017},
  year = {2017},
  eprint = {1706.02123},
  archivePrefix = "arXiv",
  primaryClass = "astro-ph.CO"
}

@article{Kreisch2020,
  title = "{Neutrino puzzle: Anomalies, interactions, and cosmological tensions}",
  author = {Kreisch, Christina D. and Cyr-Racine, Francis-Yan and Dor\'e, Olivier},
  journal = {Phys. Rev. D},
  volume = {101},
  issue = {12},
  pages = {123505},
  numpages = {34},
  year = {2020},
  month = {Jun},
  publisher = {American Physical Society},
  doi = {10.1103/PhysRevD.101.123505},
  url = {https://link.aps.org/doi/10.1103/PhysRevD.101.123505},
  eprint = {1902.00534},
  archivePrefix = "arXiv",
  primaryClass = "astro-ph.CO"
}

@article{Barenboim2019inflation,
  title = {Constraints on inflation with an extended neutrino sector},
  author = {Barenboim, Gabriela and Denton, Peter B. and Oldengott, Isabel M.},
  journal = {Phys. Rev. D},
  volume = {99},
  issue = {8},
  pages = {083515},
  numpages = {9},
  year = {2019},
  month = {Apr},
  publisher = {American Physical Society},
  doi = {10.1103/PhysRevD.99.083515},
  url = {https://link.aps.org/doi/10.1103/PhysRevD.99.083515}
}

@article{Park2019,
  title = "{$\Lambda$CDM or self-interacting neutrinos: How CMB data can tell the two models apart}",
  author = {Park, Minsu and Kreisch, Christina D. and Dunkley, Jo and Hadzhiyska, Boryana and Cyr-Racine, Francis-Yan},
  eprint = "1904.02625",
  archivePrefix = "arXiv",
  journal = {Phys. Rev. D},
  volume = {100},
  issue = {6},
  pages = {063524},
  numpages = {7},
  year = {2019},
  month = {Sep},
  publisher = {American Physical Society},
  doi = {10.1103/PhysRevD.100.063524},
  url = {https://link.aps.org/doi/10.1103/PhysRevD.100.063524},
  primaryClass = "astro-ph.CO"
}

@article{Das2021,
  year = 2021,
  month = {jul},
  publisher = {{IOP} Publishing},
  volume = {2021},
  number = {07},
  pages = {038},
  author = {Anirban Das and Subhajit Ghosh},
  title = {Flavor-specific interaction favors strong neutrino self-coupling in the early universe},
  journal = {JCAP},
  doi = {10.1088/1475-7516/2021/07/038},
  url = {https://doi.org/10.1088/1475-7516/2021/07/038},
  eprint={2011.12315},
  archivePrefix = "arXiv",
  primaryClass = "astro-ph.CO"
}

@article{Mazumdar2022flavour,
  author = {Arindam Mazumdar and Subhendra Mohanty and Priyank Parashari},
  title = "{Flavour specific neutrino self-interaction: $H_0$ tension and IceCube}",
  journal = {JCAP},
  year = {2022},
  month = {oct},
  publisher = {IOP Publishing},
  volume = {2022},
  number = {10},
  pages = {011},
  doi = {10.1088/1475-7516/2022/10/011},
  url = {https://dx.doi.org/10.1088/1475-7516/2022/10/011},
  eprint = "2011.13685",
  archivePrefix = "arXiv",
  primaryClass = "hep-ph"
}

@article{Choudhury2021,
  title = "{Updated constraints on massive neutrino self-interactions from cosmology in light of the $H_0$ tension}",
  author = "Roy Choudhury, Shouvik and Hannestad, Steen and Tram, Thomas",
  journal = "JCAP",
  volume={2021},
  number={03},
  pages = "084",
  year = "2021",
  doi = "10.1088/1475-7516/2021/03/084",
  url={https://iopscience.iop.org/article/10.1088/1475-7516/2021/03/084},
  eprint = "2012.07519",
  archivePrefix = "arXiv",
  primaryClass = "astro-ph.CO"
}

@article{Brinckmann2020self,
  author = {Brinckmann, Thejs and Chang, Jae Hyeok and LoVerde, Marilena},
  title = "{Self-interacting neutrinos, the Hubble parameter tension, and the cosmic microwave background}",
  journal = {Phys. Rev. D},
  volume = {104},
  issue = {6},
  pages = {063523},
  numpages = {40},
  year = {2021},
  month = {Sep},
  publisher = {American Physical Society},
  doi = {10.1103/PhysRevD.104.063523},
  url = {https://link.aps.org/doi/10.1103/PhysRevD.104.063523},
  eprint = "2012.11830",
  archivePrefix = "arXiv",
  primaryClass = "astro-ph.CO"
}

@article{Kumar_2022,
  title = "{Updating non-standard neutrinos properties with Planck-CMB data and full-shape analysis of BOSS and eBOSS galaxies}",
  journal = {JCAP},
  year = {2022},
  month = {sep},
  publisher = {IOP Publishing},
  volume = {2022},
  number = {09},
  pages = {060},
  author = {Kumar, Suresh and Nunes, Rafael C. and Yadav, Priya},
  doi = {10.1088/1475-7516/2022/09/060},
  url = {https://dx.doi.org/10.1088/1475-7516/2022/09/060},
  eprint = {2205.04292},
  archivePrefix = "arXiv",
  primaryClass = "astro-ph.CO"
}

@article{Kreisch:2022zxp,
  author = "Kreisch, Christina D. and others",
  title = "{Atacama Cosmology Telescope: The persistence of neutrino self-interaction in cosmological measurements}",
  doi = "10.1103/PhysRevD.109.043501",
  journal = "Phys. Rev. D",
  volume = "109",
  number = "4",
  pages = "043501",
  year = "2024",
  eprint = "2207.03164",
  archivePrefix = "arXiv",
  primaryClass = "astro-ph.CO"
}

@article{Das:2023npl,
  author = "Das, Anirban and Ghosh, Subhajit",
  title = "{The magnificent ACT of flavor-specific neutrino self-interaction}",
  doi = "10.1088/1475-7516/2023/09/042",
  journal = "JCAP",
  volume={2023},
  number={09},
  pages={042},
  year={2023},
  eprint = "2303.08843",
  archivePrefix = "arXiv",
  primaryClass = "astro-ph.CO"
}

@article{Camarena2023confronting,
  title={Confronting self-interacting neutrinos with the full shape of the galaxy power spectrum},
  author={Camarena, David and Cyr-Racine, Francis-Yan and Houghteling, John},
  journal={Phys. Rev. D},
  volume={108},
  number={10},
  pages={103535},
  year={2023},
  publisher={APS},
  doi={10.1103/PhysRevD.108.103535},
  url={https://doi.org/10.1103/PhysRevD.108.103535},
  eprint = "2309.03941",
  archivePrefix = "arXiv",
  primaryClass = "astro-ph.CO"
}

@article{He2024self,
  title={Self-interacting neutrinos in light of large-scale structure data},
  author={He, Adam and An, Rui and Ivanov, Mikhail M and Gluscevic, Vera},
  journal={Phys. Rev. D},
  volume={109},
  number={10},
  pages={103527},
  year={2024},
  publisher={APS},
  doi= {10.1103/PhysRevD.109.103527},
  url= {https://link.aps.org/doi/10.1103/PhysRevD.109.103527},
  eprint = "2309.03956",
  archivePrefix = "arXiv",
  primaryClass = "astro-ph.CO"
}

@article{Camarena2025strong,
  title={Strong constraints on a simple self-interacting neutrino cosmology},
  author={Camarena, David and Cyr-Racine, Francis-Yan},
  journal={Phys. Rev. D},
  volume={111},
  number={2},
  pages={023504},
  year={2025},
  publisher={APS},
  doi={10.1103/PhysRevD.111.023504},
  url={https://link.aps.org/doi/10.1103/PhysRevD.111.023504},
  eprint = "2403.05496",
  archivePrefix = "arXiv",
  primaryClass = "astro-ph.CO"
}

@article{Poudou:2025qcx,
  author = "Poudou, Ad\`ele and Simon, Th\'eo and Montandon, Thomas and Teixeira, Elsa M. and Poulin, Vivian",
  title = "{Self-interacting neutrinos in light of recent CMB and LSS data}",
  journal = {Phys. Rev. D},
  volume={112},
  number={10},
  pages={103535},
  year={2025},
  publisher={APS},
  doi={10.1103/mljb-42fm},
  url={https://doi.org/10.1103/mljb-42fm},
  eprint = "2503.10485",
  archivePrefix = "arXiv",
  primaryClass = "astro-ph.CO"
}

@article{Calabrese2025atacama,
  title="{The Atacama Cosmology Telescope: DR6 constraints on extended cosmological models}",
  author={Calabrese, Erminia and Hill, J Colin and Jense, Hidde T and La Posta, Adrien and Abril-Cabezas, Irene and Addison, Graeme E and Ade, Peter AR and Aiola, Simone and Alford, Tommy and Alonso, David and others},
  journal={JCAP},
  volume={2025},
  number={11},
  pages={063},
  year={2025},
  publisher={IOP Publishing},
  doi={10.1088/1475-7516/2025/11/063},
  url={https://doi.org/10.1088/1475-7516/2025/11/063},
  eprint = "2503.14454",
  archivePrefix = "arXiv",
  primaryClass = "astro-ph.CO"
}

@article{He:2025jwp,
  author = "He, Adam and Ivanov, Mikhail M. and Bird, Simeon and An, Rui and Gluscevic, Vera",
  title = "{A Fresh Look at Neutrino Self-Interactions With the Lyman-$\alpha$ Forest: Constraints from EFT and PRIYA}",
  journal = {Phys. Rev. D},
  volume={112},
  number={6},
  pages={063540},
  year={2025},
  publisher={APS},
  doi={10.1103/wzpy-p7w8},
  url={https://doi.org/10.1103/wzpy-p7w8},
  eprint = "2503.15592",
  archivePrefix = "arXiv",
  primaryClass = "astro-ph.CO"
}

@article{Noriega:2026uhw,
    author = "Noriega, Hern{\'a}n E. and De-Santiago, Josue and Garcia-Arroyo, Gabriela and P{\'e}rez-Castro, Ivan and Venzor, Jorge",
    title = "{Neutrino self-interactions in light of DESI and CMB-SPA}",
    journal={},
    eprint = "2610.26654",
    archivePrefix = "arXiv",
    primaryClass = "astro-ph.CO",
    month = "10",
    year = "2026"
}

@article{Taule2022global,
  title="{Global view of neutrino interactions in cosmology: The free streaming window as seen by Planck}",
  author={Taule, Petter and Escudero, Miguel and Garny, Mathias},
  journal={Phys. Rev. D},
  volume={106},
  number={6},
  pages={063539},
  year={2022},
  publisher={APS},
  doi={10.1103/PhysRevD.106.063539},
  url={https://doi.org/10.1103/PhysRevD.106.063539},
  eprint = {2207.04062},
  archivePrefix = "arXiv",
  primaryClass = "astro-ph.CO"
}

@article{RoyChoudhury2022,
    author = "Roy Choudhury, Shouvik and Hannestad, Steen and Tram, Thomas",
    title = "{Massive neutrino self-interactions and inflation}",
    journal = "JCAP",
    volume = "2022",
    number={10},
    pages = "018",
    year = "2022",
    doi = "10.1088/1475-7516/2022/10/018",
    url={https://iopscience.iop.org/article/10.1088/1475-7516/2022/10/018},
    eprint = "2207.07142",
    archivePrefix = "arXiv",
    primaryClass = "astro-ph.CO"
}

@article{Bostan_2024,
  year = {2024},
  month = {jul},
  publisher = {IOP Publishing},
  volume = {2024},
  number = {07},
  pages = {032},
  author = {Bostan, Nilay and Choudhury, Shouvik Roy},
  title = "{First constraints on non-minimally coupled Natural and Coleman-Weinberg inflation and massive neutrino self-interactions with Planck+BICEP/Keck}",
  journal = {JCAP},
  doi = {10.1088/1475-7516/2024/07/032},
  url = {https://doi.org/10.1088/1475-7516/2024/07/032},
  eprint = "2310.01491",
  archivePrefix = "arXiv",
  primaryClass = "astro-ph.CO"
}

@article{Pal_2025,
  title = "{Exploring neutrino interactions in light of present and upcoming galaxy surveys}",
  journal = {JCAP},
  year = {2025},
  month = {mar},
  publisher = {IOP Publishing},
  volume = {2025},
  number = {03},
  pages = {047},
  author = {Pal, Sourav and Samanta, Rickmoy and Pal, Supratik},
  doi = {10.1088/1475-7516/2025/03/047},
  url = {https://dx.doi.org/10.1088/1475-7516/2025/03/047},
  eprint = {2409.03712},
  archivePrefix = "arXiv",
  primaryClass = "astro-ph.CO"
}

@article{Libanore2025,
  title = "{Joint 21-cm and CMB forecasts for constraining self-interacting massive neutrinos}",
  author = {Libanore, Sarah and Ghosh, Subhajit and Kovetz, Ely D. and Boddy, Kimberly K. and Raccanelli, Alvise},
  journal = {Phys. Rev. D},
  volume = {112},
  issue = {6},
  pages = {063502},
  numpages = {28},
  year = {2025},
  month = {Sep},
  publisher = {American Physical Society},
  doi = {10.1103/tdms-6n76},
  url = {https://link.aps.org/doi/10.1103/tdms-6n76},
  eprint = {2504.15348},
  archivePrefix = "arXiv",
  primaryClass="astro-ph.CO"
}

@article{Montefalcone:2025ibh,
  title="{Directly probing neutrino interactions through CMB phase shift measurements}",
  author={Montefalcone, Gabriele and Ghosh, Subhajit and Boddy, Kimberly K and Ho, Daven Wei Ren and Tsai, Yuhsin},
  journal={Phys. Rev. D},
  volume={113},
  number={2},
  pages={023540},
  year={2026},
  publisher={APS},
  doi={10.1103/lylf-3xyq},
  url={https://journals.aps.org/prd/abstract/10.1103/lylf-3xyq},
  eprint = "2509.20363",
  archivePrefix = "arXiv",
  primaryClass = "astro-ph.CO",
}

@article{Whitford:2025dmq,
  title="{Limits on self-interacting neutrinos from the BAO and CMB phase shift}",
  author={Whitford, Abb{\'e} M and Howlett, Cullan and Davis, Tamara M and Camarena, David and Cyr-Racine, Francis-Yan},
  journal={JCAP},
  volume={2026},
  number={03},
  pages={064},
  year={2026},
  publisher={IOP Publishing},
  doi={10.1088/1475-7516/2026/03/064},
  url={https://iopscience.iop.org/article/10.1088/1475-7516/2026/03/064},
  eprint = "2511.00800",
  archivePrefix = "arXiv",
  primaryClass = "astro-ph.CO"
}

@article{Pal2026neutrino,
  title="{Neutrino self-interactions in post-reionization era: Lyman-$\alpha$, 21-cm and cross-spectra}",
  author={Pal, Sourav and Pal, Supratik},
  journal={JCAP},
  volume = {2026},
  number = {08},
  pages = {036},
  year={2026},
  publisher={IOP Publishing},
  doi = {10.1088/1475-7516/2026/08/036},
  url = {https://doi.org/10.1088/1475-7516/2026/08/036},
  eprint = "2604.15287",
  archivePrefix = "arXiv",
  primaryClass = "astro-ph.CO"
}

@article{Parashari:2026dxo,
    author = "Parashari, Priyank and Gluscevic, Vera and Zhang, Yue and Bird, Simeon and Ivanov, Mikhail M. and He, Adam",
    title = "{Ly{\ensuremath{\alpha}} forest bounds on sterile neutrino production via neutrino self-interactions}",
    reportNumber = "MIT-CTP/6004",
    doi = "10.1103/s1gj-d111",
    journal = "Phys. Rev. D",
    volume = "114",
    number = "6",
    pages = "063512",
    year = "2026",
    url={https://doi.org/10.1103/s1gj-d111},
    eprint = "2602.17821",
    archivePrefix = "arXiv",
    primaryClass = "astro-ph.CO",
}

@article{Mogensen2026self,
  title={Self-interacting neutrinos in cosmological perturbation theory--integrating the collision kernel},
  author={Mogensen, Jakob K and Hannestad, Steen and Tram, Thomas},
  journal={},
  year={2026},
  eprint = "2606.03395",
  archivePrefix = "arXiv",
  primaryClass = "astro-ph.CO"
}

@article{Barenboim:2026jef,
    author = "Barenboim, Gabriela and Drewes, Marco and Oldengott, Isabel M.",
    title = "{Consistent treatment of flavour-specific neutrino self-interactions in cosmology}",
    month = "9",
    year = "2026",
    journal={},
    eprint = "2609.13821",
    archivePrefix = "arXiv",
    primaryClass = "astro-ph.CO"
}

@article{Mondol:2026irr,
    author = "Mondol, Ranjini and Parashari, Priyank and Das, Anirban and Gluscevic, Vera and Laha, Ranjan",
    title = "{Latest $\nu$s from the Earliest Galaxies: Probing neutrino self-interactions with JWST}",
    journal={},
    eprint = "2611.07013",
    archivePrefix = "arXiv",
    primaryClass = "astro-ph.CO",
    reportNumber = "IISc/CHEP/2026/2",
    month = "11",
    year = "2026"
}

@article{Uggerhoj:2026jet,
    author = "Uggerh{\o}j, Magnus and Johannessen, Tobias and Thybo, Jeppe Pagaard and Hannestad, Steen and Tram, Thomas",
    title = "{Self-Interacting Neutrinos in NEDE and DRMD Cosmologies: Posteriors, profiles and evidence using CLiENT}",
    journal={},
    eprint = "2611.13756",
    archivePrefix = "arXiv",
    primaryClass = "astro-ph.CO",
    month = "11",
    year = "2026"
}

@article{Chikashige1981there,
  title="{Are there real Goldstone bosons associated with broken lepton number?}",
  author={Chikashige, Y and Mohapatra, Rabindra N and Peccei, Roberto D},
  journal={Phys. Lett. B},
  volume={98},
  number={4},
  pages={265--268},
  year={1981},
  publisher={Elsevier},
  doi = {10.1016/0370-2693(81)90011-3},
  url = {https://doi.org/10.1016/0370-2693(81)90011-3}
}

@article{Gelmini1981left,
  title={Left-handed neutrino mass scale and spontaneously broken lepton number},
  author={Gelmini, Graciela B and Roncadelli, Marco},
  journal={Phys. Lett. B},
  volume={99},
  number={5},
  pages={411--415},
  year={1981},
  publisher={Elsevier},
  doi = {10.1016/0370-2693(81)90559-1},
  url = {https://www.sciencedirect.com/science/article/abs/pii/0370269381905591?via%3Dihub}
}

@article{Berryman2018,
  title = {Lepton-number-charged scalars and neutrino beamstrahlung},
  author = {Berryman, Jeffrey M. and de Gouv\^ea, Andr\'e and Kelly, Kevin J. and Zhang, Yue},
  journal = {Phys. Rev. D},
  volume = {97},
  issue = {7},
  pages = {075030},
  numpages = {16},
  year = {2018},
  month = {Apr},
  publisher = {American Physical Society},
  doi = {10.1103/PhysRevD.97.075030},
  url = {https://link.aps.org/doi/10.1103/PhysRevD.97.075030},
  eprint={1802.00009},
  archivePrefix = "arXiv",
  primaryClass = "hep-ph"
}

@article{Beacom2004,
  title = "{Neutrinoless Universe}",
  author = {Beacom, John F. and Bell, Nicole F. and Dodelson, Scott},
  journal = {Phys. Rev. Lett.},
  volume = {93},
  issue = {12},
  pages = {121302},
  numpages = {4},
  year = {2004},
  month = {Sep},
  publisher = {American Physical Society},
  doi = {10.1103/PhysRevLett.93.121302},
  url = {https://link.aps.org/doi/10.1103/PhysRevLett.93.121302},
  
}

@article{De2020dodelson,
  title="{Dodelson-Widrow mechanism in the presence of self-interacting neutrinos}",
  author={De Gouv{\^e}a, Andr{\'e} and Sen, Manibrata and Tangarife, Walter and Zhang, Yue},
  journal={Phys. Rev. Lett.},
  volume={124},
  number={8},
  pages={081802},
  year={2020},
  publisher={APS},
  doi={10.1103/PhysRevLett.124.081802},
  url={https://doi.org/10.1103/PhysRevLett.124.081802},
  eprint={1910.04901},
  archivePrefix = "arXiv",
  primaryClass = "hep-ph"
}

@Article{Dutta:2023fdt,
  author={Dutta, Bhaskar
  and Ghosh, Sumit
  and Li, Tianjun
  and Thompson, Adrian
  and Verma, Ankur},
  title={Non-standard neutrino interactions in light mediator models at reactor experiments},
  journal={JHEP},
  year={2023},
  month={Mar},
  day={22},
  volume={2023},
  number={3},
  pages={163},
  issn={1029-8479},
  doi={10.1007/JHEP03(2023)163},
  url={https://doi.org/10.1007/JHEP03(2023)163},
  eprint = "2209.13566",
  archivePrefix = "arXiv",
  primaryClass = "hep-ph"
}

@article{Berbig2020,
  title = "{The Hubble tension and a renormalizable model of gauged neutrino self-interactions}",
  author = {Berbig, Maximilian and Jana, Sudip and Trautner, Andreas},
  journal = {Phys. Rev. D},
  volume = {102},
  issue = {11},
  pages = {115008},
  numpages = {10},
  year = {2020},
  month = {Dec},
  publisher = {American Physical Society},
  doi = {10.1103/PhysRevD.102.115008},
  url = {https://link.aps.org/doi/10.1103/PhysRevD.102.115008},
  eprint = {2004.13039},
  archivePrefix = "arXiv",
  primaryClass = "hep-ph"
}

@article{Ghosh2024hubble,
  title="{Hubble tension and cosmological imprints of $U(1)_X$ gauge symmetry: $U(1)_{B_3-3 Li}$ as a case study}",
  author={Ghosh, Dilip Kumar and Ghosh, Purusottam and Jeesun, Sk and Srivastava, Rahul},
  journal={Eur. Phys. J. C},
  volume={84},
  number={8},
  pages={853},
  year={2024},
  publisher={Springer},
  url={https://link.springer.com/article/10.1140/epjc/s10052-024-13220-8},
  doi={10.1140/epjc/s10052-024-13220-8},
  eprint = {2312.16304},
  archivePrefix = "arXiv",
  primaryClass = "hep-ph"
}

@article{He2020,
  year = {2020},
  month = {nov},
  publisher = {IOP Publishing},
  volume = {2020},
  number = {11},
  pages = {003},
  author = {Hong-Jian He and Yin-Zhe Ma and Jiaming Zheng},
  title = "{Resolving Hubble tension by self-interacting neutrinos with Dirac seesaw}",
  journal = {JCAP},
  doi = {10.1088/1475-7516/2020/11/003},
  url = {https://dx.doi.org/10.1088/1475-7516/2020/11/003},
  eprint = {2003.12057},
  archivePrefix = "arXiv",
  primaryClass = "hep-ph"
}

@article{Gerda2015results,
  title="{Results on $\beta \beta$ decay with emission of two neutrinos or Majorons in 76 Ge from GERDA Phase I}",
  author = {{GERDA Collaboration} and Agostini, M. and others},
  journal={Eur. Phys. J. C},
  volume={75},
  number={9},
  pages={416},
  year={2015},
  publisher={Springer},
  doi={10.1140/epjc/s10052-015-3627-y},
  url={https://link.springer.com/article/10.1140/epjc/s10052-015-3627-y},
  eprint = "1501.02345",
  archivePrefix = "arXiv",
  primaryClass = "nucl-ex"
}

@article{Blum2018neutrinoless,
  title={Neutrinoless double-beta decay with massive scalar emission},
  author={Blum, Kfir and Nir, Yosef and Shavit, Michal},
  journal={Phys. Lett. B},
  volume={785},
  pages={354--361},
  year={2018},
  publisher={Elsevier},
  doi={10.1016/j.physletb.2018.08.022},
  url={https://www.sciencedirect.com/science/article/pii/S0370269318306282?via%3Dihub},
  eprint = "1802.08019",
  archivePrefix = "arXiv",
  primaryClass = "hep-ph"
}

@article{Krnjaic2020probing,
  title={Probing muonphilic force carriers and dark matter at kaon factories},
  author={Krnjaic, Gordan and Marques-Tavares, Gustavo and Redigolo, Diego and Tobioka, Kohsaku},
  journal={Phys. Rev. Lett.},
  volume={124},
  number={4},
  pages={041802},
  year={2020},
  publisher={APS},
  doi={10.1103/PhysRevLett.124.041802},
  url={https://doi.org/10.1103/PhysRevLett.124.041802},
  eprint = "1902.07715",
  archivePrefix = "arXiv",
  primaryClass = "hep-ph"
}

@article{Blinov2019,
  title = "{Constraining the Self-Interacting Neutrino Interpretation of the Hubble Tension}",
  author = {Blinov, Nikita and Kelly, Kevin J. and Krnjaic, Gordan and McDermott, Samuel D.},
  journal = {Phys. Rev. Lett.},
  volume = {123},
  issue = {19},
  pages = {191102},
  numpages = {7},
  year = {2019},
  month = {Nov},
  publisher = {American Physical Society},
  doi = {10.1103/PhysRevLett.123.191102},
  url = {https://link.aps.org/doi/10.1103/PhysRevLett.123.191102},
  eprint={1905.02727},
  archivePrefix = "arXiv",
  primaryClass = "astro-ph.CO"
}

@article{Deppisch2020,
  title = {Neutrino self-interactions and double beta decay},
  author = {Deppisch, Frank F. and Graf, Lukas and Rodejohann, Werner and Xu, Xun-Jie},
  journal = {Phys. Rev. D},
  volume = {102},
  issue = {5},
  pages = {051701},
  numpages = {6},
  year = {2020},
  month = {Sep},
  publisher = {American Physical Society},
  doi = {10.1103/PhysRevD.102.051701},
  url = {https://link.aps.org/doi/10.1103/PhysRevD.102.051701},
  eprint = "2004.11919",
  archivePrefix = "arXiv",
  primaryClass = "hep-ph"
}

@article{Lyu2021self,
  title="{Self-interacting neutrinos: Solution to Hubble tension versus experimental constraints}",
  author={Lyu, Kun-Feng and Stamou, Emmanuel and Wang, Lian-Tao},
  journal={Phys. Rev. D},
  volume={103},
  number={1},
  pages={015004},
  year={2021},
  publisher={APS},
  doi={10.1103/PhysRevD.103.015004},
  url={https://doi.org/10.1103/PhysRevD.103.015004},
  eprint = "2004.10868",
  archivePrefix = "arXiv",
  primaryClass = "hep-ph"
}

@article{Demirci2024,
  title = {Solar neutrino constraints on light mediators through coherent elastic neutrino-nucleus scattering},
  author = {Demirci, M. and Mustamin, M. F.},
  journal = {Phys. Rev. D},
  volume = {109},
  issue = {1},
  pages = {015021},
  numpages = {17},
  year = {2024},
  month = {Jan},
  publisher = {American Physical Society},
  doi = {10.1103/PhysRevD.109.015021},
  url = {https://link.aps.org/doi/10.1103/PhysRevD.109.015021},
  eprint = "2312.17502",
  archivePrefix = "arXiv",
  primaryClass = "hep-ph"
}

@article{Dev2025new,
  title={New laboratory constraints on neutrinophilic mediators},
  author={Dev, PS Bhupal and Kim, Doojin and Sathyan, Deepak and Sinha, Kuver and Zhang, Yongchao},
  journal={Phys. Lett. B},
  pages={139765},
  year={2025},
  publisher={Elsevier},
  doi={10.1016/j.physletb.2025.139765},
  url={https://www.sciencedirect.com/science/article/pii/S037026932500526X?via%3Dihub},
  eprint = "2407.12738",
  archivePrefix = "arXiv",
  primaryClass = "hep-ph"
}

@article{Foroughi2026precision,
  title="{Precision electroweak constraints on neutrinophilic scalars}",
  author={Foroughi-Abari, Saeid and Mupo, Camilla and Vatsyayan, Drona and Zhang, Yue},
  journal = "JHEP",
  number = "09",
  pages = "163",
  year = "2026",
  doi = "10.1007/JHEP09(2026)163",
  url={doi.org/10.1007/JHEP09(2026)163},
  eprint = "2606.02704",
  archivePrefix = "arXiv",
  primaryClass = "hep-ph"
}

@article{Foroughi2025enabling,
  title="{Enabling Strong Neutrino Self-Interaction with an Unparticle Mediator}",
  author={Foroughi-Abari, Saeid and Kelly, Kevin J and Rai, Mudit and Zhang, Yue},
  journal={Phys. Rev. Lett.},
  volume={134},
  number={18},
  pages={181001},
  year={2025},
  publisher={APS},
  doi={10.1103/PhysRevLett.134.181001},
  url={https://doi.org/10.1103/PhysRevLett.134.181001},
  eprint = {2501.02049},
  archivePrefix = "arXiv",
  primaryClass = "hep-ph"
}

@article{Das2025impostor,
  title = "{Impostor among Neutrinos: Dark Radiation Masquerading as Self-Interacting Neutrinos}",
  author = {Das, Anirban and Dev, P. S. Bhupal and Gao, Christina and Ghosh, Subhajit and Kim, Taegyun},
  journal = {Phys. Rev. Lett.},
  volume = {136},
  issue = {13},
  pages = {131003},
  numpages = {10},
  year = {2026},
  month = {Apr},
  publisher = {American Physical Society},
  doi = {10.1103/jprg-jll6},
  url = {https://link.aps.org/doi/10.1103/jprg-jll6},
  eprint={2506.08085},
  archivePrefix={arXiv},
  primaryClass={hep-ph}
}

@article{Farzan2016neutrinos,
  title="{Neutrinos secretly converting to lighter particles to please both KATRIN and the cosmos}",
  author={Farzan, Yasaman and Hannestad, Steen},
  journal={JCAP},
  volume={2016},
  number={02},
  pages={058--058},
  year={2016},
  doi={10.1088/1475-7516/2016/02/058},
  url={https://iopscience.iop.org/article/10.1088/1475-7516/2016/02/058},
  eprint={1510.02201},
  archivePrefix = "arXiv",
  primaryClass = "hep-ph"
}

@article{Escudero2023seesaw,
  title={A seesaw model for large neutrino masses in concordance with cosmology},
  author={Escudero, Miguel and Schwetz, Thomas and Terol-Calvo, Jorge},
  journal={JHEP},
  volume={2023},
  number={2},
  pages={142},
  year={2023},
  publisher={Springer},
  doi={10.1007/JHEP02(2023)142},
  url={https://link.springer.com/article/10.1007/JHEP02(2023)142},
  eprint = "2211.01729",
  archivePrefix = "arXiv",
  primaryClass = "hep-ph"
}

@article{Gariazzo2023quantifying,
  title={Quantifying the tension between cosmological and terrestrial constraints on neutrino masses},
  author={Gariazzo, Stefano and Mena, Olga and Schwetz, Thomas},
  journal={Phys. Dark Univ.},
  volume={40},
  pages={101226},
  year={2023},
  publisher={Elsevier},
  doi= {10.1016/j.dark.2023.101226},
  url={https://doi.org/10.1016/j.dark.2023.101226},
  eprint = {2302.14159},
  archivePrefix = "arXiv",
  primaryClass = "hep-ph"
}

@article{Naredo2024critical,
  title={Critical look at the cosmological neutrino mass bound},
  author={Naredo-Tuero, Daniel and Escudero, Miguel and Fern{\'a}ndez-Mart{\'\i}nez, Enrique and Marcano, Xabier and Poulin, Vivian},
  journal={Phys. Rev. D},
  volume={110},
  number={12},
  pages={123537},
  year={2024},
  publisher={APS},
  doi={10.1103/PhysRevD.110.123537},
  url={https://doi.org/10.1103/PhysRevD.110.123537},
  eprint = {2407.13831},
  archivePrefix = "arXiv",
  primaryClass = "astro-ph.CO"
}

@article{Elbers2025constraints,
  title="{Constraints on neutrino physics from DESI DR2 BAO and DR1 full shape}",
  author={Elbers, Willem and Avil{\'e}s, Alejandro and Noriega, HE and Chebat, D and Menegas, Alexander and Frenk, Carlos S and Garcia-Quintero, C and Gonzalez, D and Ishak, Mustapha and Lahav, O and others},
  journal={Phys. Rev. D},
  volume={112},
  number={8},
  pages={083513},
  year={2025},
  publisher={APS},
  doi= {10.1103/w9pk-xsk7},
  url={https://doi.org/10.1103/w9pk-xsk7},
  eprint = {2503.14744},
  archivePrefix = "arXiv",
  primaryClass = "astro-ph.CO"
}

@article{Di2025cosmoverse,
  title="{The CosmoVerse White Paper: Addressing observational tensions in cosmology with systematics and fundamental physics}",
  author={Di Valentino, Eleonora and Said, Jackson Levi and Riess, Adam and Pollo, Agnieszka and Poulin, Vivian and G{\'o}mez-Valent, Adri{\`a} and Weltman, Amanda and Palmese, Antonella and others},
  journal={Phys. Dark Univ.},
  volume={49},
  pages={101965},
  year={2025},
  publisher={Elsevier},
  doi={10.1016/j.dark.2025.101965},
  url={https://www.sciencedirect.com/science/article/pii/S221268642500158X?via%3Dihub},
  eprint = {2504.01669},
  archivePrefix = "arXiv",
  primaryClass = "astro-ph.CO"
}

@article{SPT_2026_CMBSPA,
  title = "{SPT-3G D1: CMB temperature and polarization power spectra and cosmology from 2019 and 2020 observations of the SPT-3G main field}",
  author = {Camphuis, E. and others},
  collaboration = {SPT-3G Collaboration},
  journal = {Phys. Rev. D},
  volume = {113},
  issue = {8},
  pages = {083504},
  numpages = {81},
  year = {2026},
  month = {Apr},
  publisher = {American Physical Society},
  doi = {10.1103/7wt3-9v2y},
  url = {https://link.aps.org/doi/10.1103/7wt3-9v2y},
  eprint={2506.20707},
  archivePrefix = "arXiv",
  primaryClass = "astro-ph.CO"
}

@article{venzor2023resonant,
  title="{Resonant neutrino self-interactions and the $H_0$ tension}",
  author={Venzor, Jorge and Garcia-Arroyo, Gabriela and De-Santiago, Josue and P{\'e}rez-Lorenzana, Abdel},
  journal={Phys. Rev. D},
  volume={108},
  number={4},
  pages={043536},
  year={2023},
  publisher={APS},
  doi= {10.1103/PhysRevD.108.043536},
  url={https://link.aps.org/doi/10.1103/PhysRevD.108.043536},
  eprint = {2303.12792},
  archivePrefix = "arXiv",
  primaryClass = "astro-ph.CO"
}

@article{Noriega2025resonant,
  title = "{Resonant neutrino self-interactions: Insights from the full shape galaxy power spectrum}",
  author = {Noriega, Hern\'an E. and De-Santiago, Josue and Garcia-Arroyo, Gabriela and Venzor, Jorge and P\'erez-Lorenzana, Abdel},
  journal = {Phys. Rev. D},
  volume = {112},
  issue = {6},
  pages = {063509},
  numpages = {12},
  year = {2025},
  month = {Sep},
  publisher = {American Physical Society},
  doi = {10.1103/b9x4-hnqn},
  url = {https://link.aps.org/doi/10.1103/b9x4-hnqn},
  eprint = {2506.07994},
  archivePrefix = "arXiv",
  primaryClass = "astro-ph.CO"
}

@article{De2016relic,
  title={Relic neutrino decoupling with flavour oscillations revisited},
  author={de Salas, Pablo F and Pastor, Sergio},
  journal={JCAP},
  volume={2016},
  number={07},
  pages={051--051},
  year={2016},
  doi={10.1088/1475-7516/2016/07/051},
  url={https://iopscience.iop.org/article/10.1088/1475-7516/2016/07/051},
  eprint = "1606.06986",
  archivePrefix = "arXiv",
  primaryClass = "hep-ph"
}

@article{Akita2020precision,
  title={A precision calculation of relic neutrino decoupling},
  author={Akita, Kensuke and Yamaguchi, Masahide},
  journal={JCAP},
  volume={2020},
  number={08},
  pages={012--012},
  year={2020},
  doi={10.1088/1475-7516/2020/08/012},
  url={https://iopscience.iop.org/article/10.1088/1475-7516/2020/08/012},
  eprint = "2005.07047",
  archivePrefix = "arXiv",
  primaryClass = "hep-ph"
}

@article{Froustey2020neutrino,
  title={Neutrino decoupling including flavour oscillations and primordial nucleosynthesis},
  author={Froustey, Julien and Pitrou, Cyril and Volpe, Maria Cristina},
  journal={JCAP},
  volume={2020},
  number={12},
  pages={015--015},
  year={2020},
  doi={10.1088/1475-7516/2020/12/015},
  url={https://iopscience.iop.org/article/10.1088/1475-7516/2020/12/015},
  eprint = "2008.01074",
  archivePrefix = "arXiv",
  primaryClass = "hep-ph"
}

@article{Bennett_2021,
  author = {Bennett, Jack J. and Buldgen, Gilles and de Salas, Pablo F. and Drewes, Marco and Gariazzo, Stefano and Pastor, Sergio and Wong, Yvonne Y.Y.},
  title = "{Towards a precision calculation of the effective number of neutrinos $N_{\rm eff}$ in the Standard Model. Part II. Neutrino decoupling in the presence of flavour oscillations and finite-temperature QED}",
  journal = {JCAP},
  year = {2021},
  month = {apr},
  publisher = {IOP Publishing},
  volume = {2021},
  number = {04},
  pages = {073},
  doi = {10.1088/1475-7516/2021/04/073},
  url = {https://doi.org/10.1088/1475-7516/2021/04/073},
  eprint = "2012.02726",
  archivePrefix = "arXiv",
  primaryClass = "hep-ph"
}

@article{Abenza2020precision,
  title="{Precision early universe thermodynamics made simple: $N_{\rm eff}$ and neutrino decoupling in the Standard Model and beyond}",
  author={Abenza, Miguel Escudero},
  journal={JCAP},
  volume={2020},
  number={05},
  pages={048},
  year={2020},
  publisher={IOP Publishing},
  doi={10.1088/1475-7516/2020/05/048},
  url={https://iopscience.iop.org/article/10.1088/1475-7516/2020/05/048},
  eprint = {2001.04466},
  archivePrefix = "arXiv",
  primaryClass = "hep-ph"
}

@article{Barenboim2021invisible,
  title={Invisible neutrino decay in precision cosmology},
  author={Barenboim, Gabriela and Chen, Joe Zhiyu and Hannestad, Steen and Oldengott, Isabel M and Tram, Thomas and Wong, Yvonne YY},
  journal={JCAP},
  volume={2021},
  number={03},
  pages={087},
  year={2021},
  publisher={IOP Publishing},
  doi ={10.1088/1475-7516/2021/03/087},
  url ={https://iopscience.iop.org/article/10.1088/1475-7516/2021/03/087},
  eprint = "2011.01502",
  archivePrefix = "arXiv",
  primaryClass = "astro-ph.CO"
}

@article{Chen2022weaker,
  title={Weaker yet again: mass spectrum-consistent cosmological constraints on the neutrino lifetime},
  author={Chen, Joe Zhiyu and Oldengott, Isabel M and Pierobon, Giovanni and Wong, Yvonne YY},
  journal={Eur. Phys. J. C},
  volume={82},
  number={7},
  pages={640},
  year={2022},
  publisher={Springer},
  doi = {10.1140/epjc/s10052-022-10518-3},
  url = {https://doi.org/10.1140/epjc/s10052-022-10518-3},
  eprint = {2203.09075},
  archivePrefix = "arXiv",
  primaryClass = "hep-ph"
}

@article{Terzaghi2025impact,
  title= "{Impact of neutrino decays on the Cosmic Neutrino Background anisotropies}",
  author={Terzaghi, Nicola and Abell{\'a}n, Guillermo Franco and Zimmer, Fabian and Ando, Shin'ichiro},
  journal={},
  year={2025},
  eprint = "2510.15818",
  archivePrefix = "arXiv",
  primaryClass = "astro-ph.CO"
}

@article{Abellan2026neutrino,
  title={Neutrino decays as a natural explanation of the neutrino mass tension},
  author={Franco Abell{\'a}n, Guillermo},
  journal={Phys. Rev. D},
  volume={113},
  number={12},
  pages={123527},
  year={2026},
  publisher={APS},
  doi={10.1103/lwdv-qrcc},
  url={ https://doi.org/10.1103/lwdv-qrcc},
  eprint = "2601.04312",
  archivePrefix = "arXiv",
  primaryClass = "astro-ph.CO"
}

@article{Perez2026towards,
  title={Towards a complete scheme of cosmological neutrino self-interactions: collision term for a wide range of mediator masses},
  author={P{\'e}rez-Castro, Ivan and De-Santiago, Josue and Garcia-Arroyo, Gabriela and Venzor, Jorge and P{\'e}rez-Lorenzana, Abdel},
  journal={JCAP},
  volume={2026},
  number={07},
  pages={021},
  year={2026},
  publisher={IOP Publishing},
  doi={10.1088/1475-7516/2026/07/021},
  url={https://iopscience.iop.org/article/10.1088/1475-7516/2026/07/021},
  eprint = "2602.12477",
  archivePrefix = "arXiv",
  primaryClass = "hep-ph"
}

@article{Blas2011cosmic,
  title="{The cosmic linear anisotropy solving system (CLASS). Part II: Approximation schemes}",
  author={Blas, Diego and Lesgourgues, Julien and Tram, Thomas},
  journal={JCAP},
  volume={2011},
  number={07},
  pages={034},
  year={2011},
  publisher={IOP Publishing},
  doi={10.1088/1475-7516/2011/07/034},
  url={https://iopscience.iop.org/article/10.1088/1475-7516/2011/07/034},
  eprint = "1104.2933",
  archivePrefix = "arXiv",
  primaryClass = "astro-ph.CO"
}

@article{Lesgourgues2011cosmic,
  title="{The Cosmic Linear Anisotropy Solving System (CLASS) IV: efficient implementation of non-cold relics}",
  author={Lesgourgues, Julien and Tram, Thomas},
  journal={JCAP},
  volume={2011},
  number={09},
  pages={032--032},
  year={2011},
  doi={10.1088/1475-7516/2011/07/034},
  url={https://iopscience.iop.org/article/10.1088/1475-7516/2011/09/032},
  eprint={1104.2935},
  archivePrefix = "arXiv",
  primaryClass = "astro-ph.CO"
}

@article{Forastieri:2015paa,
    author = "Forastieri, Francesco and Lattanzi, Massimiliano and Natoli, Paolo",
    title = "{Constraints on secret neutrino interactions after Planck}",
    doi = "10.1088/1475-7516/2015/07/014",
    journal = "JCAP",
    volume = "07",
    number={07},
    pages = "014",
    year = "2015",
    eprint = "1504.04999",
    archivePrefix = "arXiv",
    primaryClass = "astro-ph.CO"
}

@article{Forastieri:2019cuf,
    author = "Forastieri, Francesco and Lattanzi, Massimiliano and Natoli, Paolo",
    title = "{Cosmological constraints on neutrino self-interactions with a light mediator}",
    doi = "10.1103/PhysRevD.100.103526",
    journal = "Phys. Rev. D",
    volume = "100",
    number = "10",
    pages = "103526",
    year = "2019",
    eprint = "1904.07810",
    archivePrefix = "arXiv",
    primaryClass = "astro-ph.CO"
}

@article{Venzor2022,
  title = {Massive neutrino self-interactions with a light mediator in cosmology},
  author = {Venzor, Jorge and Garcia-Arroyo, Gabriela and P\'erez-Lorenzana, Abdel and De-Santiago, Josue},
  journal = {Phys. Rev. D},
  volume = {105},
  issue = {12},
  pages = {123539},
  numpages = {10},
  year = {2022},
  month = {Jun},
  publisher = {American Physical Society},
  doi = {10.1103/PhysRevD.105.123539},
  url = {https://link.aps.org/doi/10.1103/PhysRevD.105.123539},
  eprint = "2202.09310",
  archivePrefix = "arXiv",
  primaryClass = "astro-ph.CO"
}

@article{Huang2018,
  title = {Observational constraints on secret neutrino interactions from big bang nucleosynthesis},
  author = {Huang, Guo-yuan and Ohlsson, Tommy and Zhou, Shun},
  journal = {Phys. Rev. D},
  volume = {97},
  issue = {7},
  pages = {075009},
  numpages = {12},
  year = {2018},
  month = {Apr},
  publisher = {American Physical Society},
  doi = {10.1103/PhysRevD.97.075009},
  url = {https://link.aps.org/doi/10.1103/PhysRevD.97.075009},
  eprint={1712.04792},
  archivePrefix = "arXiv",
  primaryClass = "hep-ph"
}

@article{Bernstein1985cosmological,
  title={Cosmological heavy-neutrino problem},
  author={Bernstein, Jeremy and Brown, Lowell S and Feinberg, Gerald},
  journal={Phys. Rev. D},
  volume={32},
  number={12},
  pages={3261},
  year={1985},
  publisher={APS},
  doi={10.1103/PhysRevD.32.3261},
  url={https://doi.org/10.1103/PhysRevD.32.3261}
}

@article{Denner1999predictions,
  title="{Predictions for all processes $e^{+} e^{-} \to 4$ fermions + $\gamma$}",
  author={Denner, Ansgar and Dittmaier, S and Roth, M and Wackeroth, D},
  journal={Nucl. Phys. B},
  volume={560},
  number={1-3},
  pages={33--65},
  year={1999},
  publisher={Elsevier},
  doi={10.1016/S0550-3213(99)00437-X},
  url={https://linkinghub.elsevier.com/retrieve/pii/S055032139900437X}
}

@article{Willenbrock2024mass,
  title={Mass and width of an unstable particle},
  author={Willenbrock, Scott},
  journal={Eur. Phys. J. Plus},
  volume={139},
  number={6},
  pages={523},
  year={2024},
  publisher={Springer},
  doi={10.1140/epjp/s13360-024-05301-0},
  url={https://link.springer.com/article/10.1140/epjp/s13360-024-05301-0},
  eprint={2203.11056},
  archivePrefix = "arXiv",
  primaryClass = "hep-ph"
}

@misc{venzor_2026_res_class,
  author       = {Venzor, Jorge and Garcia-Arroyo, Gabriela},
  title        = "{JorgeVenzor/class\_resonant\_self-interacting\_neutrinos: CLASS: Resonant neutrino
                   self-interactions
                  }",
  month        = apr,
  year         = 2026,
  publisher    = {Zenodo},
  version      = {v1.0},
  doi          = {10.5281/zenodo.19802610},
  url          = {https://doi.org/10.5281/zenodo.19802610},
  swhid        = {swh:1:dir:b9baa9d9abc24221501f1928f7f0bceba2155fdf
                   ;origin=https://doi.org/10.5281/zenodo.19802609;vi
                   sit=swh:1:snp:69f784d2eee7d2e784ff75062f1451c81643
                   4166;anchor=swh:1:rel:a3704f1e51847c3c75fee8acf1fc
                   d91252f199a0;path=JorgeVenzor-class\_resonant\_self-
                   interacting\_neutrinos-2227434
                  }
}

@article{Escudero2020CMBsearch,
  title = "{A CMB search for the neutrino mass mechanism and its relation to the Hubble tension}",
  author = {Escudero, Miguel and Witte, Samuel J},
  doi = {10.1140/epjc/s10052-020-7854-5},
  issn = {1434-6052},
  journal = {Eur. Phys. J. C},
  number = {4},
  pages = {294},
  url = {https://doi.org/10.1140/epjc/s10052-020-7854-5},
  volume = {80},
  year = {2020},
  eprint={1909.04044},
  archivePrefix = "arXiv",
  primaryClass = "astro-ph.CO"
}

@article{Li2023neff,
  title="{$N_{\rm eff}$ constraints on light mediators coupled to neutrinos: the dilution-resistant effect}",
  author={Li, Shao-Ping and Xu, Xun-Jie},
  journal={JHEP},
  volume={2023},
  number={10},
  pages={1--26},
  year={2023},
  publisher={Springer},
  doi={10.1007/JHEP10(2023)012},
  url={https://link.springer.com/article/10.1007/JHEP10(2023)012},
  eprint={2307.13967},
  archivePrefix = "arXiv",
  primaryClass = "hep-ph"
}

@article{Sandner2023precision,
  title="{Precision CMB constraints on eV-scale bosons coupled to neutrinos}",
  author={Sandner, Stefan and Escudero, Miguel and Witte, Samuel J},
  journal={Eur. Phys. J. C},
  volume={83},
  number={8},
  pages={709},
  year={2023},
  publisher={Springer},
  doi={10.1140/epjc/s10052-023-11864-6},
  url={https://link.springer.com/article/10.1140/epjc/s10052-023-11864-6},
  eprint={2305.01692},
  archivePrefix = "arXiv",
  primaryClass = "hep-ph"
}

@article{Perez2026Addressing,
  title={Addressing the on-shell mediator paradigm: resonant neutrino self-interactions in cosmology and astrophysics},
  author={P{\'e}rez-Castro, Ivan and Venzor, Jorge and P{\'e}rez-Lorenzana, Abdel},
  journal={(in preparation)},
  year={2026},
  archivePrefix = "arXiv",
  primaryClass = "hep-ph"
}

@article{Berlin2018thermal,
  title="{Thermal Dark Matter Below a MeV}",
  author={Berlin, Asher and Blinov, Nikita},
  journal={Phys. Rev. Lett.},
  volume={120},
  number={2},
  pages={021801},
  year={2018},
  publisher={APS},
  doi={10.1103/PhysRevLett.120.021801},
  url={https://doi.org/10.1103/PhysRevLett.120.021801},
  eprint = "1706.07046",
  archivePrefix = "arXiv",
  primaryClass = "hep-ph"
}

@article{Berlin2019thermal,
  title="{Thermal neutrino portal to sub-MeV dark matter}",
  author={Berlin, Asher and Blinov, Nikita},
  journal={Phys. Rev. D},
  volume={99},
  number={9},
  pages={095030},
  year={2019},
  publisher={APS},
  doi={10.1103/PhysRevD.99.095030},
  url={https://doi.org/10.1103/PhysRevD.99.095030},
  eprint = "1807.04282",
  archivePrefix = "arXiv",
  primaryClass = "hep-ph"
}

@article{Kelly2021intimate,
  title="{Intimate relationship between sterile neutrino dark matter and $\Delta N_{\rm eff}$}",
  author={Kelly, Kevin J and Sen, Manibrata and Zhang, Yue},
  journal={Phys. Rev. Lett.},
  volume={127},
  number={4},
  pages={041101},
  year={2021},
  publisher={APS},
  doi={10.1103/PhysRevLett.127.041101},
  url={https://doi.org/10.1103/PhysRevLett.127.041101},
  eprint = "2011.02487",
  archivePrefix = "arXiv",
  primaryClass = "hep-ph"
}

@article{Aloni2023dark,
  title={Dark radiation from neutrino mixing after big bang nucleosynthesis},
  author={Aloni, Daniel and Joseph, Melissa and Schmaltz, Martin and Weiner, Neal},
  journal={Phys. Rev. Lett.},
  volume={131},
  number={22},
  pages={221001},
  year={2023},
  publisher={APS},
  doi={10.1103/PhysRevLett.131.221001},
  url={https://doi.org/10.1103/PhysRevLett.131.221001},
  eprint = "2301.10792",
  archivePrefix = "arXiv",
  primaryClass = "astro-ph.CO"
}

@article{Chacko2004cmb,
  title="{CMB signals of neutrino mass generation}",
  author={Chacko, Z and Hall, Lawrence J and Okui, Takemichi and Oliver, Steven J},
  journal={Phys. Rev. D},
  volume={70},
  number={8},
  pages={085008},
  year={2004},
  publisher={APS},
  doi={10.1103/PhysRevD.70.085008},
  url={https://doi.org/10.1103/PhysRevD.70.085008},
  eprint ={hep-ph/0312267},
  archivePrefix = "arXiv",
}

@article{Escudero2019cosmology,
  title="{Cosmology with a very light $L_\mu$-$L_\tau$ gauge boson}",
  author={Escudero, Miguel and Hooper, Dan and Krnjaic, Gordan and Pierre, Mathias},
  journal={JHEP},
  volume={2019},
  number={3},
  pages={1--29},
  year={2019},
  publisher={Springer},
  doi={10.1007/JHEP03(2019)071},
  url={https://link.springer.com/article/10.1007/JHEP03(2019)071},
  eprint = "1901.02010",
  archivePrefix = "arXiv",
  primaryClass = "hep-ph"
}

@article{Dror2020discovering,
  title={Discovering leptonic forces using nonconserved currents},
  author={Dror, Jeff A},
  journal={Phys. Rev. D},
  volume={101},
  number={9},
  pages={095013},
  year={2020},
  publisher={APS},
  doi={10.1103/PhysRevD.101.095013},
  url={https://doi.org/10.1103/PhysRevD.101.095013},
  eprint = "2004.04750",
  archivePrefix = "arXiv",
  primaryClass = "hep-ph"
}

@article{Kelly2020origin,
  title={Origin of sterile neutrino dark matter via secret neutrino interactions with vector bosons},
  author={Kelly, Kevin J and Sen, Manibrata and Tangarife, Walter and Zhang, Yue},
  journal={Phys. Rev. D},
  volume={101},
  number={11},
  pages={115031},
  year={2020},
  publisher={APS},
  doi={10.1103/PhysRevD.101.115031},
  url={https://journals.aps.org/prd/abstract/10.1103/PhysRevD.101.115031},
  eprint = "2005.03681",
  archivePrefix = "arXiv",
  primaryClass = "hep-ph"
}

@article{Arias2021neutrino,
  title="{Neutrino masses and Hubble tension via a Majoron in MFV}",
  author={Arias-Arag{\'o}n, Fernando and Fern{\'a}ndez-Mart{\'\i}nez, Enrique and Gonz{\'a}lez-L{\'o}pez, Manuel and Merlo, Luca},
  journal={Eur. Phys. J. C},
  volume={81},
  number={1},
  pages={28},
  year={2021},
  publisher={Springer},
  doi={10.1140/epjc/s10052-020-08825-8},
  url={https://link.springer.com/article/10.1140/epjc/s10052-020-08825-8},
  eprint = "2009.01848",
  archivePrefix = "arXiv",
  primaryClass = "hep-ph"
}

@article{Araki2021resolving,
  title="{Resolving the Hubble tension in a $U(1)_{L_\mu -L_\tau}$ model with the Majoron}",
  author={Araki, Takeshi and Asai, Kento and Honda, Kei and Kasuya, Ryuta and Sato, Joe and Shimomura, Takashi and Yang, Masaki JS},
  journal={PTEP},
  volume={2021},
  number={10},
  pages={103B05},
  year={2021},
  publisher={Oxford University Press},
  doi={10.1093/ptep/ptab108},
  url={https://doi.org/10.1093/ptep/ptab108},
  eprint = "2103.07167",
  archivePrefix = "arXiv",
  primaryClass = "hep-ph"
}

@article{Fernandez2021inverse,
  title="{Inverse Seesaw, dark matter and the Hubble tension}",
  author={Fernandez-Martinez, Enrique and Pierre, Mathias and Pinsard, Emanuelle and Rosauro-Alcaraz, Salvador},
  journal={Eur. Phys. J. C},
  volume={81},
  number={10},
  pages={954},
  year={2021},
  publisher={Springer},
  doi={10.1140/epjc/s10052-021-09760-y},
  url={https://link.springer.com/article/10.1140/epjc/s10052-021-09760-y},
  eprint = "2106.05298",
  archivePrefix = "arXiv",
  primaryClass = "hep-ph"
}

@article{Esseili2024cosmological,
  title="{Cosmological implications of gauged $U(1)_{B-L}$ on $\Delta N_{\rm eff}$ in the CMB and BBN}",
  author={Esseili, Haidar and Kribs, Graham D},
  journal={JCAP},
  volume={2024},
  number={05},
  pages={110},
  year={2024},
  publisher={IOP Publishing},
  doi={10.1088/1475-7516/2024/05/110},
  url={https://iopscience.iop.org/article/10.1088/1475-7516/2024/05/110},
  eprint = "2308.07955",
  archivePrefix = "arXiv",
  primaryClass = "hep-ph"
}

@article{Asai2024contribution,
  title="{Contribution of Majoron to Hubble Tension in Gauged $U(1) _{L_\mu-L_\tau}$ Model}",
  author={Asai, Kento and Asano, Tomoya and Sato, Joe and Yang, Masaki JS},
  journal={PTEP},
  volume={2024},
  number={7},
  pages={073E01},
  year={2024},
  publisher={Oxford University Press},
  doi={10.1093/ptep/ptae094},
  url={ https://doi.org/10.1093/ptep/ptae094},
  eprint={2309.01162},
  archivePrefix = "arXiv",
  primaryClass = "hep-ph"
}

@article{Li2024cosmological,
  title={A cosmological sandwiched window for lepton-number breaking scale},
  author={Li, Shao-Ping and Yu, Bingrong},
  journal={JCAP},
  volume={2024},
  number={04},
  pages={047},
  year={2024},
  publisher={IOP Publishing},
  doi={10.1088/1475-7516/2024/04/047},
  url={https://iopscience.iop.org/article/10.1088/1475-7516/2024/04/047},
  eprint = "2310.13492",
  archivePrefix = "arXiv",
  primaryClass = "hep-ph"
}

@article{De2024minimal,
  title="{The minimal massive Majoron seesaw model}",
  author={de Giorgi, Arturo and Merlo, Luca and D{\'\i}az, Xavier Ponce and Rigolin, Stefano},
  journal={JHEP},
  volume={2024},
  number={3},
  pages={1--39},
  year={2024},
  publisher={Springer},
  doi={10.1007/JHEP03(2024)094},
  url={https://link.springer.com/article/10.1007/JHEP03(2024)094},
  eprint = "2312.13417",
  archivePrefix = "arXiv",
  primaryClass = "hep-ph"
}

@article{Benso2025large,
  title="{Large neutrino mass in cosmology and keV sterile neutrino dark matter from a dark sector}",
  author={Benso, Cristina and Schwetz, Thomas and Vatsyayan, Drona},
  journal={JCAP},
  volume={2025},
  number={04},
  pages={054},
  year={2025},
  publisher={IOP Publishing},
  doi={10.1088/1475-7516/2025/04/054},
  url={https://iopscience.iop.org/article/10.1088/1475-7516/2025/04/054},
  eprint = {2410.23926},
  archivePrefix = "arXiv",
  primaryClass = "hep-ph"
}

\clearpage
\newpage
\onecolumngrid

\setcounter{equation}{0}
\setcounter{figure}{0}
\setcounter{table}{0}
\setcounter{page}{1}
\setcounter{section}{0}

\makeatletter
\renewcommand{\thesection}{S\arabic{section}}
\renewcommand{\thesubsection}{\thesection.\arabic{subsection}}

\renewcommand{\theequation}{S\arabic{equation}}
\renewcommand{\thefigure}{S\arabic{figure}}
\renewcommand{\thetable}{S\arabic{table}}

\renewcommand{\theHfigure}{S\arabic{figure}}
\renewcommand{\theHtable}{S\arabic{table}}
\renewcommand{\theHequation}{S\arabic{equation}}
\makeatother

\begin{center}
    \textbf{\large On the cosmic window for neutrino self-interactions} \\ 
    \vspace{0.5cm}
    { \it \large Supplemental Material}\\ 
    \vspace{0.5cm}
    {Ivan Pérez-Castro, Jorge Venzor, Josue De-Santiago, Pablo Roig, Abdel Pérez-Lorenzana}
\end{center}

We collect here supplementary material that supports our results and may be useful for future developments, while not being essential to the main text.

\section{Boltzmann Hierarchy equations for non-standard thermal history of neutrinos} \label{S1: ModBHES}
In this section, we present our proposed generalization of the thermal approximation to cases in which the background distribution of relativistic species evolves due to a nonzero collision term at zeroth order.

There are several ways to derive the Boltzmann hierarchy equations for relativistic species; first, we consider an FLRW universe. The background distribution function $\bar{f}_\alpha^{(0)}$ in a homogeneous and isotropic Universe satisfies
    \begin{equation}
        \frac{\partial}{\partial \eta} \bar{f}_\alpha^{(0)} - {\cal H} p \frac{\partial}{\partial p} \bar{f}_\alpha^{(0)} = \frac{Q_\alpha^{(0)}}{{\rm P}^0} \, , \label{Q-zero}
    \end{equation}
where ${\cal H}$ is the comoving Hubble parameter. The local energy $E$ measured by an observer fixed at the coordinates $x^i$, determines ${\rm P}^0$ as $E = a(1+\Psi) {\rm P}^0$ for the CN gauge, given by the line element
    \begin{equation}
        ds^2 = a^2(\eta) \left[(1+2\Psi)d\eta^2
        -(1-2\Phi)\delta_{ij}dx^i dx^j\right] \, .
    \end{equation}
At first order in perturbations,
the Boltzmann equation takes the form
    \begin{equation}
        \frac{\partial \bar{f}_\alpha}{\partial \eta} + \frac{{\bf p}}{E} \cdot \nabla \bar{f}_\alpha + p \frac{\partial \bar{f}_\alpha}{\partial p} \left[ - {\cal H} + \frac{\partial \Phi}{\partial \eta} - \frac{E}{p^2} {\bf p} \cdot \nabla \Psi \right] = \frac{Q_\alpha}{{\rm P}^0} \, , \label{Q-CN}
    \end{equation}
where $Q_\alpha= Q_\alpha^{(0)} + Q_\alpha^{(1)}$ when the background collision term is non-null. 

Using the perturbation
    \begin{equation}
        \bar{f}_{\alpha} = 
        \bar{f}_{\alpha}^{(0)}[1 + \Theta_{\alpha}] \, , \label{f_def}
    \end{equation}
we can separate the relativistic Boltzmann equation into two parts: the zeroth order part, given by Eq.~\eqref{Q-zero}, and a first-order part that depends on the gauge choice. For the CN gauge from Eq.~\eqref{Q-CN}, setting $\partial \Theta_\alpha / \partial p = 0$ (because it is a second-order derivative term within the thermal approximation that we will generalize below) and using $\nabla \bar{f}^{(0)}_\alpha =0$, from Eq.~\eqref{Q-zero} the first-order contribution is given by
    \begin{equation}
        \bar{f}^{(0)}_\alpha \left[ \frac{\partial \Theta_\alpha}{\partial \eta}  + \frac{{\bf p}}{E} \cdot \nabla \Theta_\alpha \right] + \frac{a Q_\alpha^{(0)}}{E} \Theta_\alpha + p \frac{\partial  \bar{f}^{(0)}_\alpha}{\partial p} \left[ \frac{\partial \Phi}{\partial \eta} - \frac{E}{p^2} {\bf p} \cdot \nabla \Psi \right] = \frac{a [-Q^{(0)}_\alpha \Psi + Q^{(1)}_\alpha]}{E} \, .
    \end{equation}
Using the Fourier space for the ${\bf x}$ variable, with comoving momentum ${\bf q} = a {\bf p}$ ($q= a|{\bf p}|$) and comoving energy $\epsilon = a E$, we can write $\upmu = \hat{q} \cdot \hat{k} = \hat{p} \cdot \hat{k}$, then the Boltzmann equation for massive neutrinos $\alpha = \nu_i$ in the CN gauge becomes
    \begin{equation}
        \bar{f}_{\nu_i}^{(0)}(q, \eta) 
        \left[ \frac{\partial \tilde{\Theta}_{\nu_i}}{\partial \eta} + i \frac{q}{\epsilon} \upmu \, k \tilde{\Theta}_{\nu_i} \right]
        + a C^{(0)}_{\nu_i} \tilde{\Theta}_{\nu_i} + q \frac{\partial \bar{f}_{\nu_i}^{(0)}(q, \eta)}{\partial q} 
        \left( \frac{\partial \tilde{\phi}}{\partial \eta} - i \frac{\epsilon}{q} \upmu \, k \tilde{\psi} \right) = - a C_{\nu_i}^{(0)} \tilde{\psi} + a \tilde{C}_{\nu_i}^{(1)}\, .
    \end{equation}
Here, $\tilde{\phi}$ and $\tilde{\psi}$ are the Fourier transformed variables. 

After neutrino decoupling from the primordial plasma, additional exotic particles coupled to the neutrinos may be introduced. In most scenarios, these are produced exclusively via neutrino interactions, as the neutrino distribution function is initially well approximated by its thermal form $f^{(0)}_{\nu_i}(p) = (e^{p/T} + 1)^{-1}$, where $T$ denotes the temperature of the C$\nu$B. We therefore expect the neutrino distribution to take the form 
    \begin{equation}
        \bar{f}^{(0)}_{\nu_i}(p) = \xi_{r} f^{(0)}_{\nu_i}(p) 
        \, \quad \text{with} \quad \xi_{r} = \xi_{r}(T, p) \, , \, |\xi_{r}| <1, \label{barf_def}
    \end{equation}
where $\xi_r$ encodes the effects of the neutrino background evolution and depends on the chosen parametrization. In the following, we comment on some parametrizations:
\begin{itemize}
    \item One can choose $\xi_r = \bar{n}_\nu/n_\nu^{eq} < 1$ that accounts for a possible depletion of the neutrino number density relative to its equilibrium value, analogous to the parametrization of a dark sector produced from neutrinos as in Ref.~\cite{Escudero2023seesaw}.
    \item In scenarios involving thermalization via decays and inverse neutrino decays, in the Maxwell--Boltzmann approximation, the relative population fraction $\xi_r \to r$ defined in~\eqref{fr_gen} can be used to determine the modified distribution function of mediators in Eq.~\eqref{number density-parametrized}.
    \item In thermal dark sector scenarios~\cite{Berlin2018thermal, Das2025impostor}, the background evolution is studied using the integrated Boltzmann equation while retaining the form of the energy density $\rho_i = (\pi^2/30) g_{*}^i T_i^4$, this allows one to determine the evolution of $T_i$ normalized to the photon temperature $T_\gamma$, where $i$ labels the species, together with conserved quantities such as $N_{\rm eff}$. In this work, we use this parametrization in Eq.~\eqref{Xi_chi}.
\end{itemize}

To construct the neutrino multipole moments, one uses the temperature perturbation $\delta T$, so expanding around a small temperature perturbation, for the FD case denoted by $f(p,t)$, one gets
    \begin{equation}
        f(p,t) = f^{(0)} (p) - f^{(0)}(p) \frac{d\ln f^{(0)}(p)}{d \ln p} \frac{\delta T}{T} \quad \text{because} \quad T \frac{\partial f^{(0)}(p)}{\partial T} = - f^{(0)}(p) \frac{d\ln f^{(0)}(p)}{d \ln p} \, .
    \end{equation}
So, one can compare with Eq.~\eqref{f_def} to obtain the thermal approximation~\cite{Perez2026towards}.
    
In general, $\partial \xi_r/\partial T \neq 0 $, and $\xi_r$ evolves due to processes that affect the neutrino distribution. Then, we can determine similar changes with Eq.~\eqref{barf_def},
    \begin{equation}
        T \frac{\partial (\xi_r  f^{(0)})}{\partial T} = -\bar{f}^{(0)} \left[ \frac{\partial \ln \bar{f}^{(0)}}{\partial\ln p} - \Delta(p,T) \right] \quad \text{with} \quad \Delta(p,T) = \frac{\partial \ln \xi_r}{\partial \ln p} + \frac{\partial \ln \xi_r}{\partial \ln T} \, . \label{Delta(p,T)}
    \end{equation}
We thus introduce the \emph{temperature fluctuation} variable for each neutrino mass state, which, in the limit of the FD distribution function ($\xi_r \to 1$ or $\xi_r \to {\rm constant}$), reduces to a well-known result~\cite{Kreisch2020, Perez2026towards}. In this case, however, a factor $\Delta_i = \Delta_i(p, T)$ appears with $\xi_r \to \xi_{i,r}$ and $q \to q_i$, which depends on the chosen parametrization of the neutrino distribution function evolution as
    \begin{equation}
        \tilde{\Xi}_i ({\bf k}, {\bf p}, \eta) = 4 \frac{\delta T_i}{T}  = \sum_{\ell = 0} (-i)^{\ell} (2\ell + 1) \nu_{i,\ell} {\cal P}_\ell(\upmu) \quad \text{with} \quad \nu_{i,\ell} = - \frac{4\Psi_{i,\ell}}{\left[\frac{d \ln \bar{f}^{(0)}_{\nu_i}}{d \ln q_i} -\Delta_i \right]}\, . \label{Multipoles}
    \end{equation}
This generalizes the so-called \emph{thermal approximation}: the neutrino multipole moments $\nu_{i,\ell}$ are momentum-independent and are related to the Legendre multipoles $\Psi_\ell$ appearing in the expansion of $\tilde{\Theta}_{\nu_i}={\cal F}[\Theta_{\nu_i}]$ (see Eq.~\eqref{f_def}) when the neutrino distribution function takes the exact Fermi–Dirac (FD) thermal form; we obtain the usual case used in Refs.~\cite{Kreisch2020, Perez2026towards}.
    
Under the approximation that neglects second-order derivatives as $\frac{\partial}{\partial \eta} \left( \frac{\partial \ln \bar{f}_{\nu_i}^{(0)}}{\partial \ln q_i} - \Delta_i \right)^{-1}$, we can write the modified BHEs as
    \begin{equation}
        \frac{\partial \nu_{i,\ell}}{\partial \eta} + \frac{q_i}{\epsilon_i} k \left( \frac{(\ell +1)}{(2 \ell +1)} \nu_{i, \ell +1} - \frac{\ell}{(2\ell +1)} \nu_{i, \ell -1} \right) - \frac{ a C_{\nu_i}^{(0)}}{\bar{f}_{\nu_i}^{(0)}} \nu_{i,\ell} - 4 \frac{\partial \ln \bar{f}_{\nu_i}^{(0)}}{\partial \ln q_i} \Lambda_i f_{g,\ell}^{i} - 4 a \tilde{\psi} \frac{C_{\nu_i}^{(0)}} {\bar{f}_{\nu_i}^{(0)}} \Lambda_i \delta_{0\ell} = -\frac{4a} {\bar{f}_{\nu_i}^{(0)}} \Lambda_i \tilde{C}_{i,\ell} \, , \label{BHEs-multipoles}
    \end{equation}
where
    \begin{equation}
        \Lambda_i := \left( \frac{\partial \ln \bar{f}_{\nu_i}^{(0)}}{\partial \ln q_i} - \Delta_i \right)^{-1} \, , \quad f_{g,\ell}^{i} := \frac{\partial \tilde{\phi}}{\partial \eta} \delta_{0\ell} + \frac{k}{3} \frac{\epsilon_i}{q_i} \tilde{\psi} \delta_{1\ell} \quad \text{and} \quad \tilde{C}_{i,\ell} := \frac{1}{2(-i)^\ell}\int_{-1}^{1}\! d\upmu {\cal P}_\ell(\upmu)\tilde{C}_{\nu_i}^{(1)} \, . \label{C_ell}
    \end{equation}
We expect that in the thermal approximation (see Ref.~\cite{Perez2026towards}) under the approximation $\nu_\ell \approx \nu_{i,\ell}$ for each $i$ mass eigenstate, the first-order collision term will be proportional to the neutrino moment: $\tilde{C}_{i,\ell} \approx \tilde{C}_{\ell} = C_\ell \nu_\ell$. But in the general case, it will take the form
    \begin{equation}
        \tilde{C}_{i,\ell} = \sum_j C_{ij,\ell} \,  \nu_{j,\ell} \,.
    \end{equation}
For the effectively massless neutrinos $\nu_{i,\ell} \to {\cal F}_{i,\ell}$, we use the modified energy density per degree of freedom to average over this energy density the Eqs.~\eqref{BHEs-multipoles}; we get the interaction rates $\Gamma_{ij,\ell}$,
    \begin{equation}
        \bar{\rho}_{\nu_i} = \frac{1}{2\pi^2} \int_{0}^\infty dp p^3 \bar{f}^{(0)}_{\nu_i} \Rightarrow  \frac{2} {\pi^2 \bar{\rho}_{\nu_i}} \int dp p^3 \Lambda_i \tilde{C}_{i,\ell} = \sum_j \Gamma_{ij,\ell} {\cal F}_{j,\ell} \, ,
    \end{equation}
and we obtain Eq.~\eqref{BHE-0-CN-S}. It is straightforward to see that this argument can be extended to the synchronous gauge, with $ f_{g,\ell}^{i} \to f_{g,\ell}$ analogous to Ref.~\cite{Perez2026towards} (see also Eq.~\eqref{f_g}).
\subsection{Effective relaxation-time approximation}
A particular case of the BHEs for $\nu$SI with a negligible zeroth-order collision term can be obtained using the RTA, with the perturbation presented in Eq.~\eqref{f_def}, $\bar{f}_{\nu_i}^{(0)} \to f_{\nu_i}^{(0)}$, taking the FD form of the neutrino distribution function, and the expansion in terms of Legendre polynomials analogous to Eq.~\eqref{Multipoles}. Following the standard derivation of the BHEs, for the LHS (see, e.g., Ref.~\cite{Archidiacono2014}), one gets, for massive neutrinos (for $\ell=0,1$ one takes $\Gamma \to 0$),
    \begin{equation}
        \frac{\partial \Psi_{\ell}}{\partial \eta} + \frac{q}{\epsilon} k \left( \frac{(\ell +1)}{(2 \ell +1)} \Psi_{\ell +1} - \frac{\ell}{(2\ell +1)} \Psi_{\ell -1} \right)
        + \frac{d \ln f_{\nu_i}^{(0)}}{d \ln q} 
        f_{g,\ell} = -a \Gamma \Psi_\ell \, ,
        \label{BHE-RTA}
    \end{equation}
where in the two principal gauges, we have defined $f_{g,\ell}$ as
    \begin{equation}
        f_{g,\ell} = \begin{cases}
             \frac{\partial \tilde{\phi}}{\partial \eta} \delta_{0\ell} + \frac{k}{3} \frac{\epsilon}{q} \tilde{\psi} \delta_{1\ell} & \text{for the CN gauge} \\
             \frac{\dot{\tilde{h}}+ 6\dot{\tilde{\eta}}}{15}\delta_{2\ell} - \frac{\dot{\tilde{h}}}{6}\delta_{0\ell} & \text{for the synchronous gauge}
        \end{cases} \, .
        \label{f_g}
    \end{equation}
In the RHS of Eq.~\eqref{BHE-RTA}, we have used the RTA, $f_{\nu_i}-f_{\nu_i}^{(0)} = f_{\nu_i}^{(0)} \Theta_{\nu_i}$, and the Legendre orthogonality condition.

After averaging over the neutrino energy density $\rho_\nu$, one obtains the BHEs for the effectively massless neutrinos in terms of the neutrino multipole moments ${\cal F}_\ell$~\cite{Forastieri:2015paa}, which are comparable with the thermal approximation~\cite{Perez2026towards} when the background evolution is neglected:
    \begin{equation}
        \frac{\partial {\cal F}_{\ell}}{\partial \eta}  + k \left( \frac{(\ell +1)}{(2 \ell +1)} {\cal F}_{\ell +1} - \frac{\ell}{(2\ell +1)} {\cal F}_{\ell -1} \right) - 4 f_{g,\ell} = - a \begin{cases}
            \Gamma {\cal F}_{\ell} & \text{RTA}\\
            \Gamma_\ell {\cal F}_{\ell} & \text{thermal approximation}
        \end{cases}  \, .
        \label{Boltzmann-eq-relativistic neutrinos}
    \end{equation}
According to these results, the exact effect of the collision term requires $\Gamma \to \Gamma_\ell$~\cite{Perez2026towards}, and now the instantaneous decoupling is determined by $\Gamma_\ell/H \sim 1$. This is valid when $T_\nu \gtrsim 1 \, {\rm eV} \gg m_\nu$, where $m_\nu$ is a neutrino mass. Moreover, given that $\Gamma_\ell$ is determined by the contribution of relativistic neutrinos, we propose a simple but effective replacement $\Gamma\Psi_{i,\ell} \to \sum_j \Gamma_{ij,\ell} \Psi_{j,\ell}$ for massive neutrinos with the $i$ mass eigenstate as a reference, taking the BHEs in Eq.~\eqref{BHE-RTA} as a basis. We refer to this as the \emph{effective relaxation-time approximation}:
    \begin{equation}
        \frac{\partial \Psi_{i,\ell}}{\partial \eta} + \frac{q_i}{\epsilon_i} k \left( \frac{(\ell +1)}{(2 \ell +1)} \Psi_{i,\ell +1} - \frac{\ell}{(2\ell +1)} \Psi_{i,\ell -1} \right)
        + \frac{d \ln f_{\nu_i}^{(0)}}{d \ln q_i} f_{g,\ell}^i = -a \sum_j \Gamma_{ij,\ell} \Psi_{j,\ell} \, .
        \label{BHE-RTA-massless}
    \end{equation}
Here, the label $i$ indicates a specific neutrino mass state. This means that we obtain a set of BHEs for each neutrino mass, and in Eq.~\eqref{f_g} the change is $\epsilon \to \epsilon_i = (a^2 m_i^2 + q_i^2)^{1/2}$.
\section{Interaction rates} \label{IR}
In this section, we present the estimated interaction rates for the relevant processes with the interaction~\eqref{L-scalar}.

Consider the scattering $b + c \to d + f$; we can choose the $b$ particle as our reference particle. Then, within the pseudo-chemical approximation, the corresponding integrated Boltzmann equation is
    \begin{equation}
        \frac{d n_b}{d t} + 3 Hn_b = (n_b^{eq}- n_b) \Gamma_{bc \to df} \quad \text{where} \quad \Gamma_{bc \to df} = \langle \sigma v\rangle_{bc \to df} n_c^{eq} \, . \label{BE-scattering}
    \end{equation}
Here, $n_b$ is the number density of the $b$ particle, which deviates from its thermal equilibrium form $n_b^{eq}$, $\Gamma_{bc \to df}$ is the interaction rate of the process, and it is determined in terms of the thermally averaged cross-section defined by 
    \begin{equation}
        \langle \sigma v\rangle_{bc \to df} = \frac{{\rm g}_{s,b} {\rm g}_{s,c}}{n_b^{eq} n_c^{eq}} \int \frac{d^3 p_b}{(2\pi)^3} \frac{d^3 p_c}{(2\pi)^3} f_b^{eq} f_c^{eq} v \,  \sigma(s) \quad \text{with the M{\o}ller velocity} \quad v =  \frac{ \sqrt{(p_b \cdot p_c)^2 - m_b^2 m_c^2}}{E_b E_c}  \, , \label{Thermal CS}
    \end{equation}
where ${\rm g}_{s,b}$ denotes the internal degrees of freedom of the $b$ particle, and $\sigma(s)$ is the cross-section computed in the center-of-mass frame and averaged over initial spin states, as in collider physics. 

\begin{figure}[ht!]
    \centering
    \subfigure[]{\includegraphics[width=50mm]{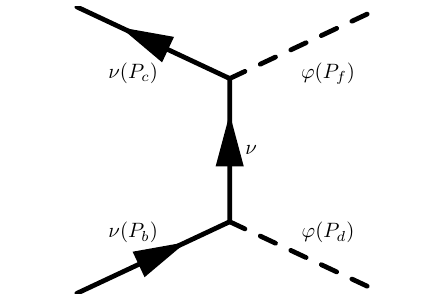}}
    \subfigure[]{\includegraphics[width=50mm]{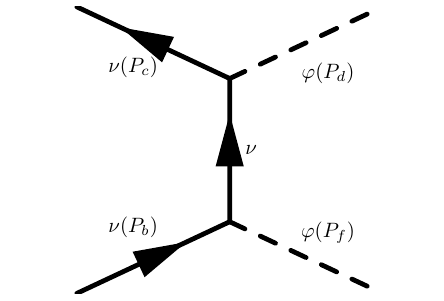}}
    \caption{Feynman diagrams at tree level involved in the process $\nu\nu \rightarrow \varphi\varphi$: a) channel t , b) channel u.}
    \label{Fig:Pair-annihilation}
\end{figure}

Expression~\eqref{Thermal CS} is derived with the MB forms of the distribution functions $f_b^{eq}$ and, of course, ignoring Pauli blocking or Bose enhancement. Here, we choose the FD distribution to improve the estimation. In the relativistic regime of the $b$ and $c$ particles, the incident angle $\theta_{bc}$ determines the M{\o}ller velocity $v=1 - \cos\theta_{bc}$,
we can write $\langle \sigma v\rangle_{bc \to df} = {\rm g}_{s,b} {\rm g}_{s,c} I/n_b^{eq} n_c^{eq}$, and in this case:
    \begin{equation}
        I = \int \frac{d^3 p_b}{(2\pi)^3} \int \frac{d^3 p_c}{(2\pi)^3} f_b^{eq}f_c^{eq} \sigma(s) v = \frac{1}{2 (2\pi)^4} \int_0^{\infty} ds \, s \, \sigma(s) \int_0^\infty dE_b  \int_0^\infty dE_c f_b^{eq} f_c^{eq} H(4E_bE_c - s) \, ,
    \end{equation}
where $H(4E_bE_c - s)$ is the Heaviside step function. The integral in energies can be written as
    \begin{equation}
        \int_0^\infty dE_b  \int_0^\infty dE_c f_b^{eq} f_c^{eq} H(4E_bE_c - s) = T_\nu^2 \int_{\sqrt{\tilde{ s}}}^\infty dx \frac{2e^{-x}}{1 - e^{-x}} \left[A + \ln\left( \frac{1 + e^{-(x/2 + A)}}{1+ e^{-(x/2 - A)}}\right) \right] \quad \text{with} \quad A=\frac{\sqrt{x^2 - \tilde{s}}}{2} \, ,
    \end{equation}
where we have used the scaling $\tilde{s} = s/T_\nu^2$. And for neutrinos ($b=\nu=c$), it is straightforward to obtain the result for the neutrino as the reference particle:
    \begin{equation}
        \Gamma_{\nu\nu \to df}  =  \frac{T_\nu^6}{4 \pi^4 n_\nu} \int_0^\infty \tilde{s} \, \sigma(\tilde{s}) F(\tilde{s}) \, d\tilde{s} \, , \quad
        F(\tilde{s}) = \int_{\sqrt{\tilde{s}}}^\infty \frac{dx e^{-x}}{1 - e^{-x}} G(x) \, , \quad
        G(x) = \frac{\sqrt{x^2 - \tilde{s}}}{2} + \ln \left( \frac{1+e^{-(x + \sqrt{x^2 - \tilde{s}})/2}}{1+e^{-(x - \sqrt{x^2 - \tilde{s}})/2}} \right) \, .  \label{Rate_neutrino}
    \end{equation}
\begin{enumerate}
    \item {\bf Resonant $\nu$SI}. We compute the resonant cross-section in the complex-mass scheme (CMS), where the propagator in the $s$-channel is modified according to $m_\varphi^2 \rightarrow m_\varphi^2 - i m_\varphi \Gamma_\varphi$, valid for $\Gamma_\varphi \ll m_\varphi$~\cite{Denner1999predictions, Willenbrock2024mass}. In our particular case, the total decay width includes a factor $f_M = 3/2$, which accounts for all possible decay channels, as well as an additional factor of $1/2$ for indistinguishable particles. Therefore, the cross-section in the center-of-mass frame is given by
        \begin{equation}
            \sigma^{\rm resonant}(\tilde{s}) = \frac{1}{T_\nu^2} \frac{g^4 f_M}{16\pi} \frac{\tilde{s}}{(\mu_\varphi^2 - \tilde{s})^2 + \mu_\varphi^2 \gamma_{\varphi}^2} \quad \text{with} \quad \gamma_{\varphi} = \frac{\Gamma_\varphi}{T_\nu} = \frac{g^2}{8\pi} \frac{m_\varphi}{T_\nu}f_M \, . \label{CS-resonant}
        \end{equation}
    The difference from the expressions in Refs.~\cite{Creque2021,venzor2023resonant} is the replacement $g \to g/\sqrt{2}$; however, our expression accounts for all allowed scattering channels and also includes the total decay width according to the CMS prescription.

    Eq.~\eqref{CS-resonant} produces a resonant contribution when we choose a reference neutrino. For example, for $i=1$, the corresponding processes are $\nu_1 \nu_1 \to \nu_1 \nu_1, \nu_2 \nu_2, \nu_3 \nu_3$:
        \begin{equation}
            \Gamma_{\nu{\rm SI}}^{\rm R} = \frac{g^4 T_\nu f_M}{96 \pi^3 \zeta(3)} \int_{0}^\infty \frac{\tilde{s}^2 F(\tilde{s})}{(\mu_\varphi^2 - \tilde{s})^2 + \mu_\varphi^2 \gamma_{\varphi}^2} \, d\tilde{s} \,. \label{Rate_NSI}
        \end{equation}
    On the other hand, using the NWA, the expression in Eq.~\eqref{CS-resonant} becomes
        \begin{equation}
            \sigma_{\rm NWA}^{\rm resonant}(\tilde{s}) = \frac{1}{T_\nu^2} \frac{g^2 f_M}{16\pi} \lim_{g \to 0}\frac{g^2 \tilde{s}}{(\mu_\varphi^2 - \tilde{s})^2 + \mu_\varphi^2 \gamma_{\varphi}^2} = \frac{g^2}{2 T_\nu^2} \pi \delta(\tilde{s} - \mu_\varphi^2) \, .
        \end{equation}
    Then, we obtain the result presented in Eq.~\eqref{Resonat-NSI}. 
    
    Notice that cosmological perturbation theory accounts for the process $\nu \nu \to \nu \nu$ by including the back-reaction through the distortion of final-state distributions, which leads to non-null terms ${\cal D}_\ell^{(r)}$ in Eq.~\eqref{ctilde_1_res}, whereas in the RTA for the BHEs this back-reaction is not captured, or its contribution is considered negligible.
    \item {\bf Pair-annihilation}. We consider the process $\nu \nu \to \varphi \varphi$ (see Fig.~\ref{Fig:Pair-annihilation}), where $\varphi$ is a massive particle with $m_\varphi \gg m_\nu$. Neglecting only the neutrino masses, the squared amplitude that neglects only the neutrino masses in terms of Mandelstam variables $s,t$ is:
        \begin{equation}
        \begin{aligned}
            \overline{|{\cal M}_A|^2} &= -\frac{g^4}{2} \bigg\{ \frac{[st + (t - m_\varphi^2)^2]}{t^2} + \frac{s[2m_\varphi^2 - (s+t)] + [m_\varphi^2 - (s+t)]^2}{[2m_\varphi^2 - (s+t)]^2} + \frac{ 2\left[ [m_\varphi^2 - (s+t)](t-m_\varphi^2) - sm_\varphi^2 \right]}{t [2m_\varphi^2 - (s+t)]}  \bigg\} \\
            &= -\frac{g^4}{2} \bigg\{ 4 + \left[\frac{s^2 - 4m_\varphi^2 s + 6m_\varphi^4}{(s - 2m_\varphi^2)} \right] \left(\frac{1}{t} + \frac{1}{u} \right) + \frac{m_\varphi^4}{t^2} + \frac{m_\varphi^4}{u^2} \bigg\} \quad \text{with} \quad u= 2m_\varphi^2 - (s+t) \, . \label{Pair-annihilation-amplitude}
        \end{aligned}
        \end{equation}
    From Eq.~\eqref{Pair-annihilation-amplitude}, we obtain the well-known result for $m_\varphi \to 0$ (see Refs.~[16,17] in Ref.~\cite{Beacom2004}). 

    In the center-of-mass frame, the limits for the $t$ variable are: $t_{\pm} =  - \frac{s}{4} (1 \mp \beta_\varphi)^2$ with $\beta_\varphi = (1- 4m_\varphi^2/s)^{1/2}$, which imposes the threshold $s \geq 4m_\varphi^2$. The cross-section scaled with the temperature $T_\nu$ takes the form
        \begin{equation}
            \sigma_{\nu\nu \to \varphi \varphi} (\tilde{s}) = \frac{1}{T_\nu^2} \frac{g^4}{16\pi \tilde{s}} \left\{\left[\frac{\tilde{s}^2 - 4\mu_\varphi^2 \tilde{s} + 6\mu_\varphi^4}{\tilde{s}(\tilde{s} - 2\mu_\varphi^2)} \right] \ln \frac{(1+\beta_\varphi)}{(1-\beta_\varphi)} - \beta_\varphi \right\} H(\tilde{s}-4 \mu_\varphi^2) \, ,
        \end{equation}
    where $H(\tilde{s}-4 \mu_\varphi^2)$ is the Heaviside step function. Then, we obtain the interaction rate~\eqref{Rate-pair-annihilation}.
\end{enumerate}
Note that we have neglected the neutrino masses; therefore, the same expressions also apply to a pseudoscalar $\phi$.
\section{Effective resonant behavior}
In this part, we present detailed arguments showing when the resonant $\nu$SI approach becomes more natural than out-of-equilibrium decays that produce thermalization via decays (see Refs.~\cite{Escudero2020CMBsearch, Li2023neff}). For this purpose, in Sec.~S3.1, we demonstrate that the resonant regime can be consistently described within first-order cosmological perturbation theory. In Sec.~S3.2, we discuss the equivalence between the descriptions of these phenomena and provide detailed calculations supporting the arguments presented in the main text. Finally, in Sec.~S3.3, we discuss how the assumption of instantaneous thermalization in the relativistic regime affects boson production through inverse decays, and its consequences for $N_{\rm eff}^{\rm CMB}$.
\subsection{Resonant neutrino self-interactions: events and contribution}
In the recent work~\cite{Perez2026towards}, we obtained the collision term within the thermal approximation when the background is irrelevant; one non-resonant contribution to the collision term at first order for $\nu_i \nu_i \to \nu_i \nu_i$ scattering (without degeneracy factors) is given by
     \begin{equation}
        \tilde{C}^{(1)}_{\nu_i} = \frac{1}{4} \frac{g^4 T_\nu}{128 \pi^3 z_1}
        \frac{d \ln f^{(0)}(p_1)}{d \ln p_1} \sum_{\ell=0 }^\infty (-i)^{\ell}(2\ell+1) \nu_{\ell} {\cal P}_\ell(\upmu) \,  \left[ {\cal A}(z_1)+ {\cal B}_\ell(z_1) - 2{\cal D}_\ell(z_1) \right] \, ,
        \label{ctilde_1}
    \end{equation}
where the local momentum of the reference neutrino ${\bf p}_1$ with the vector wavenumber ${\bf k}$ defines the argument in the Legendre polynomials as $\upmu = \hat{\bf p}_1 \cdot \hat{\bf k}$, the energies and neutrino masses are scaled
$z_i = E_i/T_\nu$ and $\mu_i = m_i/T_\nu$ with respect to the temperature of the neutrino background, $T_\nu$.

In Eq.~\eqref{ctilde_1}, the ${\cal A}(z_1),  {\cal B}_\ell(z_1)$, and ${\cal D}_\ell(z_1)$ functions depend on the $J$-integrals~\cite{Perez2026towards}. The first two correspond to contributions from the multipoles of incoming neutrinos, while the latter term corresponds to final-state neutrinos (in the relativistic regime well before neutrino free-streaming due to $\nu$SI, $\nu_\ell \approx \nu_{i,\ell}$); the latter contributes as a back-reaction of the process. 

\begin{figure}
\subfigure[]
{\includegraphics[width=0.4\linewidth]{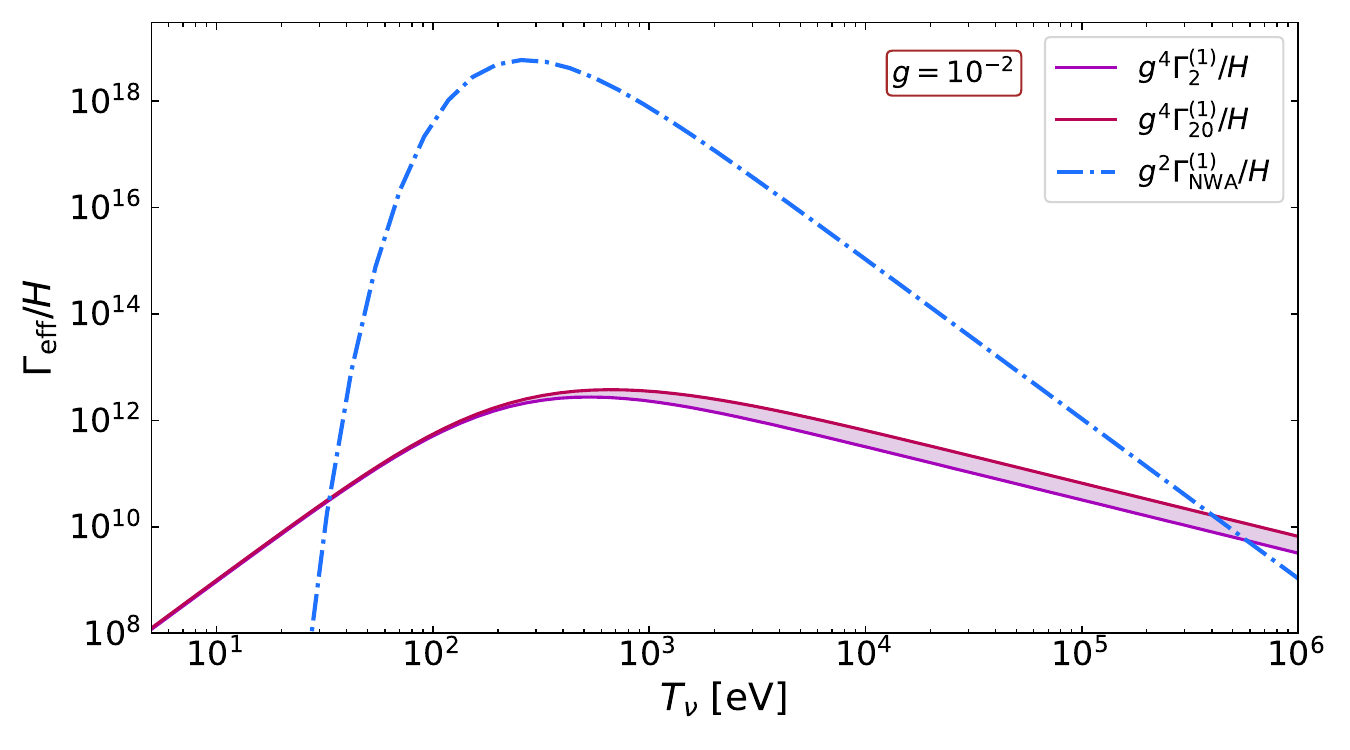}}
\subfigure[]
{\includegraphics[width=0.4\linewidth]{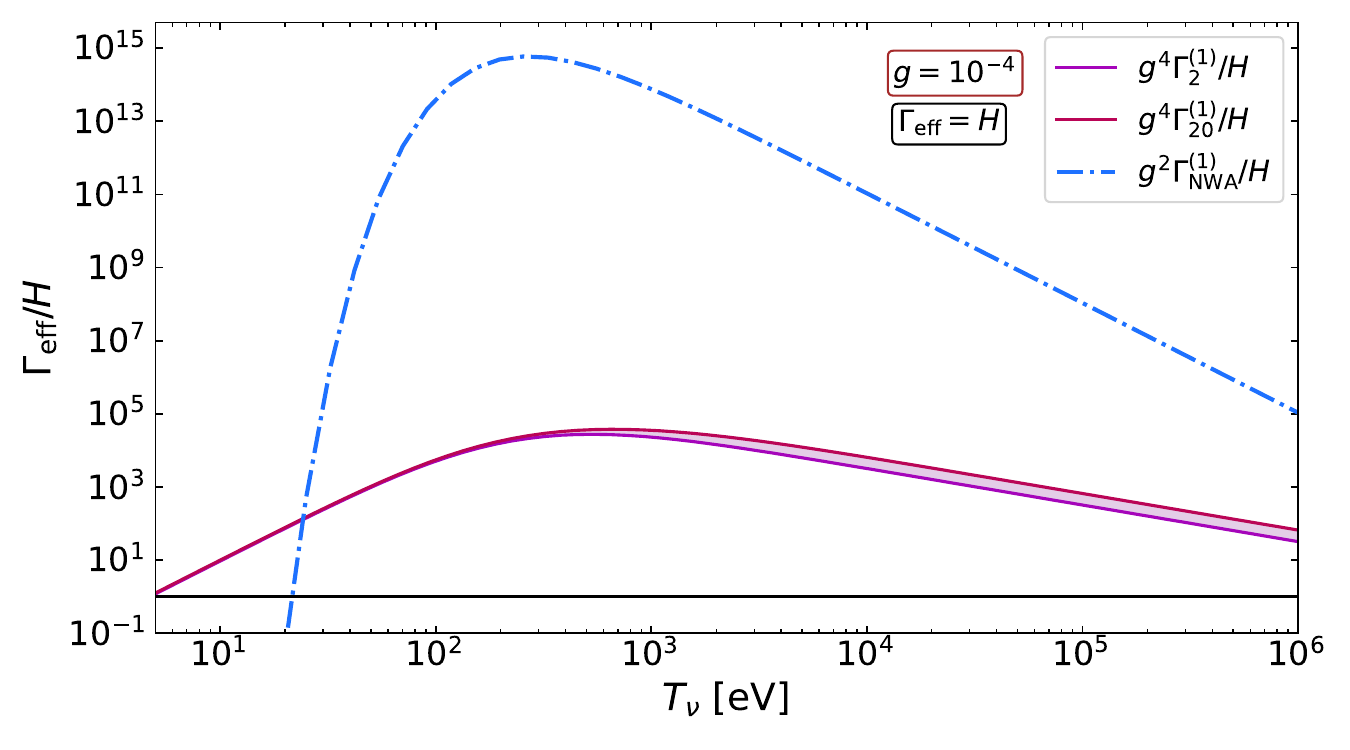}}
\subfigure[]
{\includegraphics[width=0.4\linewidth]{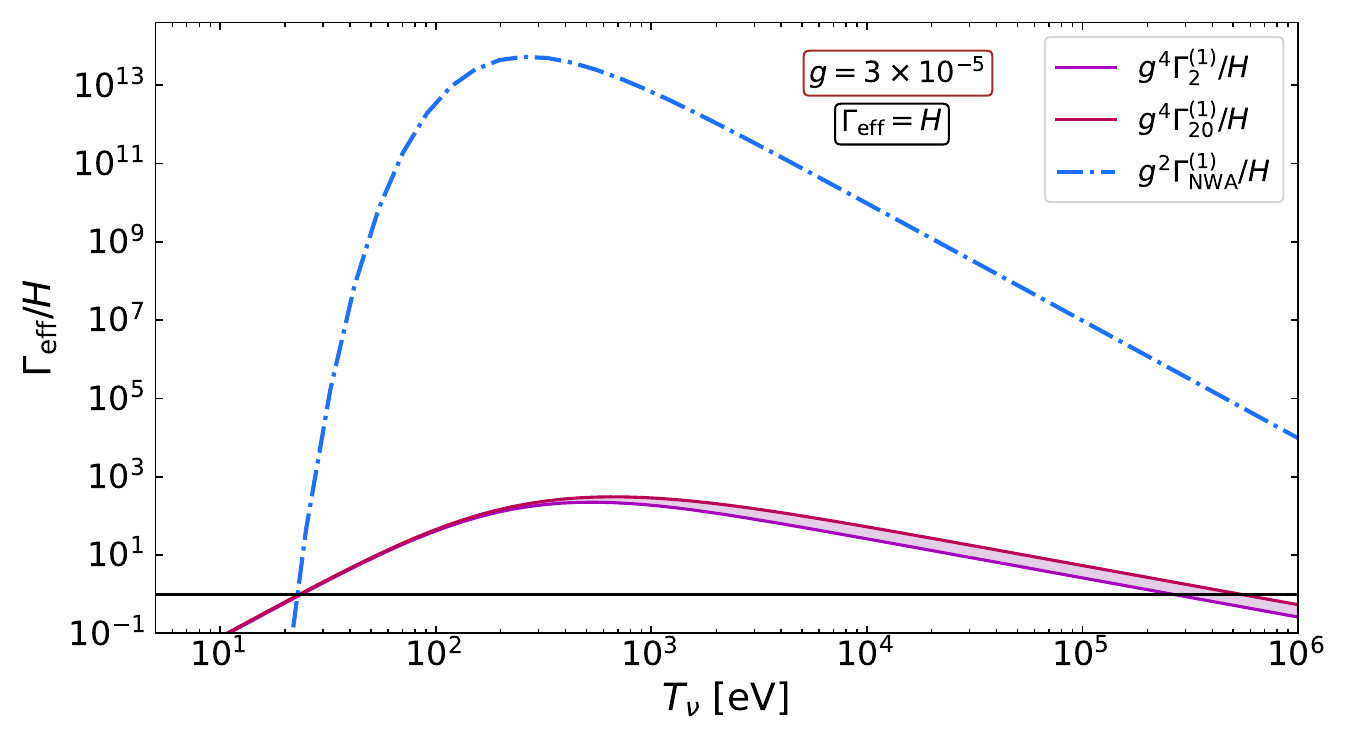}}
\subfigure[]
{\includegraphics[width=0.4\linewidth]{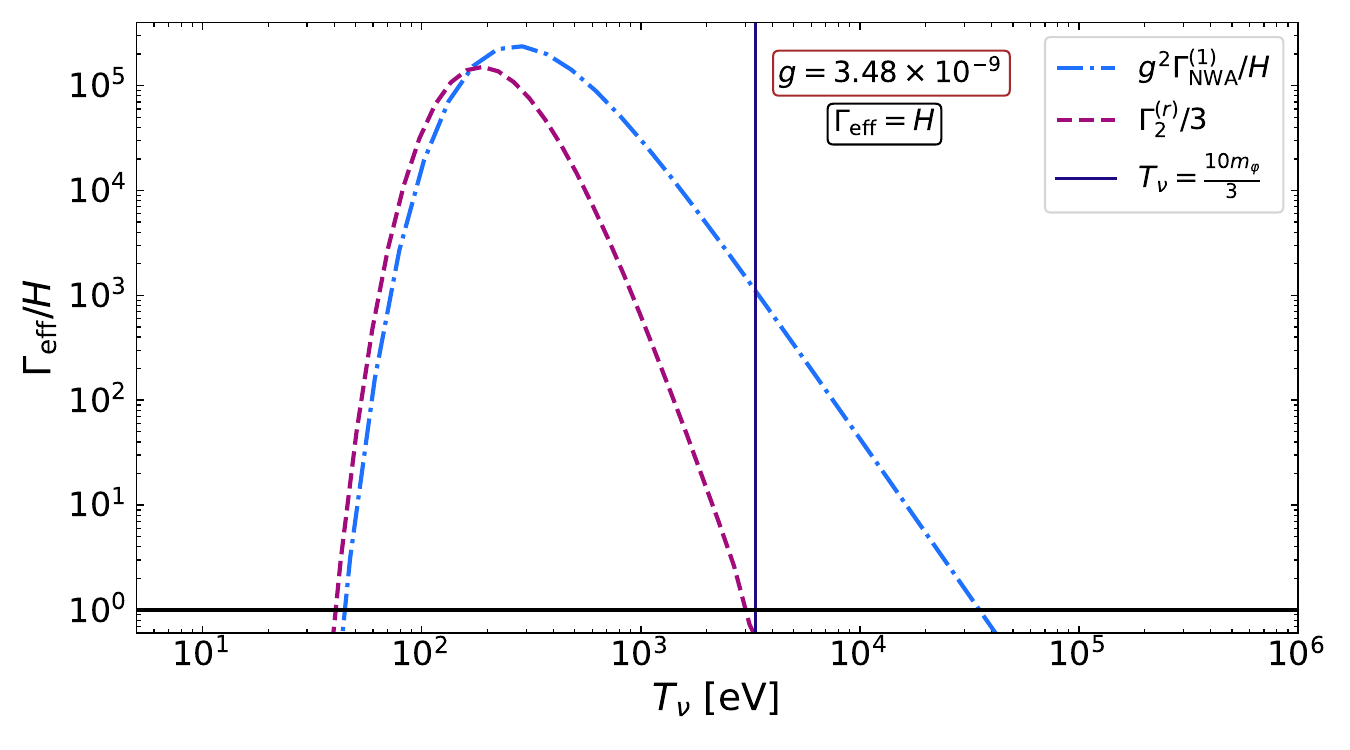}}
\caption{\label{fig:Comparative_rates} Resonant versus non-resonant contributions to $\nu$SI in the diagonal coupling scenario. Comparative plots (real scalar case) of the interaction rates with $m_\varphi = 1 \, {\rm keV}$ for a single contribution using the $t$- and 
$u$-channel processes neglecting the $s$-channel contribution for $\ell =2$ up to $\ell=20$ are shown against the resonant contribution in the narrow-width approximation, $\Gamma_{\rm NWA}^{(1)}$: (a) $g=10^{-2}$, (b) $g=10^{-4}$, (c) $g=3 \times 10^{-5}$. Due to this scaling in $g$, as $g$ decreases, the resonant $\nu$SI contribution is within the tightly coupled regime and only dominates the cosmic $\nu$SI scenario once the non-resonant contributions become Hubble-diluted ($g^4 \Gamma_\ell^{(1)} < H$). For $g<3 \times 10^{-5}$, only the resonance is relevant; in panel (d), we compare the estimated interaction rate with the full computation of the interaction rate in the first-order collision term for a single contribution $\Gamma_2^{(r)}/3$~\cite{Perez2026Addressing}.}
\end{figure}

We can determine the interaction rate for one contribution like $\nu_1 \nu_1 \to \nu_1 \nu_1$ as
    \begin{equation}
        \Gamma_\ell = \frac{g^4 \beta_\ell(\mu_\varphi)}{16}= g^4 \Gamma_\ell^{(1)} \, , \label{Rate_nNSI}
    \end{equation}
where the form of $\beta_\ell(\mu_\varphi)$ is determined by the first order collision term~\eqref{ctilde_1} see details in Ref.~\cite{Perez2026towards}.

In Fig.~\ref{fig:Comparative_rates} for the scalar case,  we compare the different contributions; resonant $\nu$SI in the estimation~\eqref{Resonat-NSI}, $\Gamma_{\nu\nu \to \nu\nu} = 3 g^2 \Gamma_{\rm NWA}^{(1)}$, against a single contribution using the $t$- and 
$u$-channel processes from~\cite{Perez2026towards} neglecting the $s$-channel contribution, written as Eq.~\eqref{Rate_nNSI} for $\ell =2$ up to $\ell=20$. This illustrates a key feature of resonant $\nu$SI: the coupling $g$ determines whether the process is faster than the Hubble expansion rate.
Notice that the coupling plays an important role in determining when the interaction is relevant against non-resonant contributions, as shown in panels (a), (b) and (c) of Fig.~\ref{fig:Comparative_rates}. From panel (c), resonant $\nu$SI does not interfere with the HML, which is valid for $T_\nu < 100 \, m_\varphi$~\cite{Perez2026towards}.

In panel (d) of Fig.~\ref{fig:Comparative_rates}, we see that the exact computation of the resonant contribution in the thermal approximation~\cite{Perez2026Addressing} only slightly modifies the estimation at low temperatures. However, the relevant feature at high temperatures $T_\nu \gg m_\varphi/3$ (we highlight the line $\mu_\varphi = 0.3$) is that the estimation of the interaction rate for resonant $\nu$SI fails~\eqref{Resonat-NSI}, as studied in detail in Ref.~\cite{Perez2026Addressing}. This is because, near the mediator mass, the resonance is more likely to be produced.

In order to explore the on-shell production in a model-independent way, we first characterize the fraction of collisions associated with the resonant invariant energy, $s=M^2$ or $\tilde{s}=\mu_M^2$, where $\mu_M= M/T_\nu$, by adopting the delta-function approximation for the reduced cross-section, $\sigma_{\rm red}(\tilde{s}) = \delta(\tilde{s}-\mu_M^2)$. Then, the corresponding fraction of collisions with the on-shell condition is
    \begin{equation}
        {\rm Frac}_{\tilde{s}} = \int_0^\infty \tilde{s} \,\sigma_{\rm red}(\tilde{s})F(\tilde{s}) d\tilde{s} = \mu_M^2 F(\mu_M^2) \, .
    \end{equation}
This expression quantifies the resonant contribution under the adopted normalization of the reduced cross-section.

To assess how the distribution is spread across different invariant-energy scales, we introduce the distribution per logarithmic interval as
    \begin{equation}
        \frac{dN}{d\tilde{s}} = \tilde{s}F(\tilde{s}) \Rightarrow \frac{dN}{d\log\tilde{s}} = \tilde{s}\frac{dN}{d\tilde{s}} = \tilde{s}^2F(\tilde{s})  \, .
    \end{equation}
Evaluating this distribution at the resonance yields
    \begin{equation}
    \frac{dN}{d\log\tilde{s}}\bigg|_{\tilde{s}=\mu_M^2}
    = \mu_M^4F(\mu_M^2) \, . \label{logarithmic distribution}
    \end{equation}
While ${\rm Frac}_{\tilde{s}}$ characterizes the resonant fraction, the logarithmic distribution describes the relative weight per logarithmic interval in invariant energy, facilitating comparison of contributions across energy scales. In Fig.~\ref{fig:Events}, we show the distribution per logarithmic interval normalized to obtain the maximum as $1$.

Notice also that the expression~\eqref{logarithmic distribution} is derived using the interaction rate that arises at the background level and is inaccurate for $T_\nu \gg M$ (see the scalar case in panel (d) of Fig.~\ref{fig:Comparative_rates}). This estimation, in fact, is comparable with the effective interaction rates constructed with the first-order collision term within the thermal approximation.

In the limit of equal multipole moments for a scalar mediator ($\mu_M = \mu_\varphi$), the first-order collision term for resonant $\nu$SI has the form
\begin{equation}
        \tilde{C}^{(r, 1)}_{\nu_i} = \frac{1}{4} \frac{g^2 T_\nu}{128 \pi^3 q_1}
        \frac{d \ln f^{(0)}(p_1)}{d \ln p_1} \sum_{\ell=0 }^\infty (-i)^{\ell}(2\ell+1) {\cal P}_\ell(\upmu) [{\cal A}^{(r)}(q_1) + {\cal B}_\ell^{(r)}(q_1) - 2{\cal D}_\ell^{(r)} (q_1)] \, {\cal F}_{\ell} \, .
        \label{ctilde_1_res}
    \end{equation}
The term ${\cal A}^{(r)}(q_1)$ is multipole independent. In particular, in the MB approximation for the distribution function, together with the NWA, after averaging over the neutrino energy density, we need to perform the integral
    \begin{equation}
        \int_0^\infty dq_1 q_1^2 {\cal A}^{(r)} (q_1) = \frac{4\pi^2 \mu_\varphi^2}{f_M} \int_0^\infty dq_1 q_1 \int_0^\infty dq_2 \frac{ (q_1 + q_2) e^{-(q_1 + q_2)}}{\sqrt{(q_1 + q_2)^2 - \mu_\varphi^2}}H(1-|x_\delta|) =  \frac{2\pi^2 \mu_\varphi^2}{f_M} \Gamma(3,\mu_\varphi) \, ,
    \end{equation}
where $x_\delta = 1 - \mu_\varphi^2/(2q_1 q_2)$ is fixed from the on-shell condition, and $\Gamma(3, \mu_\varphi)$ is the incomplete gamma function; therefore, neglecting the back-reaction effect, which depends on the multipole moment and other $\ell$-dependent moments
    \begin{equation}
        \Gamma_{\nu{\rm SI}}^{(r)} \propto g^2 T_\nu \mu_\varphi^2 \Gamma(3, \mu_\varphi) \, .
    \end{equation}
Here, the incomplete gamma function $\Gamma(3, \mu_\varphi)$, plays the role of $F(\mu_\varphi^2)$, and then we also show in Fig.~\ref{fig:Events} the corresponding distribution per logarithmic interval; we highlight the band 
    \begin{equation}
        \mu_\varphi \sim (0.3, 20) \,. \label{Resonant prod}
    \end{equation}
The choice of this temperature range is motivated by the requirement that $T_\nu \gg m_\varphi/3$ since, as shown in panel (d) of Fig.~\ref{fig:Comparative_rates}, the effective interaction rate $\Gamma_2^{(r)}$~\cite{Perez2026Addressing}, which characterizes the collisional damping of the quadrupole moment of the neutrino distribution and, hence, its anisotropic stress, is overestimated by orders of magnitude when using the simplest estimate in Eq.~\eqref{Resonat-NSI}. The corresponding fraction per logarithmic interval remains below ${\cal O}(10^{-4})$ for $T_\nu\gtrsim 10 \, m_\varphi/3$, indicating that the realistic contribution is strongly suppressed at high temperatures. In the following subsection, we build on this insight to investigate realistic on-shell production over cosmological timescales.

\subsection{Effective resonant behavior: Resonant neutrino self-interactions vs thermalization via decays}
From now on, we use the symbol $A$ to denote the new mediator ($A=\varphi, \phi, Z^\prime, {\rm etc.}$) to keep the discussion as general as possible. If a light mediator has a mass below the neutrino decoupling temperature, $T_\nu \sim 2 \, {\rm MeV}$, and the coupling is sufficiently small (below the blue line (A$\nu$) in Fig.~\ref{fig:PS}), the production via inverse decays $\nu \nu \to A$ is possible in the freeze-in regime. As the Universe cools, the decay process $A \to \nu \nu$ becomes allowed, leading to thermalization. Once production becomes negligible, the mediator population is depleted through decays, effectively removing the $A$ particles. In the following, we examine the limitations of this framework and establish the conditions under which the system can be effectively described as resonant $\nu$SI.

A key observation is that for ${\rm q} \sim m_A$, both the decay and inverse-decay channels are effectively captured by the resonant $\nu\nu \to A \to \nu\nu$ scattering. In the NWA, this resonant scattering reduces to the corresponding decay amplitude (not spin-averaged) as
    \begin{equation}
        |{\cal M}_{\nu \nu \to \nu\nu}|^2 \bigg|_{{\rm q} \sim m_A} = \frac{|{\cal M}_{\nu \nu \to A}|^2 \, |{\cal M}_{A \to \nu\nu}|^2}{({\rm q}^2 - m_A^2)^2 + m_A^2 \Gamma_{A}^2} \quad \underbrace{\;\Longrightarrow\;}_{\rm NWA} \quad |{\cal M}_{\nu \nu \to \nu\nu}|^2 = \frac{|{\cal M}_{\nu \nu \to A}|^2}{m_A \Gamma_A} |{\cal M}_{A \to \nu\nu}|^2 \pi \delta({\rm q}^2-m_A^2) \, . \label{Amplitudes-factorization}
    \end{equation}
We expect that the kinematics allows the production of the particle over a set of available energies within the collision terms for the $s$-channel, then $s={\rm q}^2$.  The equivalence in Eq.~\eqref{Amplitudes-factorization} is exact for the real (pseudo)scalar case, but it does not hold for the vector resonance. In general, the replacement~\eqref{Amplitudes-factorization} is valid only if spin correlations between the resonance's production and decay processes can be neglected.

For $\nu \bar{\nu} \to A$ and $A \to \nu\bar{\nu}$, the Boltzmann equation for the $A$ particle is given by (using conventions that are equivalent to those adopted in Ref.~\cite{Li2023neff}),
    \begin{equation}
        \frac{1}{a^3(t)}\frac{d [(N_A \bar{n}_A)a^3(t)]}{dt} = \frac{dn_A}{dt} + 3Hn_A = \Gamma_A^{\rm p} - \Gamma_A^{\rm d} = N_\nu |{\cal M}_{A \to \nu\bar{\nu}}|^2 [C^{\rm p}_A - C^{\rm d}_A] \, , \label{A-Eq}
    \end{equation}
where the number density of the $A$ particle is $n_A = N_A \bar{n}_A$, with $N_A$ denoting the number of internal degrees of freedom and $\bar{n}_A$ the number density per degree of freedom. The factor $N_\nu$ accounts for the sum over all neutrino contributions in the limit of negligible neutrino masses, including the appropriate symmetry factors for indistinguishable particles. For a diagonal coupling, one obtains $N_\nu=3$. The quantity $|{\cal M}_{A \to \nu\bar{\nu}}|^2$ denotes the squared matrix element (not spin-averaged) evaluated in the rest frame of the $A$ particle.

Assuming CP invariance, the same amplitude describes both the decay process and its inverse. Then, the explicit expressions for $C^{\rm p}_A$ and $C^{\rm d}_A$ are given by
    \begin{align}
        C^{\rm p}_A &= \int d\Pi_{A} d\Pi_{\nu}d\Pi_{\bar\nu} (2\pi)^4 \delta^4(P_1 + P_2 - P_A) f_\nu f_{\bar\nu} [1 + f_A] \, , \label{C_p}\\
        C^{\rm d}_A &= \int d\Pi_{A} d\Pi_{\nu}d\Pi_{\bar\nu} (2\pi)^4 \delta^4(P_A - P_1 
        -P_2)  f_A [1 - f_\nu][1 - f_{\bar\nu}] \, , \label{C_d}
    \end{align}
where $d\Pi_{b} = d^3 p_b/[2E_b(2\pi)^3]$, $f_b$ is the distribution function of particle $b=A, \nu, \bar\nu$, and $p$ stands for production as $d$ stands for depletion.

In this scenario, $C_A^{\rm p} = C_\nu^{\rm d}$ and $C_A^{\rm d} = C_\nu^{\rm p}$; these equalities are also valid for antineutrinos $\bar{\nu}$. For this reason, it is important to clarify any ambiguities regarding the nature of neutrinos.
\begin{itemize}
    \item {\bf Dirac neutrinos}. We have $N_\nu = N_{\bar\nu}$,
        \begin{align}
            \frac{1}{a^3(t)}\frac{d [(N_\nu \bar{n}_\nu)a^3(t)]}{dt} &= N_\nu |{\cal M}_{A \to \nu\bar{\nu}}|^2 [C^{\rm p}_\nu - C^{\rm d}_\nu] = -N_\nu |{\cal M}_{A \to \nu\bar{\nu}}|^2 [C^{\rm p}_A - C^{\rm d}_A] \, , \label{D-nu}\\
            \frac{1}{a^3(t)}\frac{d [(N_{\nu} \bar{n}_{\bar\nu})a^3(t)]}{dt} &= N_\nu |{\cal M}_{A \to \nu\bar{\nu}}|^2 [C^{\rm p}_{\bar\nu} - C^{\rm d}_{\bar\nu}] = -N_\nu |{\cal M}_{A \to \nu\bar{\nu}}|^2 [C^{\rm p}_A - C^{\rm d}_A] \, . \label{D-antinu}
        \end{align}
    Then, we have the following conserved quantities
        \begin{equation}
            (N_A \bar{n}_A + N_\nu \bar{n}_\nu)a^3(t) = {\rm constant} \quad \text{and} \quad (N_A \bar{n}_A + N_\nu \bar{n}_{\bar\nu})a^3(t) = {\rm constant} \, . \label{C-law}
        \end{equation}
    Recall that $\bar{n}_\nu$ is defined per degree of freedom.
    \item {\bf Majorana neutrinos}. In the inverse decay $\nu \nu \to A$, we include a factor of $1/g_\nu = 1/2$ because the reference Majorana particle has two degrees of freedom, as is standard in collision terms. This implies that $N_\nu \to N_\nu/2$:
        \begin{equation}
            \frac{1}{a^3(t)}\frac{d [(N_\nu \bar{n}_\nu)a^3(t)]}{dt} = \frac{1}{2}N_\nu |{\cal M}_{A \to \nu\nu}|^2 [C^{\rm p}_\nu - C^{\rm d}_\nu] = - \frac{1}{2} N_\nu |{\cal M}_{A \to \nu\nu}|^2 [C^{\rm p}_A - C^{\rm d}_A] \,.  \label{M-L}
        \end{equation}
    As the final states are indistinguishable in $A \to \nu\nu$, we need to add a $1/2$ factor in the RHS of Eq.~\eqref{A-Eq}. The conservation law then applies to a single chirality in this case,
        \begin{equation}
            (N_A \bar{n}_A + N_\nu \bar{n}_\nu)a^3(t) = {\rm constant} \, .  \label{C-law-Majorana}
        \end{equation}
\end{itemize}
The generalization of the previous conservation laws requires that in the RHS of Eqs.~\eqref{A-Eq}, \eqref{D-nu}, \eqref{D-antinu}, and \eqref{M-L}, $N_\nu \to N_\nu^\prime$, which in general is not equal to $3$ due to neutrino mixing $g_{ij} = g^{\alpha\beta} U_{\alpha i}^{*} U_{\beta j}$ ($g_{ij} = g^{\alpha\beta} U_{\alpha i} U_{\beta j}$) for the Dirac (Majorana) case.

A similar conservation law arises by considering the energy densities per degree of freedom of neutrinos, $\bar{\rho}_\nu$, and for $A$ particles $\bar{\rho}_A$, when both are relativistic, as~\cite{Li2023neff}: 
    \begin{equation}
        (2N_\nu \bar{\rho}_\nu + N_A \bar{\rho}_A) a^4(t) = {\rm constant} \, . \label{ed-conservation}
    \end{equation}
The amplitude for decays takes the general form:
    \begin{equation}
        |{\cal M}_{A \to \nu \nu}|^2 = \kappa g_A^2 m_A^2 \, , \label{Amplitude-gen}
    \end{equation}
with $\kappa$ a constant of order $1-10$. This depends on the neutrino--mediator interaction, and we absorb the indistinguishability factor of $1/2$ into $\kappa$ for the diagonal-coupling case.

In the following, we will use the symbol $\nu$ without specifying the neutrino nature. Using the MB distribution function and neglecting the Pauli blocking factors, the production rate is
    \begin{equation}
        \Gamma_A^{\rm p} = N_\nu \frac{|{\cal M}_{A \to \nu\nu}|^2}{32\pi^3}  T_\nu m_A K_1(m_A/T_\nu) \, , \label{production rate}
    \end{equation}
where $K_1$ is the modified Bessel function of the second kind, part of the family $K_n$ of integer-order functions.

Given that, in the standard cosmology, the entropy density of the Standard Model (SM), ${s}_{\rm SM}$, is approximately conserved in a comoving volume $V=a^3(t)$, and that after $e^\pm$ annihilation this entropy density is determined by the photon and neutrino baths separately, while the number density of the new mediator, $n_A$, evolves according to its production rate, these relations translate, respectively, into
    \begin{equation}
        \frac{d {s}_{\gamma}}{dt} + 3H {s}_{\gamma} \approx 0 \quad \text{and} \quad \frac{dn_A}{dt} + 3Hn_A \approx \Gamma_A^{\rm p}\, ,
    \end{equation}
where $H$ is the Hubble dilution term and ${s}_{\gamma}$ is the entropy density of the photon bath. 

Since the effective entropy degrees of freedom $g_{s*}$ change only slightly after neutrino decoupling, the number density of $A$ particles can be estimated as
    \begin{equation}
        n_A = - s_p \int_{s_i}^{s_p}  \frac{\Gamma_A^{\rm p}}{3H s_{\gamma}^2} d s_{\rm \gamma} \,,
    \end{equation}
where $s_p$ ($s_i$) is the photon bath entropy density at the temperature $T_p$ ($T_i$), with $T_p < T_i$. Particularly, for the production rate given in Eq.~\eqref{production rate}, taking the neutrino bath temperature as the reference temperature:
    \begin{equation}
        n_A = - T_{\nu}^3 \frac{M_{\rm pl}}{g_H}  N_\nu \frac{|{\cal M}_{A \to \nu\nu}|^2}{32\pi^3} m_A \int_{T_{\nu,i}}^{T_{\nu}} (T_\nu^\prime)^{-5} K_1(m_A/T_\nu^\prime) \, dT_\nu^\prime \quad \text{with} \quad g_H = \frac{1}{3} \sqrt{\frac{4\pi^3}{5} g_{*}(T_\nu)} \approx {\rm constant} \, . \label{n_A-prod}
    \end{equation}
From now on, we also work with an auxiliary variable defined as the inverse of the scaled mass $\beta = T_{\nu}/m_A = 1/\mu_A$. This is convenient because it quantifies how far the temperature is from the mass $m_A$.

Using the change of variable $z= T_\nu^\prime/m_A$, the limits of integral in Eq.~\eqref{n_A-prod} are $\beta$ and $\beta_0 = T_{\nu,i}/m_A$, we obtain the evolution of the number density of $A$ particles in terms of its decay width $\Gamma_A$ as
    \begin{equation}
        n_A = \beta^3 \frac{M_{\rm pl}}{g_H} \frac{N_A \Gamma_A}{2\pi^2} m_A \, I(\beta, \beta_0) \quad \text{with} \quad I(\beta, \beta_0) := \int_{\beta}^{\beta_0} {\rm z}^{-5} K_1(1/{\rm z}) \, d{\rm z} \, . \label{A-number density}
    \end{equation}
It is worth noting that the integral, $I(\beta, \beta_0)$, is monotonically decreasing, and its asymptotic value can be obtained by taking $\beta \to 0$ and $\beta_0 \to \infty$. This asymptotic behavior provides a general bound for the production temperature, $T_{\nu}^{\rm asymp} = (9\pi/2)^{-1/3} m_A \approx 0.414 \, m_A$. This means that the maximum production is in the non-relativistic regime~\cite{Li2023neff}. 

For thermalization via decays in the BHEs, thermal particle distribution functions are assumed. If we choose to follow the $A$ particle, we obtain the interaction rate $\Gamma_{{\rm int}, A}$, while following neutrinos per degree of freedom yields a different rate $\Gamma_{{\rm int}, \nu}$. These rates are given, respectively, by  
    \begin{equation}
        \Gamma_{{\rm int}, A} = \Gamma_A \frac{K_1(m_A/T_\nu)}{K_2(m_A/T_\nu)} \bigg|_{n_A^{eq}|_{(m_A \to 0)}}= \frac{\Gamma_A}{2} \frac{m_A^2}{T_\nu^2} K_1(m_A/T_\nu) \quad \text{and} \quad  \Gamma_{{\rm int}, \nu} = \frac{\Gamma_A N_A}{2 N_\nu} \frac{m_A^2}{T_\nu^2} K_1(m_A/T_\nu) \, . \label{Gamma_int_A}
    \end{equation}
However, the expression of $\Gamma_{{\rm int}, A}$ that takes into account the non-relativistic regime of the $A$ particle does not describe the physical decoupling when the production $\nu \nu \to A$ becomes negligible, thereby breaking the usual pseudo-chemical approximation, which is one of the starting points for constructing the RTA.

\begin{figure}
\subfigure[]{
\includegraphics[width=0.4\linewidth]{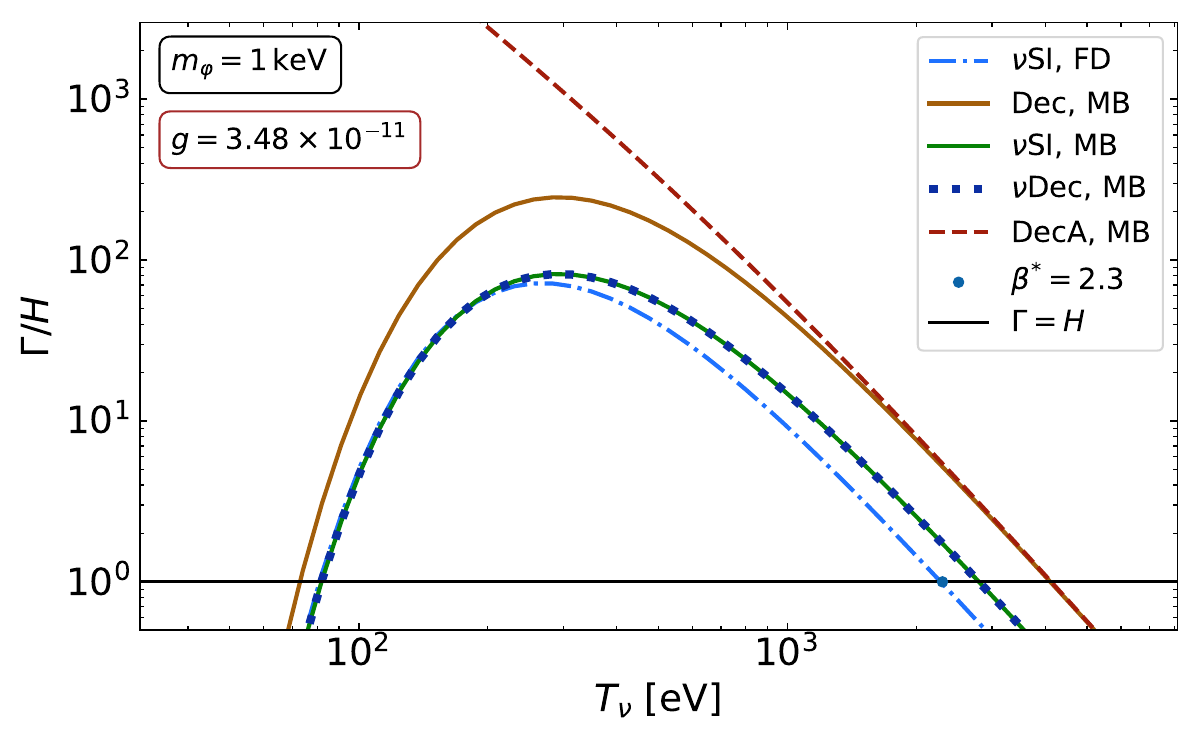}}
\subfigure[]{
\includegraphics[width=0.4\linewidth]{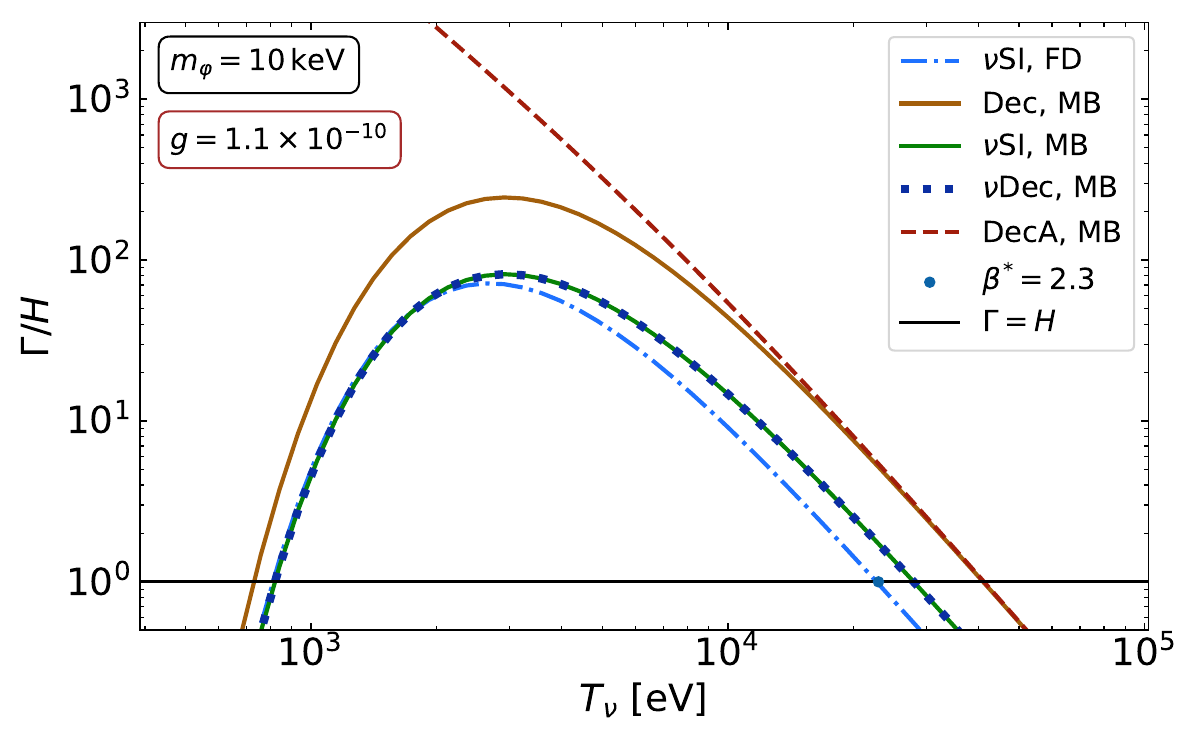}}
\caption{Comparative plot of interaction rates for the real-scalar (Majorana) case: Resonant $\nu$SI computed using Eq.~\eqref{Resonat-NSI} with Fermi–Dirac statistics ($\nu$SI, FD); resonant $\nu$SI in the MB approximation ($\nu$SI, MB); decay rates in the MB approximation assuming a massless $A$ number density (Dec, MB); decays following the neutrino number density ($\nu$Dec, MB); and decays accounting for the non-relativistic behavior of the $\varphi$ particle in the decoupling regime (DecA, MB).} \label{fig: Rates com}
\end{figure}

In Fig.~\ref{fig: Rates com}, we have plotted the interaction rates for resonant scalar $\nu$SI in the MB approximation and the exact contribution from thermalization via decays following the $\varphi$ particle, which are respectively:
    \begin{equation}
         \Gamma_{\rm resonant}^{\rm MB} = \frac{g^2 T_\nu}{32\pi} \, \mu_\varphi^3  K_1(\mu_\varphi) \quad \text{and} \quad \Gamma_{\rm Dec}^{\rm MB} = \frac{3g^2 m_\varphi}{16\pi} \,   \frac{K_1(\mu_\varphi)}{K_2(\mu_\varphi)} \quad  \, .
    \end{equation}
This illustrates the limitations of choosing the interaction rate for the RTA.

Due to this issue, some authors in the pseudoscalar (also known as Majoron $\phi$) case treat the coupled neutrino–Majoron fluid as approximately massless to model phase-space perturbations~\cite{Escudero2020CMBsearch}, adopting a specific case of the first expression in Eq.~\eqref{Gamma_int_A}, where the number density of the $A$ particle is always massless. In this scenario, ${\cal F}_\ell \to F_{\nu \phi \ell}$, and $\Gamma \to \Gamma_{{\rm int}, \phi}$ in Eq.~\eqref{Boltzmann-eq-relativistic neutrinos}. This approach necessarily considers common multipoles of $\nu$ and $\phi$ and imposes the vanishing of the $\ell=0,1$ modes by hand. Importantly, this null contribution arises naturally when particle number is conserved as in $\nu$SI scenarios (see Appendix C of Ref.~\cite{Perez2026towards}). Here, this reminds us of the conservation law~\eqref{C-law-Majorana}, which is not satisfied if we use the thermal forms of $\phi$ and $\nu$ distribution functions. Moreover, Ref.~\cite{Sandner2023precision} follows an approach analogous to relativistic massive-neutrino decay~\cite{Chen2022weaker}, $\nu_H \to \nu_l + \phi$. However, this analogy does not account for the freeze-in production of (pseudo)scalar particles: after scalar production, $\bar{f}_\nu \neq \bar{f}_\phi$, whereas Eq.~(11) of Ref.~\cite{Chen2022weaker} assumes the approximation,
    \begin{equation}
        F_{i,\ell} = - \frac{1}{4} \frac{d \bar{f}_i}{d \ln q} {\cal F}_{i,\ell} \quad \text{with} \quad i = \nu_H, \nu_l, \phi \, .
    \end{equation}
As noted in Ref.~\cite{Perez2026towards}, this approximation ($F_{i,\ell} = \bar{f}_i \Psi_{i, \ell}$) describes relativistic species and is often regarded merely as an ansatz.

In this scenario, the particle number density of neutrinos and $A$ particles evolves; it is useful to define an auxiliary variable that is easier to manipulate. We define the relative fraction of  populated $A$ particles at the temperature $T_{\nu,p} = \beta m_A$ as
    \begin{equation}
        r := \frac{N_A \bar{n}_A}{(N_A \bar{n}_A + N_\nu \bar{n}_\nu)} \, . \label{fr_gen}
    \end{equation}
Eq.~\eqref{fr_def} is a particular case for a real scalar ($A = \varphi$) field.

In principle, due to the conservation law~\eqref{C-law} (or Eq.~\eqref{C-law-Majorana}), one cannot maintain the thermal form of $\bar{n}_A$ and $\bar{n}_\nu$ during the production of $A$ particles; however, due to our ansatz to define Eqs.~\eqref{C_p}-\eqref{C_d}, we expect only small deviations from neutrino free-streaming, which implies $a(t) \propto T_\nu^{-1}$. Thus, from the conservation law~\eqref{C-law-Majorana}, the number density of $A$ particles can be parametrized in the freeze-in regime by
    \begin{equation}
        n_A(\beta) = N_A\bar{n}_A (\beta) = N_\nu r(\beta) n_{\rm MB} \quad \text{and} \quad n_{\rm MB} = \int \frac{d^3 p}{(2\pi)^3} e^{-p/T_{\nu}}
        \, . \label{number density-parametrized}
    \end{equation}
Also, the effective neutrino number density $N_\nu \bar{n}_\nu$ would change as $\bar{n}_\nu = [1-r(\beta)] n_{\rm MB}$.
    
The criterion for thermalization is that the interaction rate of the process $A \leftrightarrow \nu \nu$ will be comparable to the Hubble expansion rate, as is typically considered in the literature. Moreover, full thermalization implies that the zeroth-order collision term must vanish, which requires
$\bar{f}_A - \bar{f}_\nu \bar{f}_\nu = 0$. In this case, from Eq.~\eqref{number density-parametrized}, the modified distribution functions per single degree of freedom are: $\bar{f}_A = N_\nu r^{\rm th}e^{-p_A/T_\nu}/N_A$, and for neutrinos $\bar{f}_\nu = (1- r^{\rm th}) e^{-p/T_\nu}$. To satisfy the thermalization condition, in the relativistic regime, we need to work with the physical condition:
    \begin{equation}
        r^{\rm th} = \frac{\left[2 + \left( \frac{N_\nu}{N_A} \right) \right] - \sqrt{\left[2 + \left( \frac{N_\nu}{N_A} \right) \right]^2 - 4}}{2} = \begin{cases}
            0.382 & \text{for (axial-)vector case} \\
            0.209 & \text{for real (pseudo)scalar case}
        \end{cases} \, . \label{fr_th}
    \end{equation}
Under this condition, it is easy to show that in the relativistic regime, the interaction rate $\Gamma_{{\rm int}, A}$ given in Eq.~\eqref{Gamma_int_A} is valid, and then $\beta = \beta_{\rm th}$ defines when the thermalization starts. This condition, together with $a(t) \propto T_\nu^{-1}$ and the evolution given in Eq.~\eqref{A-number density}, allows us to estimate
    \begin{equation}
        r(\beta) =  \frac{M_{\rm pl}}{2 N_\nu g_H}  \frac{N_A \Gamma_A}{m_A^2} I(\beta, \beta_0) \Rightarrow r(\beta) = r^{\rm th} \begin{cases}
            1 & \text{for $\beta < \beta_{\rm th}$} \\
            \frac{I(\beta, \beta_0)}{I(\beta_{\rm th}, \beta_0)}  & \text{for $\beta \geq \beta_{\rm th}$}
        \end{cases} \, . \label{f-evol}
    \end{equation}
To build up a $\varphi$ population, the particle’s lifetime must be sufficiently long; otherwise, it decays too quickly. This can be safely ruled out when the produced particle is relativistic, since the on-shell production in the relativistic regime is determined by $s = 2E_\nu E_\nu^\prime (1 - \cos \alpha) = m_A^2$, where $E_\nu$ and $E_\nu^\prime$ are the energies of the interacting neutrinos, and the average relativistic neutrino energy at temperature $T_{\nu}$ in the MB approximation is $\langle E_\nu \rangle = 3 (1-r(\beta))T_\nu$.

For relativistic production ($T_\nu \gg m_A$), we obtain:
    \begin{equation}
        (1 - \cos \alpha_p) = \frac{m_A^2}{18 [1-r(\beta)]^2T_\nu^2} \quad \underset{\text{small angles}}{\Longrightarrow} \quad  \alpha_p = \frac{m_A}{3 [1-r(\beta)] T_\nu} \, . \label{Suppression}
    \end{equation}
At the end of the production, $\alpha_{\rm th} = [3 (1-r^{\rm th})\beta_{\rm th}]^{-1}$. As the production starts in $\beta_0 \gg \beta_{\rm th}$, it follows that at high energies the production is suppressed, on average, by $\alpha_p \lesssim [3 (1-r^{\rm th})\beta_{\rm th}]^{-1}$ (see panel (a) in Fig.~\ref{fig:Kin-thermalization}). A similar argument applies for the FD distribution function where $\langle E_\nu \rangle_{\rm FD} = 3.151 \, T_\nu$. In such a case, an analogous equation to Eq.~\eqref{Suppression} will be constructed with the background function evolving according to $r(\beta)$. However, when we increase the coupling $g$ (within the parameter space where perturbative quantum field theory is valid), we always achieve the fraction $r^{\rm th}$ in the relativistic regime (see panel (b) of Fig.~\ref{fig:Kin-thermalization}) necessary at $\beta_{\rm th}$ fixed by $\Gamma_{{\rm int}, A} = c_{\rm th} H$, where $c_{\rm th}$ is nearly constant.

To determine $c_{\rm th}(\beta_{\rm th})$, we choose the fraction $r^{\rm th}$ required for thermalization, which fixes the value of $\beta_{\rm th}$ and, consequently, the value of the coupling $g$ in the total decay width $\Gamma_A$ (see Eq.~\eqref{f-evol}) determined by the form of the squared amplitude~\eqref{Amplitude-gen}. 
Comparing the interaction~\eqref{Gamma_int_A} with the Hubble rate, we obtain
    \begin{equation}
       g = \left[\frac{32\pi g_H}{M_{\rm pl} \kappa} \left(\frac{r^{\rm th}}{I(\beta_{\rm th}, \beta_0)} \right) m_{A} \right]^{1/2} \Longrightarrow c_{\rm th}(\beta_{\rm th}) = \frac{K_1(1/\beta_{\rm th})}{\beta_{\rm th}^4 I(\beta_{\rm th}, \beta_0)} \frac{N_\nu}{N_A} r^{\rm th} \, . \label{c_th}
    \end{equation}
For the real (pseudo)scalar case, $c_{\rm th} \gtrsim 1.8$ in the relativistic region.

\begin{figure}
\subfigure[]{
\includegraphics[width=0.32\linewidth]{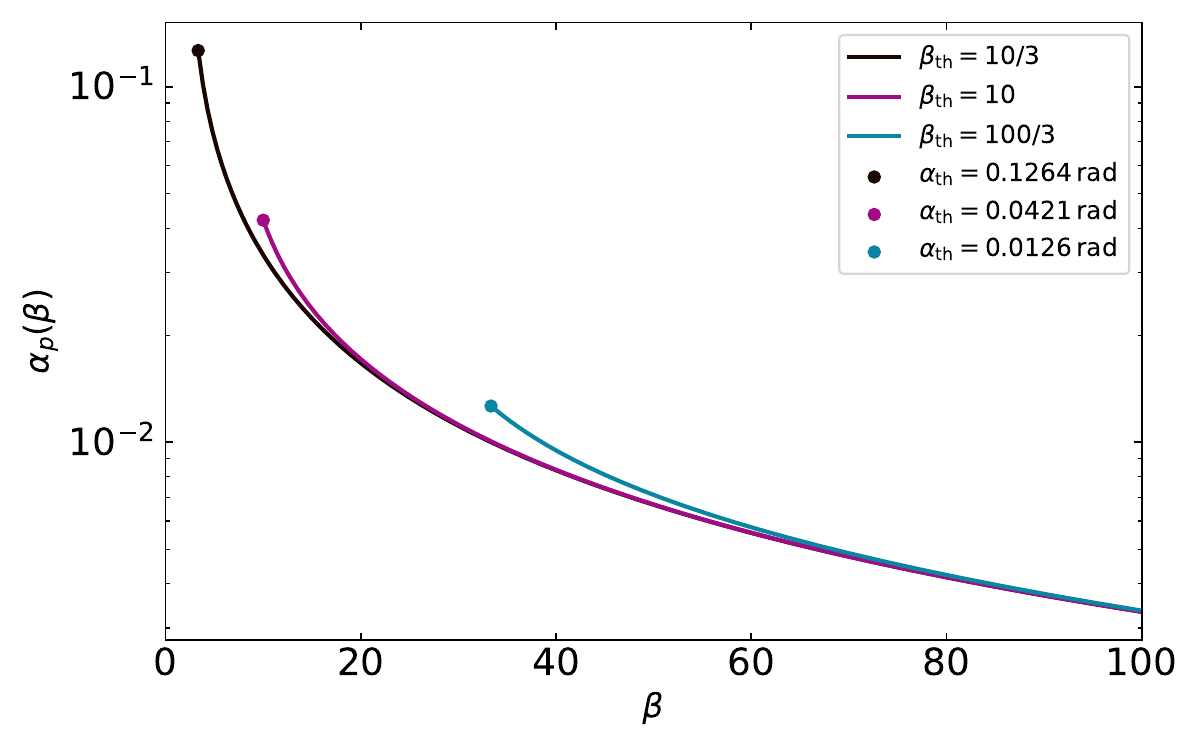}}
\subfigure[]{
\includegraphics[width=0.32\linewidth]{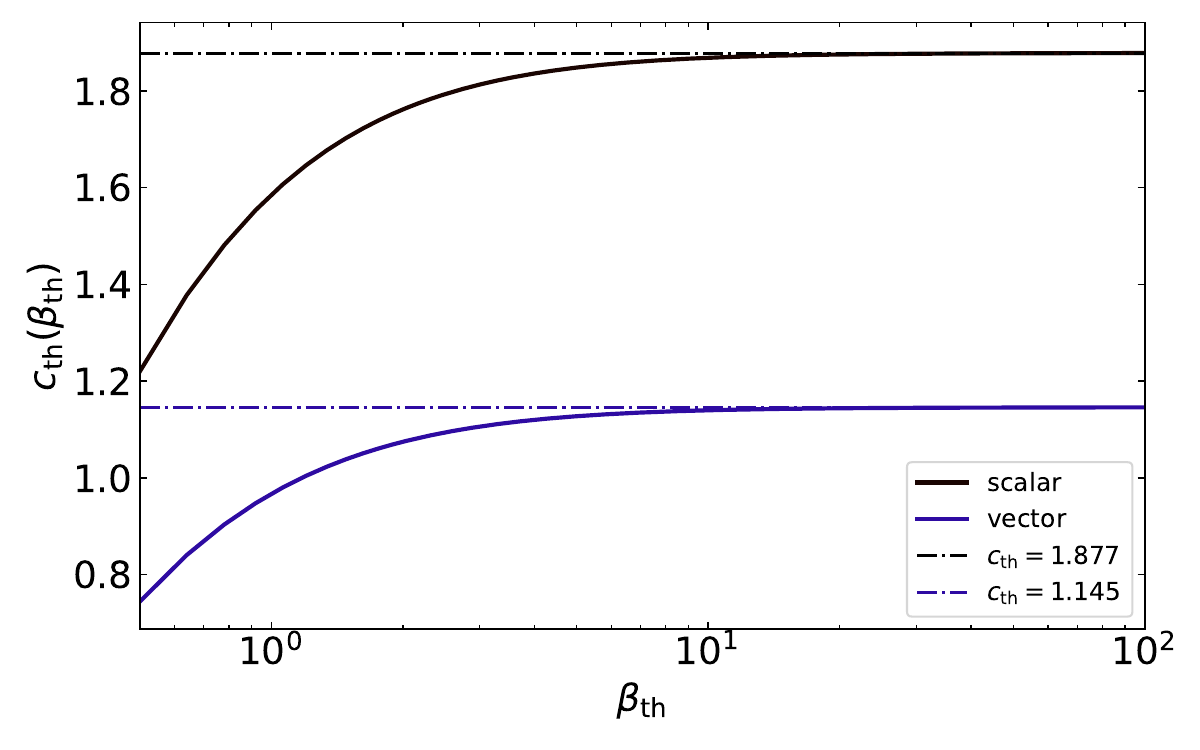}}
\subfigure[]{
\includegraphics[width=0.32\linewidth]{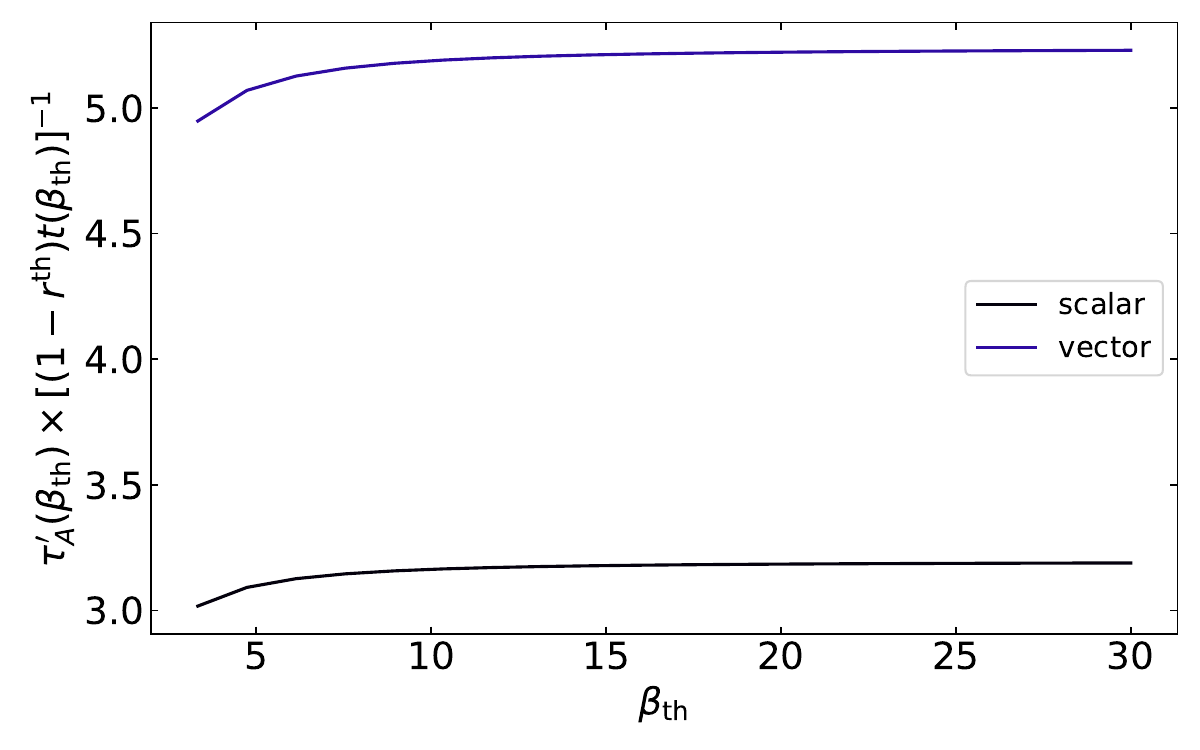}}
\caption{(a) thermal suppression of the on-shell production where $\alpha_p \lesssim \alpha_{\rm th}$. (b) The coefficient $c_{\rm th}(\beta_{\rm th})$ that indicates whether thermalization occurs at $\beta_{\rm th}$ in the MB approximation, according to the usual criterion $\Gamma_{{\rm int}, A}/H \gtrsim 1$. In the relativistic production regime, this criterion is always sufficient to achieve thermalization; (c) the lifetime of $A$ particles compared with the cosmic time $t$ at the moment when $\Gamma_{{\rm int}, A}=c_{\rm th}H$.  
} \label{fig:Kin-thermalization}
\end{figure}

\begin{figure}
\subfigure[]{
\includegraphics[width=0.32\linewidth]{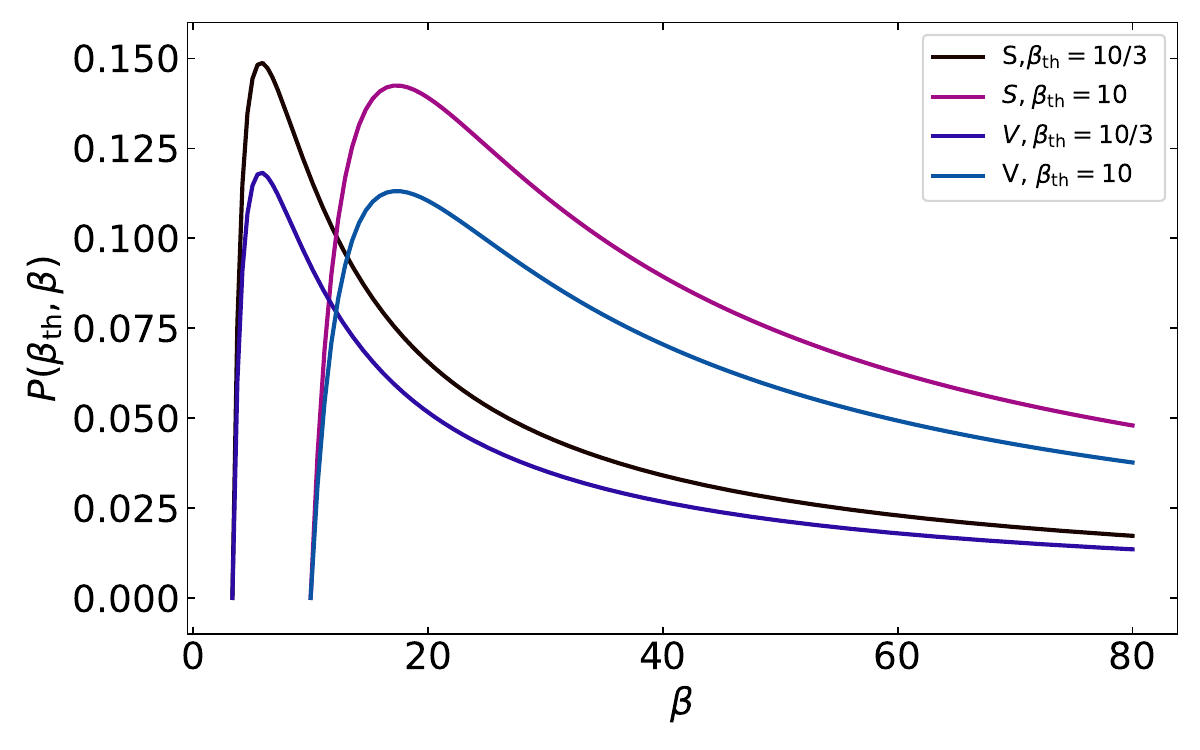}}
\subfigure[]{
\includegraphics[width=0.32\linewidth]{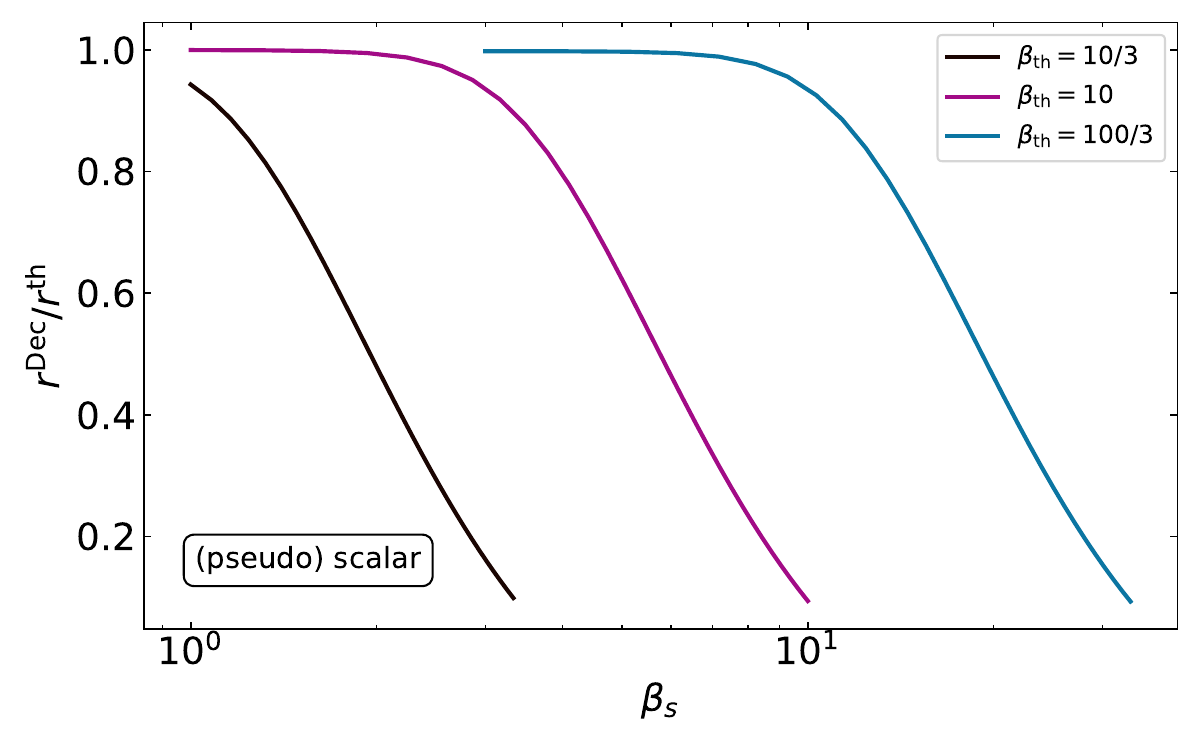}}
\subfigure[]{
\includegraphics[width=0.32\linewidth]{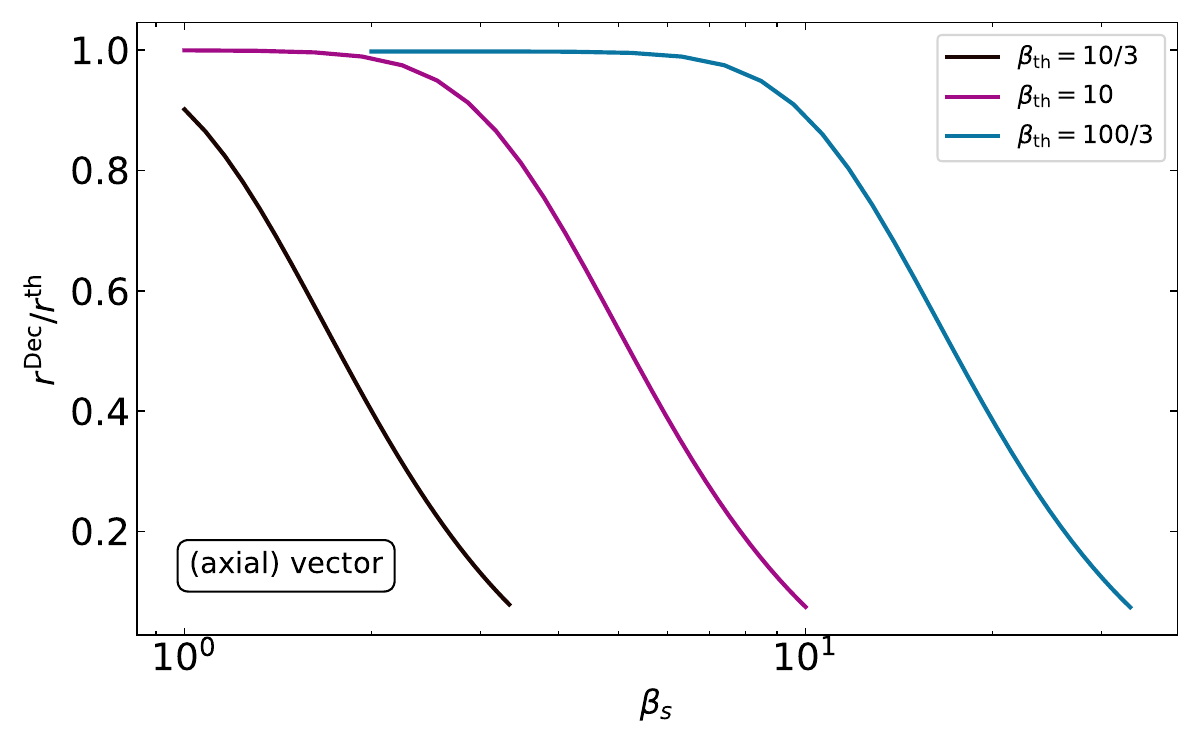}}
\caption{Decays of the mediator: (a) Decay probability of the relativistic (pseudo)scalars S and (axial-)vectors V produced via neutrino inverse decays with maximum probabilities $P_{\rm max}^{\rm S} = 0.1488, 0.1424$ and $P_{\rm max}^{\rm V} = 0.1182, 0.1131$ at the thermalization temperature determined by $\beta_{\rm th} = 10/3, 10$, respectively. (b) Evolution of the relative fraction of the (pseudo)scalar population generated through inverse decays after reaching $r^{\rm th} \approx 0.209$. Notice that the decays occur relatively late; then, in the relativistic regime, we cannot have realistic thermalization (this also applies to (c)). (c) Evolution of the relative fraction of the (axial-)vector population generated through inverse decays after reaching $r^{\rm th} \approx 0.382$.
} \label{fig:No-thermalization}
\end{figure}
    
During the radiation-dominated era, when $T_\nu > 0.57 \, {\rm eV}$, the scale factor evolves as $a(t)\propto t^{1/2}$, and the cosmological time is given by
    \begin{equation}
        t (\beta) = \frac{1}{2H} = \frac{M_{\rm pl}}{2g_H \beta^2 m_A^2} \, . \label{t(beta)}
    \end{equation}
We can explore what happens with the relativistic production in the decay picture, as $\cos\alpha_p \approx 1$ for $\alpha_p \lesssim 0.127 \, {\rm rad}$, the $A$ particle is emitted forward, and its Lorentz factor is $\gamma = 2\langle E_\nu \rangle/m_A \gtrsim 6 (1-r^{\rm th}) \beta_{\rm th}$. It is then easy to see that the boosted lifetime of the produced particle, on average, $\tau_A^\prime \gtrsim 6(1-r^{\rm th}) \beta_{\rm th} \tau_A$, satisfies $\tau_A^\prime/t(\beta_{\rm th}) > 9 (1-r^{\rm th})(0.209/r^{\rm th}) N_A/N_\nu$. As long as both $A$ particles and neutrinos remain relativistic, they must satisfy the conservation law given in Eq.~\eqref{ed-conservation}. Therefore, instantaneous $\nu\text{--}\varphi$ thermalization is not possible in the relativistic regime $T_\nu \gg m_A/3$.

To make this explicit, we compute the decay probability at the temperature $T_{\nu,s} = \beta_s m_A$, with $\beta_s \leq \beta_{\rm th}$, which is determined at $t= t(\beta_s)$, when production starts at $t_0 = t(\beta)$, as (see Fig.~\ref{fig:No-thermalization} (b) for $\beta_s =\beta_{\rm th}$)
    \begin{equation}
        P(\beta_{\rm th}, \beta_s, \beta) = 1 - \begin{cases}
            \exp\left\{ - \frac{(5 - \sqrt{21})}{ 4 [1-r(\beta)]\beta I(\beta_{\rm th}, \beta_0)} \left(\frac{1}{\beta_s^{2}} -  \frac{1}{\beta^2}\right) \right\} & \text{for the (pseudo)scalar case} \\
            \exp\left\{ - \frac{(3 - \sqrt{5})}{ 12 [1-r(\beta)] \beta I(\beta_{\rm th}, \beta_0)} \left(\frac{1}{\beta_s^{2}} -  \frac{1}{\beta^2}\right) \right\} & \text{for the (axial-)vector case}
        \end{cases}  \, .
    \end{equation}
After reaching the maximum, the production of the $A$ particles should be equilibrated with the decays. Fixing the production temperature by $\beta_{\rm th}$, we can then determine the fraction of decays at $\beta_s < \beta_{\rm th}$ as
    \begin{equation}
        \frac{r^{\rm Dec}}{r^{\rm th}} = \frac{1}{I(\beta_{\rm th}, \beta)} \int^{\beta_0}_{\beta_{\rm th}} P(\beta_{\rm th}, \beta_s, \beta) \beta^{-5} K_1(1/\beta) d \beta \, .
    \end{equation}
In Fig.~\ref{fig:No-thermalization} (b) and (c), we see that the decays of the relativistic particles immediately after reaching $r^{\rm th}$ are too small. From Eq.~\eqref{t(beta)}, one can see that the difference between the corresponding time scales is close to one order of magnitude when the fraction of particles that have decayed becomes appreciable ($\sim 80 \, \%$). Therefore, this breaks down the usual picture of thermal equilibrium by showing that decays yield a thermal-like particle distribution function after production in the freeze-in regime.

Now, for the generic amplitudes given in Eq.~\eqref{Amplitude-gen}, we use a value of $g$ that exceeds $f_r^{\rm th}$ as
    \begin{equation}
        r^{\rm th} \lesssim g^2 \frac{M_{\rm pl}}{g_H}  \frac{\kappa}{32\pi m_A} I(\beta_{\rm th}, \beta_0) \Rightarrow g \gtrsim \sqrt{\frac{g_H 32\pi}{(1.22) \kappa}} \left(\frac{r^{\rm th}}{I(\beta_{\rm th}, \beta_0)} \right)^{1/2} \times 10^{-14} \sqrt{\frac{m_A}{\rm eV}} \, . \label{g-bound}
    \end{equation}
In our numerical estimation of the resonant events, within the thermal approximation, the most important contribution is within the range determined in Eq.~\eqref{Resonant prod}, that is, the non-relativistic regime, as long as $T_\nu \sim m_A$.

To establish a threshold for resonant behavior, we choose $\beta_{\rm th} \gtrsim 10/3$. This translates into $\alpha_p \lesssim 0.127 \, {\rm rad}$ for (pseudo)scalar case and $\alpha_p \lesssim 0.162 \, {\rm rad}$ for (axial-)vector case. Under these conditions, we obtain
    \begin{equation}
        g \gtrsim \sqrt{\frac{m_A}{\rm eV}} \times \begin{cases}
        1.1 \times 10^{-12} & \text{for (pseudo)scalar case} \\
        1.49 \kappa^{-1/2}\times 10^{-12} & \text{for (axial-)vector  case} 
        \end{cases} \, . 
    \end{equation}
Given that the $A$ particle lifetime at the resonant pole satisfies
    \begin{equation}
        \tau_A = \frac{1}{\Gamma_A} \lesssim \frac{N_A}{N_\nu} \frac{M_{\rm pl}}{2 g_H m_A^2} \frac{I(\beta_{\rm th}, \beta_0)}{r^{\rm th}},
    \end{equation}
from Eq.~\eqref{fr_th}, we only need to restrict the function $R_r(\beta_{\rm th})$ with the ratio between the lifetime of $A$ particle and the cosmological time:
    \begin{equation}
        \frac{\tau_A}{t(\beta_{\rm th})} \lesssim R_r (\beta_{\rm th}) \ll 1 \quad \text{with} \quad R_r (\beta_{\rm th}) := \frac{N_A}{N_\nu} \frac{\beta_{\rm th}^2 I(\beta_{\rm th}, \beta_0)}{r^{\rm th}} \, . \label{Lifetime}
    \end{equation}
That condition is satisfied for $\beta_{\rm th} =10/3$ when $\beta_0$ is  large enough, and we obtain $R(10/3, \beta_0)=0.15$ for the (pseudo)scalar case. This is a practical criterion for identifying a resonant behavior during cosmic evolution when mediators are produced nearly non-relativistic due to $R= \tau_\varphi/t(\beta^{*}) \approx 0.072$, where $t(\beta^{*})$ is the cosmic time evaluated when $\Gamma_{\nu\nu \to \nu\nu}/H = 1$ (see Fig.~\ref{fig: Rates com} and Eq.~\eqref{Resonat-NSI}). Notice that as the Universe cools during the interaction, $R$ decreases. Here, we expect an overestimation of $\Gamma_{\nu\nu \to \nu\nu}$ at $T_\nu \gg m_\varphi/3$, which implies that the ratio $R$ will be even smaller. This is confirmed by the analysis in Ref.~\cite{Perez2026Addressing}.

Note that full thermalization in our scheme would imply
    \begin{equation}
        2 N_\nu \bar{\rho }_\nu a^4\bigg|_{T_{\nu, i}} = 2N_\nu T_{\nu, 0}^4 \frac{3}{\pi^2} \neq (2N_\nu \rho_\nu + N_A \rho_A) a^4 \bigg|_{T_{\nu, {\rm th}}} = 2 N_\nu \left( 1 - \frac{r^{\rm th}}{2} \right) T_{\nu, 0}^4 \frac{3}{\pi^2} \, .
    \end{equation}
Here, $T_{\nu, 0}$ is the neutrino temperature today. Below, we present an approach that satisfies the law~\eqref{ed-conservation}, and we argue why it still leads to incorrect evolution in the freeze-in regime.
\subsection{On entropy production and instantaneous thermalization via decays}
Another parametrization of the neutrino–(pseudo)scalar system consists of evolving the chemical potentials numerically, as done in Refs.~\cite{Abenza2020precision, Escudero2020CMBsearch, Li2023neff}. Following Ref.~\cite{Abenza2020precision}, one assumes that thermalization occurs in the relativistic regime. In this case, the conservation laws in Eqs.~\eqref{C-law-Majorana} and \eqref{ed-conservation} apply. However, previous estimations exhibit slight violations of these conservation laws, particularly in the pseudoscalar case,
    \begin{align}
        2N_\nu\bar{\rho}_\nu(T_{\nu,in}, 0) &= 2N_\nu\bar{\rho}_\nu(T_{eq}, \mu_{eq}) + \bar{\rho}_\phi (T_{eq}, 2\mu_{eq}) \, , \\
        N_\nu  \bar{n}_\nu(T_{\nu,in}, 0) &= N_\nu \bar{n}_\nu(T_{eq}, \mu_{eq}) + \bar{n}_\phi (T_{eq}, 2\mu_{eq}) \, .
    \end{align}
We remark that both sides of the equations are evaluated assuming the same scale factor, even though the LHS corresponds to the initial temperature $T_{\nu,in}$, while the RHS corresponds to $T_{eq}$, where kinetic and chemical equilibrium are established. Then, defining $\alpha= e^{\mu_{eq}/T_{eq}}$ and $x=T_{\nu,in}/T_{eq}$, using the thermal form for the distribution functions for relativistic particles, we need to solve (in terms of the polylogarithm functions ${\rm Li}_n(v)$, where $n$ denotes their order):
    \begin{equation}
        \frac{7}{80} \pi^2 x^4 = -\frac{9}{\pi^2} {\rm Li}_4(-\alpha) + \frac{3}{2 \pi^2} {\rm Li}_4(\alpha^2) \quad \text{and} \quad \frac{9\zeta(3)}{4\pi^2} x^3 = -\frac{3}{\pi^2} {\rm Li}_3(-\alpha) + \frac{1}{\pi^2} {\rm Li}_3(\alpha^2) \, . \label{Sol-1}
    \end{equation}
Thus, we obtain the same result presented in Eq.~(4.9) of Ref.~\cite{Abenza2020precision}: $T_{eq} = 1.120818 \, T_{\nu,in}$, $ \mu_{eq} = - 0.641987 \, T_{\nu,in}$.

During thermalization, the entropy density is also conserved in a comoving volume $V=a^3(t)$~\cite{Chacko2004cmb}, which implies that
    \begin{equation}
        (2 N_\nu \bar{s}_\nu + \bar{s}_\phi)a^3(t) = {\rm constant} \, .  \label{entropy-law}
    \end{equation}
With the conditions derived from Eq.~\eqref{Sol-1} and considering that the behavior of the photon bath satisfies the standard evolution $T_\gamma a = T_\gamma^\prime a^\prime$, one derives the characteristic change in the effective relativistic degree of freedom $N_{\rm eff}$ by solving
    \begin{align}
        - 6\left[ 4 {\rm Li}_4 (- \alpha) - \ln(\alpha) {\rm Li}_3 (- \alpha) \right] + \left[ 4 {\rm Li}_4 ( \alpha^2) - 2\ln(\alpha) {\rm Li}_3 (\alpha^2) \right] &= - 6y^3 \left[ 4 {\rm Li}_4 (- \omega) - \ln(\omega) {\rm Li}_3 (- \omega) \right] \, , \label{entropyB}\\
        - 3{\rm Li}_3(-\alpha) + {\rm Li}_3(\alpha^2) &= - 3y^3 {\rm Li}_3(-\omega) \,. \label{numberB}
    \end{align}
where we have used the conservation laws~\eqref{C-law-Majorana} and \eqref{entropy-law}, defining
    \begin{equation}
        y = \frac{T_\nu^\prime}{T_\gamma^\prime} \frac{T_\gamma}{T_{eq}} \quad \text{and} \quad \omega = e^{\mu_\nu/T_\nu^\prime} \, .
    \end{equation}
To derive the solution presented in~\cite{Abenza2020precision}, one needs to use the NLO-QED ratio $T_\gamma/T_{\nu,in}=1.39578$ also computed in~\cite{Abenza2020precision}. We obtain $y \approx 0.9319$ ($T_\gamma^\prime/T_\nu^\prime = 1.336322$), and $\omega \approx e^{-0.142}$, and then
    \begin{equation}
        N_{\rm eff} = -\frac{360 N_\nu {\rm g}_s}{7 \pi^4} \left( \frac{11}{4} \right)^{4/3} \left( \frac{T_\nu^\prime}{T_\gamma^\prime} \right)^4 {\rm Li_4}(-\omega) =  3.16368 \Rightarrow \Delta N_{\rm eff} = 0.1187 \, ,
    \end{equation}
where we have used $N_\nu=3$, ${\rm g}_s=2$ and $N_{\rm eff}^{\rm SM} \approx 3.045$ as in Ref.~\cite{Abenza2020precision}.

We note that, in principle, $T_\gamma$ and $T_{eq}$ are evaluated at the same scale factor in Eqs.~\eqref{entropyB}-\eqref{numberB}. However, in order to obtain an increase in $N_{\rm eff}$ by reheating the neutrino bath, one must instead assume that $T_\gamma$ is evaluated at the same scale factor as $T_\nu$ (this also conserves $N_{\rm eff}$ during Majoron production)~\cite{Abenza2020precision}; otherwise, one obtains $N_{\rm eff} = 2.005$. In this approximation, the population of $\phi$ particles is created nearly instantaneously, while the temperature changes, implying that the scale factor must also change, contrary to the initial assumption. Moreover, full numerical solutions~\cite{Abenza2020precision, Li2023neff} indicate that, in reality, these populations are produced over a range of temperatures (see panel (a) of Fig.~\ref{fig:fr-evol} for the vector case).

On the other hand, Ref.~\cite{Kelly2021intimate} uses a different interpretation, where the authors employ the standard parametrization of sterile neutrino production in the Dodelson--Widrow mechanism in the presence of $\nu$SI, namely $z= m_\phi/T_\nu^{\rm sc}$. Here, $T_\nu^{\rm sc}$ denotes the neutrino temperature in the standard cosmology without $\phi$. In this case, $x$ in Eq.~\eqref{Sol-1} is the ratio between the ``time'' $z_A$, at which neutrinos have just decoupled from weak interactions, and $z_B$, the timescale at which the $\nu\text{--}\phi$ system has just equilibrated:
    \begin{equation}
        x = \frac{T_\nu(z_A)}{T_\nu(z_B)}\frac{z_A}{z_B} = \frac{T_\nu(z_A)}{T_\nu(z_B)} \frac{T_\nu^{\rm sc}(z_B)}{T_\nu^{\rm sc}(z_A)} \, .
    \end{equation}
In the last line, one uses the definition of $z$ and notes that this is equivalent to considering the scale factor in the standard cosmology, $a = T_{\nu,0} a_0/T_\nu^{\rm sc}$ ($a_0=1$ today). In this way, the conservation laws~\eqref{C-law-Majorana} and~\eqref{ed-conservation} are consistently satisfied. For Eqs.~\eqref{entropyB}-\eqref{numberB}, we have,
    \begin{equation}
        y = \frac{T_\nu(z_c)}{T_\nu(z_B)} \frac{z_c}{z_B} = \frac{T_\nu(z_c)}{T_\nu(z_B)} \frac{T_\nu^{\rm sc}(z_B)}{T_\nu^{\rm sc}(z_c)} \, .
    \end{equation}
Here, $z_c$ is the time scale where all $\phi$'s have decayed away ($T \ll m_\phi$).

Of course, we obtain the same numerical results for $x$ and $y$, but with different interpretations. Since one starts with no population of $\phi$ particles, the neutrino bath temperature satisfies $T_\nu(z_A) = T_\nu^{\rm sc}(z_A)$. Consequently, $T_\nu(z_B)/T_\nu^{\rm sc}(z_B) = 1.120818$, while 
    \begin{equation}
        \frac{T_\nu(z_c)}{T_\nu^{\rm sc}(z_c)} = y/x \approx 1.044 \quad \text{and} \quad \frac{\mu_\nu(z_c)}{T_\nu(z_c)} \approx -0.142\, .
    \end{equation}
This implies that
    \begin{equation}
        N_{\rm eff} = -\frac{2160}{7 \pi^4} \left( \frac{11}{4} \right)^{4/3} \left( \frac{T_\nu(z_c)}{(1.39578) T_\nu^{\rm sc}(z_c)} \right)^4 {\rm Li_4}(-\omega) = 3.16368 \, .
    \end{equation}
And therefore $\Delta N_{\rm eff} = 0.1187$. 

This result is, in principle, independent of $g$; however, one must restrict $g$ to avoid a significant population of $\phi$ particles during BBN~\cite{Huang2018, Kelly2021intimate}. Notice then that the relative fraction of produced scalars $\tilde{r}^{\rm th}$ that thermalizes with neutrinos in this approach is
    \begin{equation}
        \tilde{r}^{\rm th} = \frac{ {\rm Li}_3(\alpha^2)}{- 3{\rm Li}_3(-\alpha) + {\rm Li}_3(\alpha^2)} = 0.1728 \, .
    \end{equation}
Now, we are interested in obtaining a form to compare with our previous analysis with the MB approximation; in such a case, we obtain the simple form:
    \begin{equation}
        \tilde{r}^{\rm th} = \frac{N_A \bar{\alpha}^2}{N_\nu \bar{\alpha} + N_A\bar{\alpha}^2}  \Rightarrow \bar{\alpha} = \frac{N_\nu \tilde{r}^{\rm th}}{N_A(1-\tilde{r}^{\rm th})} \quad \text{or} \quad
        \tilde{r}^{\rm th} = \frac{ N_A \bar{\alpha}^2}{N_\nu} \frac{1}{\bar{x}^3}\Rightarrow \bar{\alpha} = \sqrt{\frac{N_\nu}{N_A} \tilde{r}^{\rm th}} \, \bar{x}^{3/2} \, ,
    \end{equation}
where we have defined $\bar{x} = T_\nu^{\rm sc}/T_\nu$, and one gets $\bar{\alpha}=  0.4514 \, (0.6296)$, which implies that $\tilde{r}^{\rm th} = 0.311 \, (0.1735)$ for the vector (scalar) case with $N_\nu=3$. 

At first sight, this solution is the most convincing, as it assumes instantaneous thermalization in the relativistic regime while satisfying all conservation laws, namely Eqs.~\eqref{C-law-Majorana}, \eqref{ed-conservation}, and \eqref{entropy-law}. Nevertheless, if the produced particles are generated while still relativistic, then
    \begin{equation}
        n_A =  \beta^3 \frac{M_{\rm pl}}{g_H} \frac{N_A \Gamma_A}{2\pi^2} m_A  e^{\mu_A/T_\nu^{\rm sc}}I(\beta,\beta_0) \bigg|_{\beta=\beta_{\rm th} \gg 1} - \text{ depletion term} \approx \frac{M_{\rm pl}}{3 g_H} \frac{N_A \Gamma_A}{2\pi^2} m_A \bar{\alpha} - \text{ depletion term} \, .
    \end{equation}
Here, the ``depletion term'' must account for decays of the $A$ particles during their production. In the standard freeze-in scenario, these decays are assumed to be negligible; however, neglecting them leads to an overestimation of the production of $A$ particles. Below, we argue why this is the case within the standard pseudo-chemical approximation even when a particular form of the depletion term is included.
    
Notice that we have constructed the function $r(\beta)$ (see Eq.~\eqref{fr_gen} and Eq.~\eqref{f-evol}) to constrain the evolution of the number density with $\text{``depletion term''}=0$. In this way, the evolution can be matched to the overestimated interaction rate (first option in Eqs.~\eqref{Gamma_int_A}) in the relativistic regime through the coefficient $c_{\rm th}$~\eqref{c_th}. However, we obtain fractions $r^{\rm th}$ that differ from those derived using the chemical-potential evolution because of the different parametrization schemes. 

In Fig.~\ref{fig:fr-evol}, we show the full numerical solution in the pseudo-chemical approximation for the vector case (solid lines) using the public code of Ref.~\cite{Li2023neff}. Initially, the number density of vector particles is nearly zero; subsequently, during thermalization, the population freezes and later depletes.  Indeed, we observe that for $g_{Z^\prime} \gtrsim 10^{-7}$, the numerical solution deviates from the value $\tilde{r}^{\rm th} = 0.311$ (see panel (a) of Fig.~\ref{fig:fr-evol}) due to the stiffness of the collision terms. In principle, the depletion in this approach should be independent of the chosen coupling $g_{Z^\prime}$~\cite{Abenza2020precision}. However, obtaining the solution in the regime $T_\nu \ll m_{Z^\prime}$~\cite{Abenza2020precision, Li2023neff} where $K_1(x_{Z^\prime}) \approx (\pi/2x_{Z^\prime})^{1/2} e^{-x_{Z^\prime}}$ with $x_{Z^\prime} = m_{Z^\prime}/T_\nu$ is challenging for the solver. While the standard instantaneous-thermalization criterion predicts that increasing $g_{Z^\prime}$ delays neutrino free-streaming to lower temperatures, as expected for resonant $\nu$SI (see Fig.~\ref{fig:Comparative_rates} for the scalar case), the numerical results do not exhibit this behavior because of the approximations adopted in Ref.~\cite{Li2023neff} (see Appendix B therein).

\begin{figure}
\subfigure[]{
\includegraphics[width=0.4\linewidth]{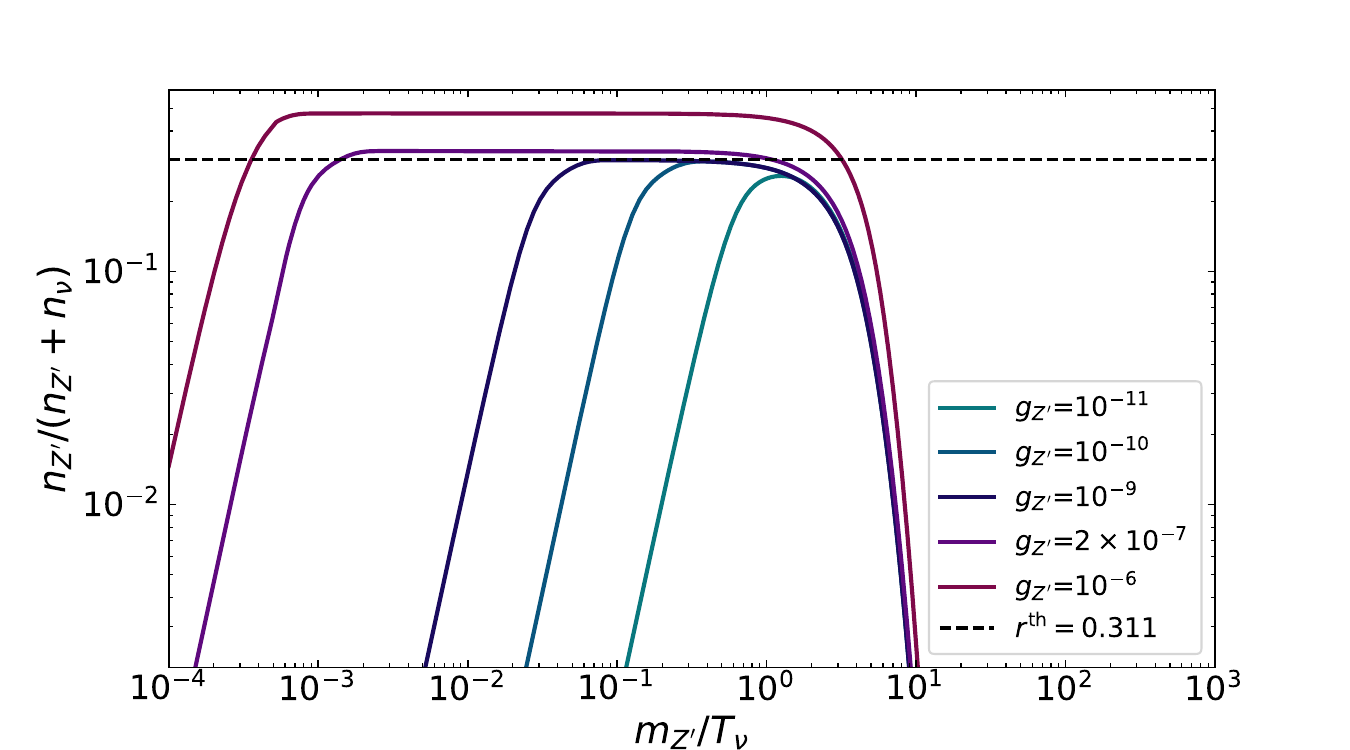}}
\subfigure[]{
\includegraphics[width=0.4\linewidth]{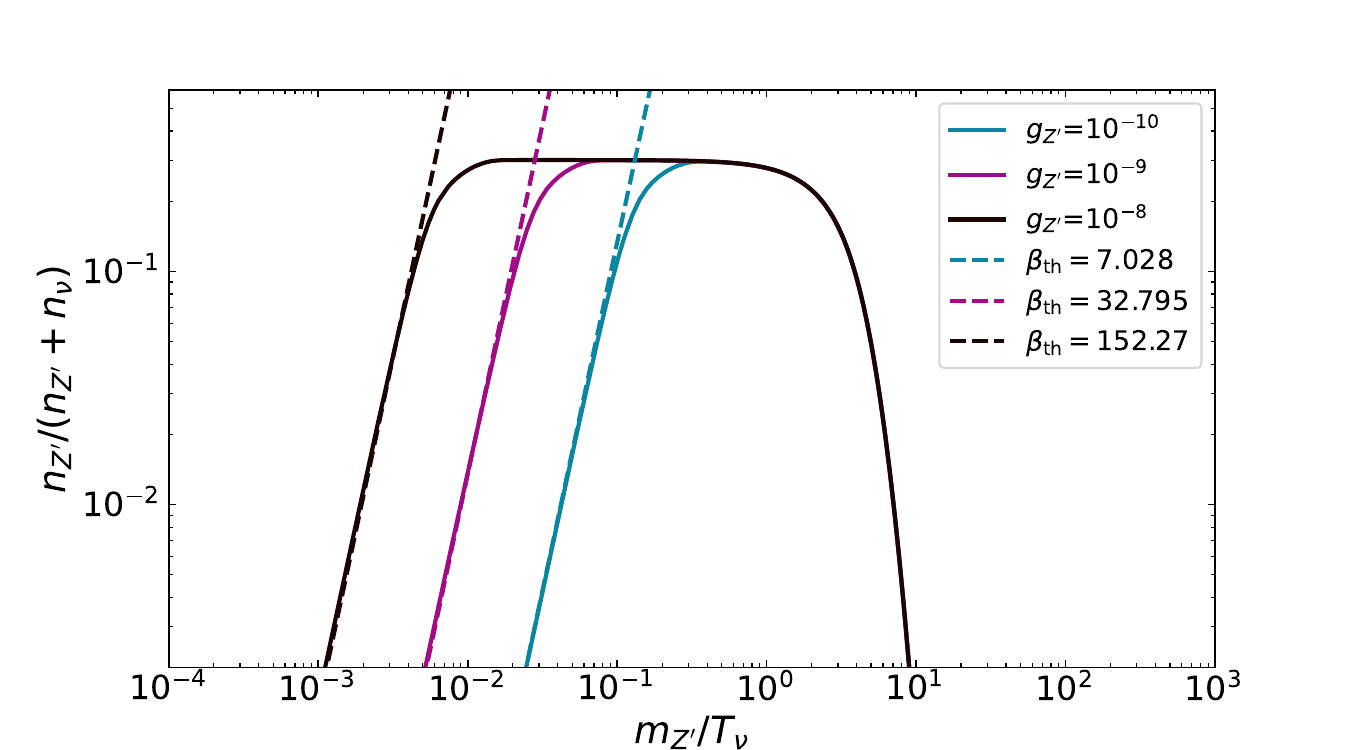}}
\caption{Plots showing the relative fraction of populated vector particles for $m_{Z^\prime} = 1 \, {\rm keV}$: (a) evolution of the $Z^\prime$ population for different couplings $g_{Z^\prime}$; (b) evolution of $r(\beta)$ with $\beta=T_\nu/m_{Z^\prime}$ (dashed lines) compared with the exact solution in the pseudo-chemical approximation. Note the excellent agreement in the small-population regime.} \label{fig:fr-evol}
\end{figure}

Moreover, in panel (b) of Fig.~\eqref{fig:fr-evol}, we compare the estimated evolution of $r(\beta)$ with the exact solution obtained in the pseudo-chemical approximation~\cite{Li2023neff} using MB distribution functions. For this purpose, we fix the coupling $g_{Z^\prime}$,
    \begin{equation}
        g_{Z^\prime} = \left[\frac{32\pi g_H}{M_{\rm pl} \kappa} \left(\frac{r^{\rm th}}{I(\beta_{\rm th}, \beta_0)} \right) m_{Z^\prime} \right]^{1/2} \, , 
    \end{equation}
where $\kappa=2$ (vector case) is the value associated with the amplitude~\eqref{Amplitude-gen}. In this way, we obtain an expression for $g_{Z^\prime}$, analogous to Eq.~\eqref{g}, for the $Z^\prime$ particles by matching the interaction rate given in Eq.~\eqref{Gamma_int_A} in the relativistic regime with the parameter $\beta_{\rm th}$. 

Near the maximum of $n_{Z^\prime}/(n_{Z^\prime} + n_\nu)$ in the pseudo-chemical parametrization, the function $r(\beta)$ fails to describe the evolution accurately and cannot reproduce the standard interaction rates given in Eq.~\eqref{Gamma_int_A}. The solution indicates that the decays of nearly non-relativistic $Z^\prime$ particles delay the attainment of the maximum population, but only when $T_{{Z^\prime}} \sim T_{\nu}$. 
Moreover, in the relativistic regime, the numerical implementation solves~\cite{Abenza2020precision}:
    \begin{align}
        \frac{\delta n_{Z^\prime}}{\delta t} &= \frac{\Gamma_{Z^\prime} m_{Z^\prime}^2}{2\pi^2} \left[ T_\nu e^{\frac{2\mu_\nu}{T_\nu}} K_1\left(\frac{m_{Z^\prime}}{T_\nu} \right) - T_{Z^\prime} e^{\frac{\mu_{Z^\prime}}{T_{Z^\prime}}}K_1\left(\frac{m_{Z^\prime}}{T_{Z^\prime}}\right)\right] \approx \frac{\Gamma_{Z^\prime} m_{Z^\prime}^2}{2\pi^2} \left[ \frac{T_\nu^2}{m_{Z^\prime}} e^{\frac{2\mu_\nu}{T_\nu}} - T_{Z^\prime} e^{\frac{\mu_{Z^\prime}}{T_{Z^\prime}}}K_1\left(\frac{m_{Z^\prime}}{T_{Z^\prime}}\right) \right]  \, ,\\
        \frac{\delta \rho_{Z^\prime}}{\delta t} &= \frac{\Gamma_{Z^\prime} m_{Z^\prime}^3}{2\pi^2} \left[ T_\nu e^{\frac{2\mu_\nu}{T_\nu}} K_2\left(\frac{m_{Z^\prime}}{T_\nu} \right) - T_{Z^\prime} e^{\frac{\mu_{Z^\prime}}{T_{Z^\prime}}}K_2\left(\frac{m_{Z^\prime}}{T_{Z^\prime}}\right)\right] \approx \frac{\Gamma_{Z^\prime} m_{Z^\prime}^3}{2\pi^2} \left[ \frac{2T_\nu^3}{m_{Z^\prime}^2} e^{\frac{2\mu_\nu}{T_\nu}} - T_{Z^\prime} e^{\frac{\mu_{Z^\prime}}{T_{Z^\prime}}}K_2\left(\frac{m_{Z^\prime}}{T_{Z^\prime}} \right) \right] \, .
    \end{align}
The effect of $Z^\prime$ decays is not captured in the fully evolving scenario when the initial condition $T_{Z^\prime,\mathrm{in}} \ll T_{\nu,\mathrm{in}}$ is imposed. For this reason, $r(\beta)$, despite neglecting the depletion term, still provides an accurate description of the solution, even when decays become kinematically allowed for $\beta_d \gtrsim T_{\nu, d} /m_{Z^\prime} = [r^{\rm th}/I(\beta_{\rm th}, \beta_0)]^{1/2}$. This behavior suggests that the approach does not properly account for the lifetime of the produced particles: \emph{as one starts with $T_{Z^\prime,in} \ll T_{\nu, in}$, the numerical solution fails to properly capture the decays of nearly non-relativistic $Z^\prime$-particles, instead adding their contribution to the relativistic population while simultaneously forcing the relativistic vectors to decay instantaneously in order to maintain thermal equilibrium}. Consequently, the assumption of instantaneous thermalization in the relativistic regime, $T_\nu \gg m_{Z^\prime}$, becomes unrealistic. The same argument naturally applies to other types of mediators described by the amplitude~\eqref{Amplitude-gen}.
\section{The modified Boltzmann hierarchy equations and the heavy mediator limit}
\subsection{Toy model: thermalization via decays}
In this subsection, we revisit thermalization via decays and examine how this approach modifies the BHEs. As argued previously, this is not a realistic scenario; nevertheless, it provides a useful toy model for revealing interesting features of the modified BHEs presented in Eqs.~\eqref{BHE-0-CN-S}.

From Eq.~\eqref{fr_gen}, we obtain the evolution of neutrino and $A$-particle distribution functions for $\beta \geq \beta_{\rm th}$. Using the thermalization criterion $\Gamma_{{\rm int}, A} \geq c_{\rm th} H$, where $c_{\rm th}$ is defined in Eq.~\eqref{c_th}, we estimate the temperature at which, in this picture, the fraction $r^{\rm th}$ begins to decrease; under these conditions, we define $\beta_{\rm dep} = T_{\nu, {\rm dep}}/ m_A$,
    \begin{equation}
        \frac{K_1(1/\beta_{\rm dep})}{\beta_{\rm dep}^4} = \frac{K_1(1/\beta_{\rm th})}{\beta_{\rm th}^4} \,. \label{beta_dep}
    \end{equation}
Notice that here we have assumed $g_H = {\rm constant}$.

After starting the depletion of $A$ particles, as our distribution functions follow a thermal-like form, we expect that the collision term, by construction, follows the form $\Gamma_A^d = \Gamma_A n_A^{\rm eq}$, and then we need to solve the following differential equation for the $A$ particles
    \begin{equation}
        \frac{d Y}{da} = - \xi a Y \Rightarrow Y \propto \exp \left[ - \frac{\xi}{2} a^2\right] \, . \label{Exp}
    \end{equation}
Here, $Y=n_Aa^3$ and $\xi = \Gamma_A M_{\rm pl}/(g_HT_{\nu,0}^2)$ with $T_{\nu,0} = aT_\nu$.

The initial condition for neutrino decoupling is $n_A^{\rm max} = n_A^{\rm th}$ given in Eq.~\eqref{number density-parametrized}. Therefore, we obtain
    \begin{equation}
        n_A = n_A^{\rm max} \begin{cases}
            \frac{\beta^3}{\beta_{\rm th}^3} \frac{I(\beta, \beta_0)}{I(\beta_{\rm th}, \beta_0)} & \text{for $\beta \geq \beta_{\rm th}$} \\
            1 & \text{for $\beta_{\rm dep} < \beta < \beta_{\rm th}$} \\
            \exp \left( - \frac{N_\nu}{N_A}\frac{r^{\rm th} [(\beta)^{-2}-(\beta_{\rm dep})^{-2}]}{I(\beta_{\rm th}, \beta_0)} \right) & \text{for $\beta \leq \beta_{\rm dep}$}
        \end{cases} ,
    \end{equation}
where we have used the identity showing that $\beta_{\rm th}$ fixes the coupling $g$,
    \begin{equation}
        \frac{\Gamma_A M_{\rm pl}}{2g_H} = \frac{N_\nu}{N_A}\frac{r^{\rm th}}{I(\beta_{\rm th}, \beta_0)} m_A^2 \, .
    \end{equation}
Using the parametrization of the modified neutrino distribution function $\bar{f}_{\nu} = \xi_r e^{-p/T_\nu}$, and given that we need to satisfy local energy conservation and describe the $A$-particles in their non-relativistic regime, we use $\bar{f}_A = \frac{N_\nu}{N_A} (1-\xi_r) e^{-E_A/T_\nu}$. This violates the conservation law Eq.~\eqref{C-law-Majorana} in the non-relativistic regime; therefore, we must enforce that the coupled $\nu$--$A$ fluid is approximately massless.

Then, the relevant quantity that determines how the neutrino number density (or, in this approximation, the energy density) evolves is $\xi_r = \bar{n}_\nu/n_\nu$, and note that it is straightforward to obtain the variation of this parameter as $\Delta$ given in Eq.~\eqref{Delta(p,T)}, 
    \begin{align}
        \xi_r &= \begin{cases}
            [1 - r(\beta)] & \text{for $\beta \geq \beta_{\rm th}$} \\
            [1 - r^{\rm th}] & \text{for $\beta_{\rm dep} < \beta < \beta_{\rm th}$} \\
            \left[ 1 - \frac{\beta_{\rm dep}^3}{\beta^3} r^{\rm th} \exp \left( - \frac{N_\nu}{N_A}\frac{r^{\rm th} [(\beta)^{-2}-(\beta_{\rm dep})^{-2}]}{I(\beta_{\rm th}, \beta_0)} \right) \right] & \text{for $\beta \leq \beta_{\rm dep}$}
        \end{cases} , \\
        \Delta &= \begin{cases}
            \frac{1}{[1-r(\beta)]} \frac{r^{\rm th}}{I(\beta_{\rm th}, \beta_0)} \beta^{-4} K_1(1/\beta) & \text{for $\beta \geq \beta_{\rm th}$} \\
            0 & \text{for $\beta_{\rm dep} < \beta < \beta_{\rm th}$} \\
            \frac{1-\xi_r}{\xi_r} \left( 3 -  \frac{N_\nu}{N_A}\frac{2r^{\rm th}}{\beta^2 I(\beta_{\rm th}, \beta_0)} \right) & \text{for $\beta \leq \beta_{\rm dep}$}
        \end{cases} \, .
    \end{align}

\begin{figure}
\subfigure[]{
\includegraphics[width=0.4\linewidth]{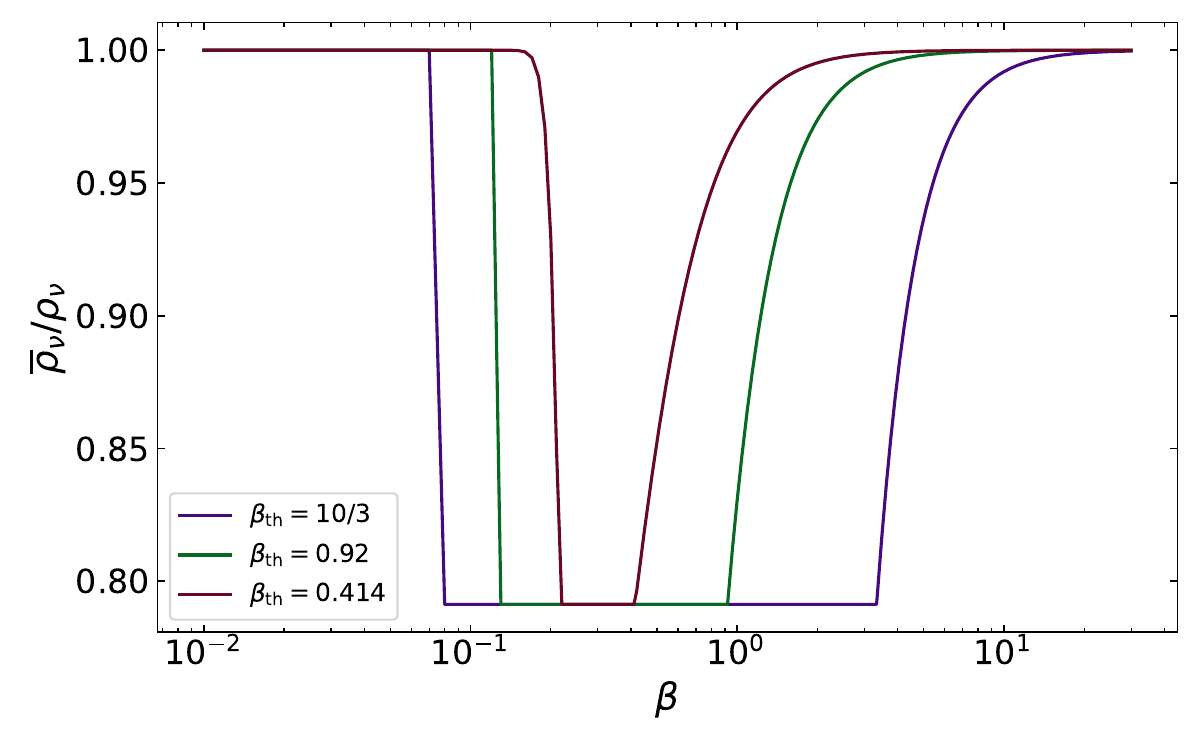}}
\subfigure[]{
\includegraphics[width=0.4\linewidth]{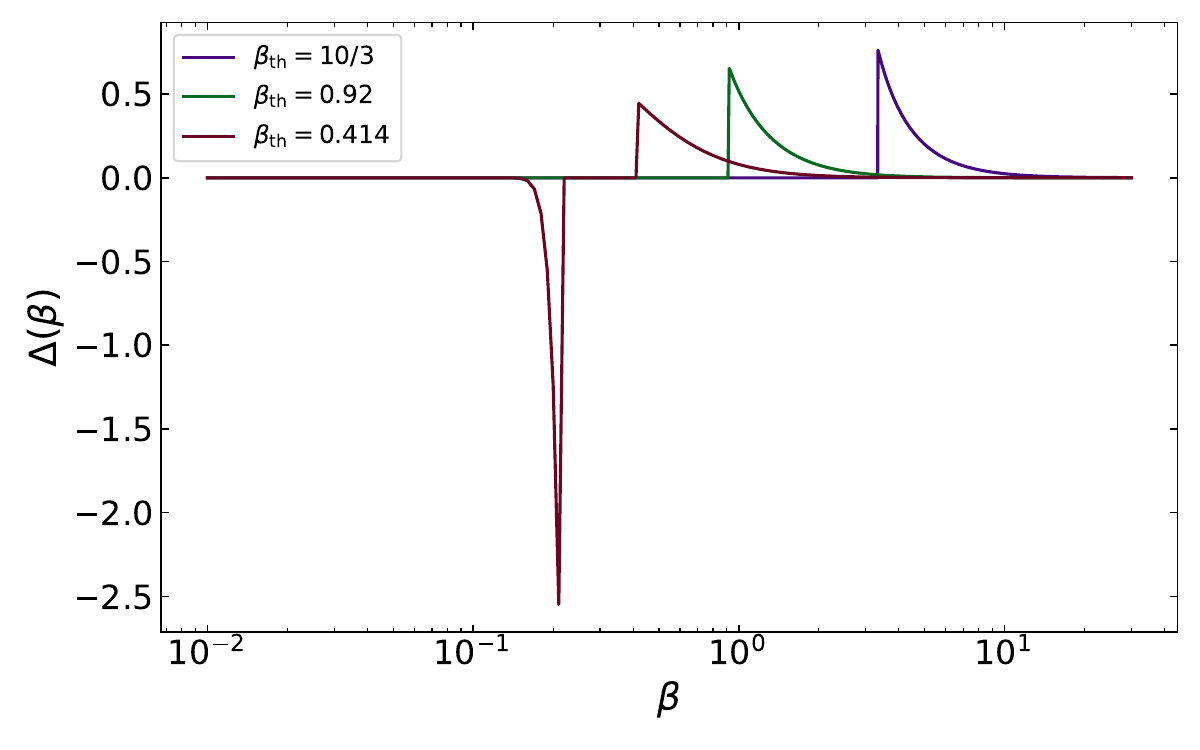}}
\caption{Plot that characterizes the neutrino-(pseudo)scalar fluid evolution: (a) evolution of neutrino energy density due to the loss of energy for creating $\varphi$ particles, (b) evolution of the $\Delta$-term that alters the Boltzmann hierarchy equations.} \label{fig:Toy model}
\end{figure}

Here, notice that as we model the decay of the $A$ particles with an exponential behavior, when the thermal equilibrium ends, they restore the neutrino energy density quickly, as shown in panel (a) of Fig.~\ref{fig:Toy model}, and this implies that $\Delta \geq 0$ for $\beta_{\rm th} \gtrsim 0.92$. Notice that in other cases, such as when the $A$-particles slowly restore energy to the neutrinos, the thermal approximation breaks down. For $\Delta < 0$,
    \begin{equation}
        \frac{\partial \ln \bar{f}_\nu^{(0)}}{\partial \ln p} - \Delta = - \frac{p}{T_\nu} - \Delta \, ,
    \end{equation}
this term generates divergences, and the relation between the Legendre multipoles $\Psi_\ell$ and ${\cal F}_\ell$ (or $\nu_\ell$) in Eq.~\eqref{Multipoles} becomes ill-defined. So for $\Delta \geq 0$, we can determine the term that appears in the BHEs that takes into account the background evolution~\eqref{BHE-0-CN-S}; in this case, for the amplitudes given in Eq.~\eqref{Amplitude-gen} we obtain (we omit the label $i$):
    \begin{align}
        R(\Delta) &= \begin{cases}
            1 & \text{for $\Delta=0$} \\
            1 - \frac{\Delta}{3} + \frac{\Delta^2}{6} - \frac{\Delta^3}{6} + \frac{\Delta^4}{6} e^{\Delta} E_1(\Delta) & \text{for $\Delta > 0 $}
        \end{cases} \quad \text{with} \quad E_1(\Delta) = \int_\Delta^{\infty} dt \frac{e^{-t}}{t} \, , \\
        {\cal I}_{\rm CN}^{(0)} &=  - T_\nu \frac{g_A^2 \kappa}{24 \pi\xi_r \beta^4} \int_{1}^\infty dx e^{-x/\beta} \left[\frac{N_\nu}{N_A}(1-\xi_r) - \xi_r^2 \right] \left(x - \frac{\Delta}{\beta} \ln \left| \frac{x + \sqrt{x^2 -1} + \Delta/\beta}{x - \sqrt{x^2 -1} + \Delta/\beta}\right| \right) = g_A^2 T_\nu \tilde{\Gamma}_{\rm CN}^{(0)} \, , \\
        {\cal I}_{\rm eff}^{(0)} &= T_\nu \frac{ g_A^2 \kappa}{192\pi} \frac{1}{\xi_r \beta^5} \int_1^{\infty} dx x \sqrt{x^2 - 1} e^{- x/\beta} \left[\frac{N_\nu}{N_A}(1-\xi_r) - \xi_r^2 \right] = g_A^2 T_\nu \tilde{\Gamma}_{\rm eff}^{(0)} \, .
    \end{align}

\begin{figure}
\subfigure[]{
\includegraphics[width=0.32\linewidth]{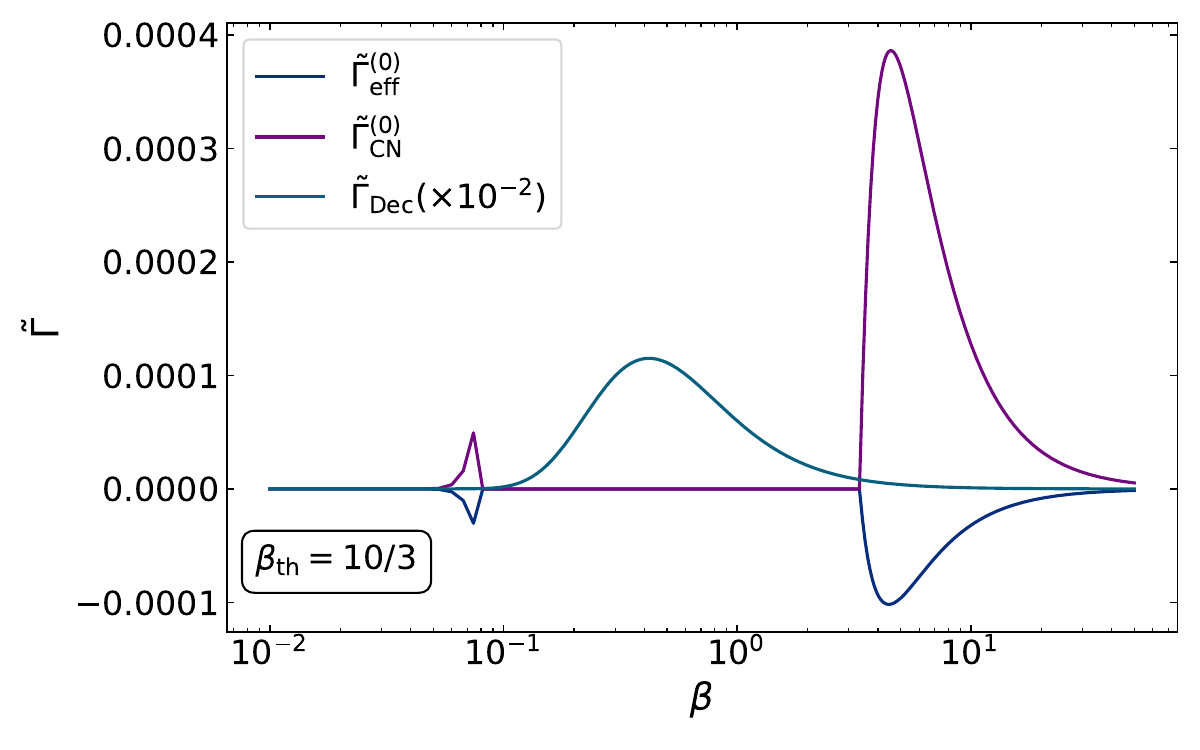}}
\subfigure[]{
\includegraphics[width=0.32\linewidth]{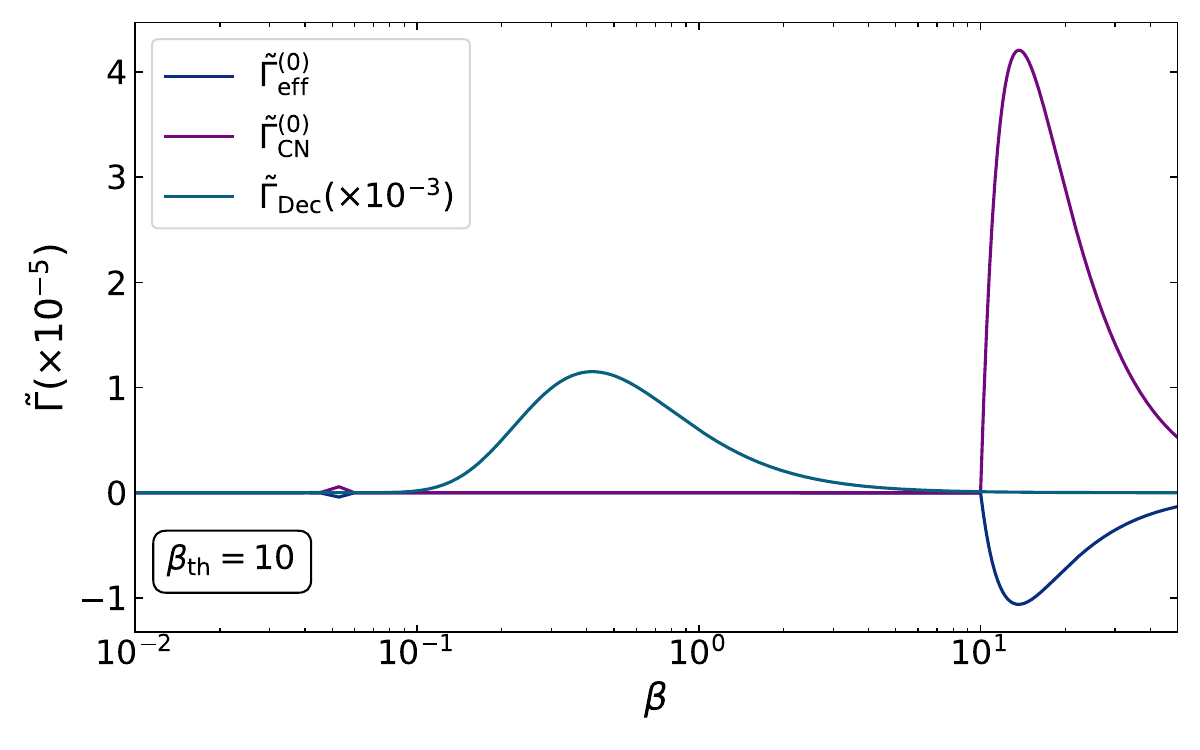}}
\subfigure[]{
\includegraphics[width=0.32\linewidth]{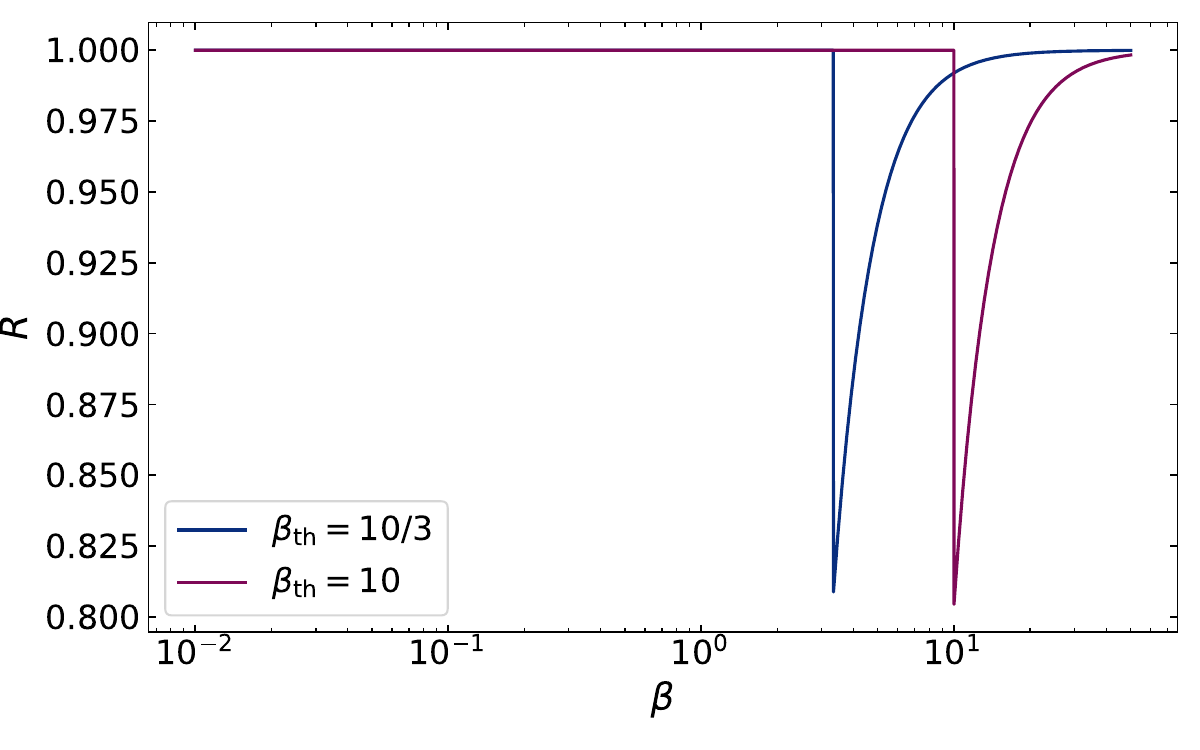}}
\caption{Plot illustrating the evolution of the additional terms in the Boltzmann hierarchy induced by the background dynamics for the real scalar case. Increasing $\beta_{\rm th}$ enhances the coupling $g$ and broadens the thermalization band, indicating that energy restoration occurs more rapidly during late cosmic evolution.} \label{fig:Terms evol}
\end{figure}

In Fig.~\ref{fig:Terms evol}, we compare the interaction rate, factoring out the common term $g^2 T_\nu$ for the scalar case. The relevant effects occur at early times; thus, increasing the coupling toward the end of thermalization causes the $A$-particles to transfer their energy to the neutrinos very rapidly. When $\Delta<0$ is appreciable, it is because the scalars are long-lived enough to restore the energy very slowly.
\subsection{On the heavy mediator limit paradigm}
Recent literature~\cite{Perez2026towards} has proposed an approach that enables us to compute the exact collision term, in which, for relativistic neutrinos, one can define the $\beta_\ell$ functions that characterize the $\nu$SI after its decoupling from the primordial plasma. This neglects the possible background evolution due to the allowed additional processes, such as $\nu\nu \to \varphi\varphi$. Under this approximation, the HML parameter space shown in Fig.~\ref{fig:PS} is determined using the parametrization $\log_{10}(G_{\rm eff} \,{\rm MeV}^2)=-p$ with $p\in[1.0,6.5]$, where $G_{\rm eff}=g^2/m_\varphi^2$. We assume that the interaction rate becomes relevant when $\Gamma_\ell = G_{\rm eff}^2 T_\nu^5 \alpha_\ell/4 \gtrsim H$, while the effective description remains valid for $T_\nu < m_\varphi/100$~\cite{Perez2026towards}. Under these conditions, the parameter space is bounded by the following three curves,
    \begin{equation}
        g_{\rm up} = 10^{-1/2} \frac{m_\varphi}{{\rm MeV}} \, , \quad g_{\rm low} = 10^{-6.5/2} \frac{m_\varphi}{{\rm MeV}} \quad \text{and} \quad g = \left(\frac{4 g_H}{(12.2) \alpha_\ell} \right)^{1/4} 10^{-3.75} \left( \frac{m_\varphi}{{\rm MeV}} \right)^{1/4} \, .
    \end{equation}
For the ``$\nu$SI $G_{\rm eff}$'' region in Fig.~\ref{fig:PS}, we have used $\alpha_{\ell \geq 2} \approx 0.19$ and $g_H \approx 6$.

\begin{figure}
\subfigure[]{
\includegraphics[width=0.4\linewidth]{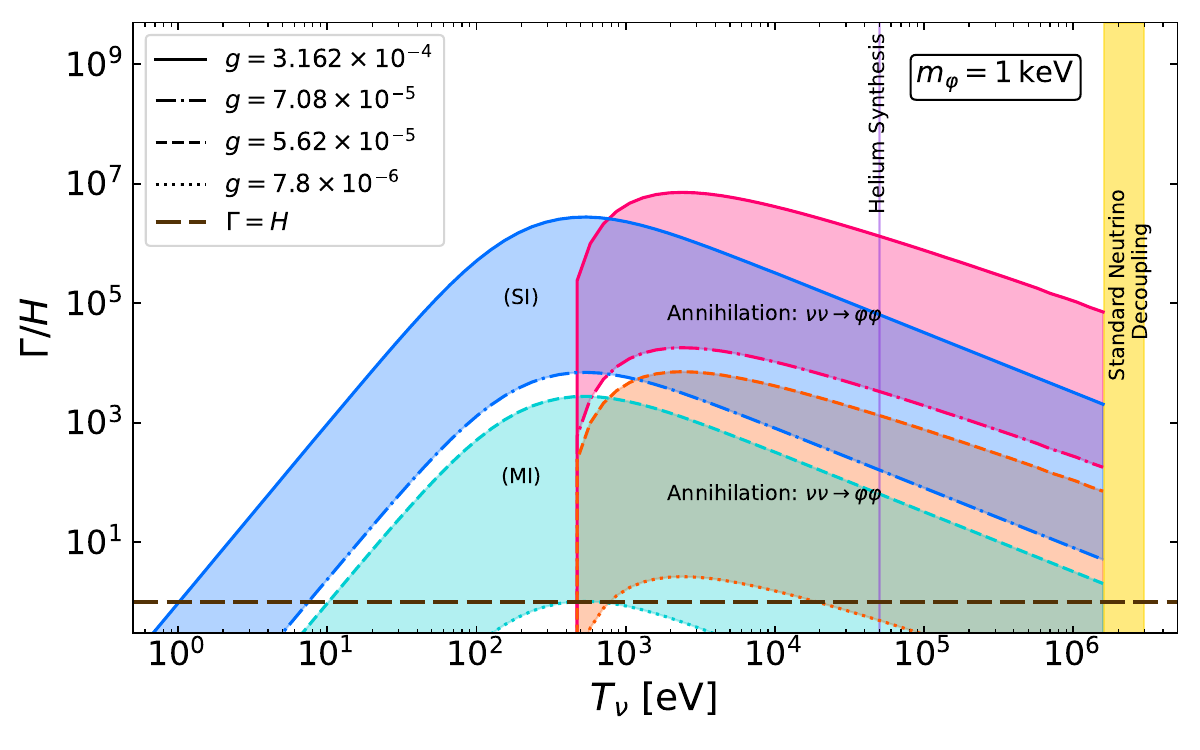}}
\subfigure[]{
\includegraphics[width=0.4\linewidth]{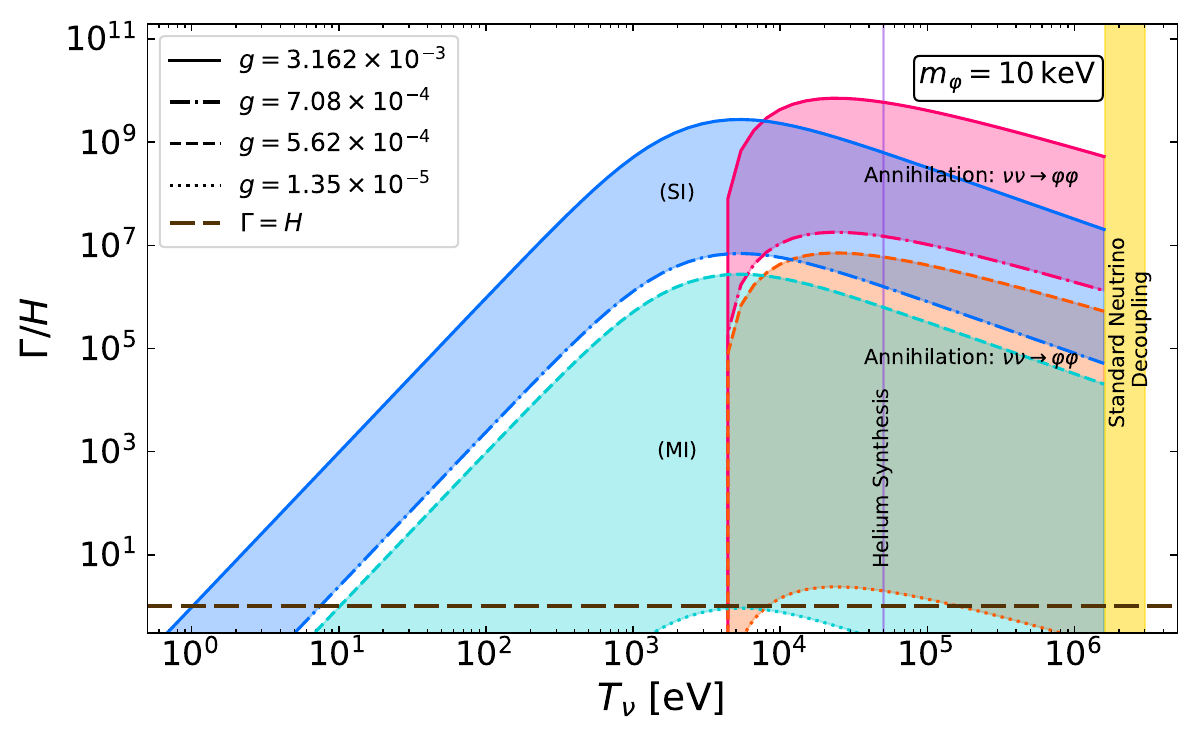}}
\caption{Comparative plot of one contribution to the $\nu$SI interaction rate for the $\ell =2$ multipole in the neutrino--scalar interaction model, excluding the $s$-channel contribution, and the interaction rate $\Gamma_{\nu\nu \to \varphi\varphi}$ responsible for producing relativistic scalars in the regime $T_\nu \gg m_\varphi$, for (a) $m_\varphi = 1\,{\rm keV}$ and (b) $m_\varphi = 10\,{\rm keV}$. The shaded regions indicate the strongly interacting (SI) mode, here defined by $\log(G_{\rm eff}\,{\rm MeV}^{2}) \in [-2.3, -1]$, and the moderately interacting (MI) mode, defined by $\log(G_{\rm eff}\,{\rm MeV}^{2}) \in [-5, -2.5]$. In the latter case, we further exclude the region where $\Gamma_2 < H$.} \label{fig:HML vs annihilation}
\end{figure}

In Fig.~\ref{fig:HML vs annihilation}, we compare the main channel of scalar production in the relativistic regime, $\nu\nu \to \varphi\varphi$. In the relativistic regime, this production must dominate the inverse decays $\nu\nu \to \varphi$; in fact, as argued previously, these decays are better understood as resonant $\nu$SI throughout cosmic evolution. Compared with the toy model in the previous subsection, we expect that the thermal approximation generates a term $\Delta > 0$ (see Eq.~\eqref{Delta(p,T)}) when we follow the neutrino background evolution associated with the energy loss of the neutrino sector due to the production of $\varphi$ particles. Modeling these effects exactly is beyond the scope of this work. However, in the most optimistic case, if a population of $\varphi$ particles is produced, this process ends around $T_\nu \sim (0.4-0.5) \, m_\varphi$; then they become non-relativistic, and their number density $n_\varphi$ decays exponentially, as in Eq.~\eqref{Exp}, rapidly restoring the neutrino energy density and driving $\Delta \to 0$. As the population of $\varphi$ particles is generated through freeze-in, an entropy density is produced and subsequently conserved together with the neutrino entropy density, as in Eq.~\eqref{entropy-law}, during thermalization. Since the reheating process is rapid, one expects a deviation of the neutrino temperature relative to the photon temperature from its standard cosmological value, and consequently a shift in $N_{\rm eff}$, denoted by $\Delta N_{\rm eff}$. Although the precise value is model-dependent, this effect generally induces only small modifications to the $\alpha_\ell$ coefficients of the HML.

\subsection{On the self-interacting dark radiation imitating strong neutrino self-interactions}
Recent literature~\cite{Das2025impostor} has proposed a model in which active neutrinos resonantly convert into self-interacting dark radiation after BBN but before the CMB epoch (using the Farzan--Hannestad mechanism~\cite{Farzan2016neutrinos}), thereby producing a strongly self-interacting dark radiation component that mimics the $\nu$SI favored by cosmological data. A simple realization is based on a type-I seesaw framework supplemented by a keV-scale scalar mediator, where the relevant interactions are
    \begin{equation}
        {\cal L} \supset -\frac{1}{2} \lambda_{\phi\nu} \phi \overline{\nu} \nu - \frac{1}{2} \lambda_{\phi\chi} \phi \bar{\chi} \chi \, .
    \end{equation}
Here $\nu$ denotes the active neutrinos with $N_\nu = 3$, while $\chi$ represents dark Majorana fermions, which can be generalized to $n_\chi$ species. As in Sec.~\ref{IR}, we compute the interaction rate for the resonant production process $\nu \nu \to \chi \chi$. Focusing on a single neutrino species, the interaction rate is given by
    \begin{equation}
        \Gamma_{\nu\nu \to \chi\chi} = \frac{\lambda_{\phi\nu}^2 \lambda_{\phi\chi}^2 T f_\chi}{96 \pi^3 \zeta(3)} \int_{0}^\infty \frac{\tilde{s}^2 F(\tilde{s})}{(\mu_\phi^2 - \tilde{s})^2 + \mu_\phi^2 \gamma_{\phi}^2} \, d\tilde{s} \quad \text{with} \quad f_\chi = \frac{n_\chi}{2} \,.
    \end{equation}
That is very similar to the expression given in Eq.~\eqref{Rate_NSI}; here $T$ is a reference temperature (e.g., the neutrino temperature) used to scale the interaction rate, and
    \begin{equation}
        \gamma_\phi = \frac{\Gamma_\phi}{T} = \frac{\lambda_{\phi\chi}^2}{8\pi} \frac{m_\phi}{T} f_\chi \quad \text{for} \quad \lambda_{\phi\chi} \gg \lambda_{\phi\nu} \, .
    \end{equation}
Particularly, in the NWA, one obtains the expression in terms of $\mu_\phi=m_\phi/T$,
    \begin{equation}
         \Gamma_{\nu\nu \to \chi\chi} = \lambda_{\phi\nu}^2\frac{ \mu_\phi^2 T}{12\pi \zeta(3)} F(\mu_\phi^2) \quad \text{or in the MB approximation} \quad \Gamma_{\nu\nu \to \chi\chi}^{\rm MB} = \frac{\lambda_{\phi\nu}^2 T}{32\pi} \, \mu_\phi^3  K_1(\mu_\phi) \, .
    \end{equation}
The latter expression differs from Eq.~(S39) presented in the \emph{Supplemental Material} of Ref.~\cite{Das2025impostor}, because they do not take into account the well-known identity in their implementation of the NWA,
    \begin{equation}
        \lim_{\epsilon \to 0} \frac{\epsilon}{x^2 + \epsilon^2} = \pi \delta(x) \, .
    \end{equation}
As the dark fermions are created via a resonant mechanism, they thermalize with the neutrino sector, implying $T_\nu = T_\chi$. One can then track the evolution of these temperatures with respect to the photon temperature $T_\gamma$. In particular, according to the analysis presented in Ref.~\cite{Das2025impostor}, and within the MB approximation for the distribution functions, we choose the estimate,
    \begin{equation}
        T_\chi \approx \frac{2}{3} T_\gamma \quad \text{for $n_\chi = 1$} \quad \text{and} \quad  T_\chi \approx \frac{4}{11} T_\gamma \quad \text{for $n_\chi = 40$} \, .
    \end{equation}
In this scenario, energy is extracted from the neutrino sector ($\Delta_i \to \Delta_\chi > 0$ in the thermal approximation for the BHEs~\eqref{BHE-0-CN-S}, $\xi_{i,r} \to \xi_\chi$). The dark fermions are produced via resonant production, which must occur rapidly for sufficiently large couplings $\lambda_{\phi\chi}$.
As a result, one obtains a thermal-like distribution function describing the dark fermions per degree of freedom, parametrized by $\bar{f}_{\chi}^{(0)} = \xi_\chi e^{-p/T_\chi}$ in the MB approximation, where $p$ is the momentum of the dark fermion under consideration.

Given that this mechanism always produces particles out of equilibrium, it leads to some entropy generation, which indeed makes $N_{\rm eff}$ slightly larger than $3.04$. However, this small difference is negligible for practical purposes~\cite{Escudero2023seesaw}. Moreover, one can obtain
    \begin{equation}
        N_{\rm eff} = N_{\rm eff}^\nu + N_{\rm eff}^\chi \quad \text{with} \quad N_{\rm eff}^\nu \approx \frac{N_\nu}{N_\nu + n_\chi} N_{\rm eff} \, , \, N_{\rm eff}^\chi \approx \frac{n_\chi}{N_\nu + n_\chi} N_{\rm eff} \, .
    \end{equation}
This enables the relaxation of the cosmological mass bound as ($N_\nu=3$):
    \begin{equation}
        \sum m_\nu^{(N_{\rm eff}^\chi, N_{\rm eff}^\nu)} < \sum m_\nu^{(\Lambda{\rm CDM})} \left(1 + \frac{n_\chi}{3}\right) \, .
    \end{equation}
As the number of relativistic species can be redefined as
    \begin{equation}
        N_{\rm eff} = \frac{\rho_\nu}{\rho_\gamma} \frac{8}{7} \left( \frac{T_\gamma}{T_\nu}\right)^4 + \frac{\rho_{\rm DS}}{\rho_\gamma} \frac{8}{7} \left( \frac{T_\gamma}{T_\chi}\right)^4 = N_{\rm eff}^\nu + N_{\rm eff}^{\chi} \, ,
    \end{equation}
where $T_\gamma$ is the photon temperature, and $\rho_\gamma = \frac{\pi^2}{15} T_\gamma^4$. Here, $T_\nu^{\rm SM} = T_\nu = \left( \frac{4}{11} \right)^{1/3} T_\gamma$ after electron–positron annihilation, for $\rho_{\rm DS} = 0$. For the dark sector, the energy density with an MB-like form is given by $\rho_{\rm DS} = \xi_\chi \frac{3}{\pi^2} (2 n_\chi) T_\chi^4$. Therefore, we obtain
    \begin{equation}
        \xi_\chi = \frac{7\pi^4}{720} \left( \frac{N_{\rm eff}}{3 + n_\chi} \right) \Rightarrow 
        \xi_\chi =\begin{cases}
            0.720 & \text{for $n_\chi =1$} \\
            0.576 & \text{for $n_\chi =2$} \\
            0.067 & \text{for $n_\chi =40$}
        \end{cases} . \label{Xi_chi}
    \end{equation}
This setup can reproduce the HML for self-interacting dark radiation in the $\chi\chi \to \chi\chi$ scattering channel. Following the strategy of Ref.~\cite{Perez2026towards}, the first-order collision term takes the form in Eq.~\eqref{ctilde_1}, with the only modification occurring in the functions $\Psi_{S, i}$. From local energy conservation, one obtains
    \begin{equation}
        \Psi_{S,i} = \xi_\chi^2 e^{-(p_1+ p_2)/T_\chi} \, .
    \end{equation}
And, as in the HML, we have the relation~\cite{Perez2026towards}
    \begin{equation}
        {\cal A}(z_1) \approx \frac{4 T_\nu^4}{m_\phi^4} 8\pi^3 A^{\rm MB}(q_1) \, , \quad 
        {\cal B}_\ell(z_1) \approx \frac{4 T_\nu^4}{m_\phi^4}  8\pi^3 B_\ell^{\rm MB}(q_1) \quad \text{and} \quad 
        {\cal D}_\ell(z_1) \approx \frac{4 T_\nu^4}{m_\phi^4} 8\pi^3 D_\ell^{\rm MB}(q_1) \, .\label{Heavy-approx}
    \end{equation}
with
    \begin{align}
        A^{\rm MB}(q_1) &= \frac{e^{-q_1}}{8\pi^3}\int_0^\infty dq_2 e^{-q_2} \int_0^{q_1+q_2} dq_3 q_3 \int_{\rho_{+}}^{1}\,dy\frac{{|\cal M}_y(q_i)|^2}{\sqrt{q_1^2 + q_3^2 -2q_1 q_3 y}} ~, \label{A}\\
        B_\ell^{\rm MB} (q_1)&= \frac{e^{-q_1}}{8\pi^3 q_1}\int_0^\infty dq_2 q_2^2 e^{-q_2}\int_0^{q_1+q_2} dq_3 \int_{\eta_{+}}^{1}\,dx\frac{{|\cal M}_x(q_i)|^2{\cal P}_\ell(x)}{\sqrt{q_1^2 + q_2^2 +2q_1 q_2 x}} \, , \label{Bl}\\
        D_\ell^{\rm MB}(q_1)&= \frac{e^{-q_1}}{8\pi^3 q_1}\int_0^\infty dq_2 e^{-q_2} \int_0^{q_1+q_2} dq_3 q_3^2 \int_{\rho_{+}}^{1}\,dy\frac{{|\cal M}_y(q_i)|^2{\cal P}_\ell(y)}{\sqrt{q_1^2 + q_3^2 -2q_1 q_3 y}} \, . \label{Dl}
    \end{align}
where $q_i = p_i / T_\chi$ is the dimensionless momentum, defined as the local three-momentum of the particles involved in the process. For further details on the notation, see Appendix~E.2 of Ref.~\cite{Perez2026towards}. 

Notice that we can also write the $A^{\rm MB}(q_1)$ term as
    \begin{equation} 
        A^{\rm MB}(q_1) = \frac{e^{-q_1}}{8\pi^3}\int_0^\infty dq_2 q_2 e^{-q_2} \int_0^{q_1+q_2} dq_3 \int_{\eta_{+}}^{1}\,dx\frac{{|\cal M}_x(q_i)|^2}{\sqrt{q_1^2 + q_2^2 +2q_1 q_2 x}} \, .
    \end{equation}
Particularly, in our approach, we have used
    \begin{equation}
        \int_{\eta_{+}}^{1}\,dx\frac{{|\cal M}_x(q_i)|^2 {\cal P}_\ell(x)}{\sqrt{q_1^2 + q_2^2 +2q_1 q_2 x}} = \int_{-1}^{1}\,dx\frac{{|\cal M}_x(q_i)|^2 {\cal P}_\ell(x)}{\sqrt{q_1^2 + q_2^2 +2q_1 q_2 x}} H(\Delta_x) \, .
    \end{equation}
with
    \begin{equation}
        \Delta_x = 4q_3^2 q_2^2(1-x^2) {\cal P}(x, q_3, q_2) \quad \text{and} \quad {\cal P}(x, q_3, q_2) = - [(q_1 + q_2)q_3 - q_1 q_2 (1-x)]^2 + (q_1^2+q_2^2+2q_1 q_2 x)q_3^2 \, .
    \end{equation}
In the reduction~\cite{Kreisch2020}, one gets the roots $x_{\pm}$, but notice that we can obtain the roots $R_{1,2}$ for the quadratic polynomial ${\cal P}(x, q_3, q_2)$ in $q_3$:
    \begin{equation}
        R_1 = \frac{q_1 + q_2 - |{\bf q}_1 + {\bf q}_2|}{2} \quad \text{and} \quad R_2 = \frac{q_1 + q_2 + |{\bf q}_1 + {\bf q}_2|}{2} \, .
    \end{equation}
From the triangle inequality, $|{\bf q}_1 + {\bf q}_2| \leq q_1 + q_2$, one obtains $0 \leq R_1 \leq R_2 \leq  q_1 + q_2$. We can then first perform the integral over the $q_3$ variable and subsequently integrate over the variable $x$, and we obtain
    \begin{align}
        A^{\rm MB}(q_1) &= \frac{e^{-q_1}}{8\pi^3} \frac{5q_1^2}{6} \int_{-1}^{1}dx (1-x)^2\int_0^\infty dq_2 q_2^3 e^{-q_2} = \frac{e^{-q_1}}{(2\pi)^3} \frac{40q_1^2}{3}  \, , \label{A_MB_red} \\
        B_\ell^{\rm MB}(q_1) &= \frac{5q_1 e^{-q_1}}{6(2\pi)^3} \int_0^\infty dq_2 q_2^4 e^{-q_2} \int_{-1}^{1}\,dx \, {\cal P}_\ell(x) (1-x)^2 = \frac{80 q_1 e^{-q_1}}{3(2\pi)^3} \left(2 \delta_{0\ell} -  \delta_{1\ell} + \frac{1}{5} \delta_{2\ell} \right) \, . \label{Bl_MB_red}
    \end{align}
If we follow the strategy presented in Ref.~\cite{Oldengott2015}, we must reduce the phase-space integral of ${\cal J}_3 \equiv {\cal C}_1^m/(2a)$, defined in Ref.~\cite{Perez2026towards}, using the identity
    \begin{equation}
        \frac{d^3p_2}{2E_2} =  d^4P_2\,\delta(P_2^2) H(P_2^0) \, .
    \end{equation}
In this case, the squared amplitude (over initial spin states) can be expressed in terms of the scaled three-momentum magnitudes, $q_i=p_i/T$, defined with respect to a reference temperature $T$, as (later, we will replace $g \to \lambda_{\phi\chi}$)
    \begin{equation}
        \overline{|{\cal M}|^2} = \frac{g^4}{2m_\phi^4} (s^2 + t^2 + u^2) = \frac{g^4}{m_\phi^4} (t^2 + u^2 + tu) = \frac{4g^4 T^4}{m_\phi^4} [\bar{\Delta}_2(y)z^2 - \bar{\Delta}_1(y) z + \bar{\Delta}_0(y)] \, , \label{Amplitude-HML}
    \end{equation}
where we have used the scattering angle to define $z=\hat{\bf p}_1 \cdot \hat{\bf p}_4 = \cos\theta^\prime$, and
    \begin{equation}
        \bar{\Delta}_2(y) = q_1^2q_4^2 \, , \quad \bar{\Delta}_1(y) = q_1^2q_4[2q_4 + q_3(1-y)] \quad \text{and} \quad \bar{\Delta}_0(y) = q_1^2[q_4^2 +q_3q_4(1-y) + q_3^2(1-y)^2]\, .
    \end{equation}
Then, following the strategy of Ref.~\cite{Perez2026towards}, one can reduce the ${\cal J}_3$ integral by integrating over $z$, which yields the kinematic constraint
$\bar{\Delta}_y \geq 0$, namely
    \begin{equation}
        \bar{\Delta}_y = 8q_3^2q_4^2(1-y^2) q_1q_3(1-y)\left[q_4^2 - (q_1-q_3) q_4 - \frac{1}{2} q_1q_3(1-y)\right] \, .
    \end{equation}
That has two null roots, and the remaining quadratic polynomial in $q_4$ has the roots:
    \begin{equation}
        R_{-} = \frac{(q_1-q_3) - |{\bf q}_1 - {\bf q_3}|}{2} \leq 0 \quad \text{and} \quad R_{+} = \frac{(q_1-q_3) + |{\bf q}_1 - {\bf q_3}|}{2} \geq 0 \, .
    \end{equation}
We have the kinematic lower bound $q_4^{\rm low} \geq {\rm max} (q_1 - q_3, 0)$. In this case, however, $R_{+} \geq q_4^{\rm low}$, which follows from the inverse triangle inequality, $|{\bf q}_1-{\bf q}_3| \geq |q_1- q_3|$. Thus, the ${\cal J}_3$-integral takes the form
    \begin{equation*}
    \begin{aligned}
        {\cal J}_3 &= \frac{4\pi G_{\rm eff}^2 T^5}{8 (2\pi)^4 q_1} \sum_{\ell =0}^\infty
        (-i)^\ell(2\ell+1) {\cal P}_\ell(\upmu)\int\! dq_3 q_3 dq_4 q_4 \Psi_{S} H(p_3 + p_4 - p_1) \Psi_\ell(q_3) \int dy \frac{|{\cal M}_y^\prime|^2 {\cal P}_\ell(y) H(\bar{\Delta}_y)}{q_4\sqrt{q_1^2 + q_3^2 - 2q_1q_3 y}} \\
        &= \frac{ 4G_{\rm eff}^2 T^5}{16 (2\pi)^3 q_1} \sum_{\ell =0}^\infty
        (-i)^\ell(2\ell+1) {\cal P}_\ell(\upmu)\int_{0}^\infty dq_3 q_3 \Psi_\ell(q_3) \int_{-1}^{1} \frac{dy {\cal P}_\ell(y)}{\sqrt{q_1^2 + q_3^2 - 2q_1q_3 y}} \int_{R_{+}}^{\infty}dq_4 \Psi_{S} |{\cal M}_y^\prime|^2 \, ,
    \end{aligned}
    \end{equation*}
where 
    \begin{equation}
        |{\cal M}_y^\prime|^2 = \bar{\Delta}_2(y)\left(\frac{3\bar{b}_y^2 - 4\bar{a}_y\bar{c}_y}{8\bar{a}_y^2}\right) - \bar{\Delta}_1(y)\left(\frac{\bar{b}_y}{2\bar{a}_y}\right) + \bar{\Delta}_0(y) \, ,
    \end{equation}
with the coefficients 
    \begin{equation}
        \bar{a}_y = - q_4^2(q_1^2 + q_3^2 - 2q_1q_3 y) \, , \, \bar{b}_y = -2q_4(q_1-q_3y)[(q_1-q_3)q_4 + q_1q_3(1-y)] \, , \, \bar{c}_y = -[(q_1-q_3)q_4 + q_1q_3(1-y)]^2 + q_3^2 q_4^2(1-y^2) \, .
    \end{equation}
Within the MB approximation, $\Psi_S = e^{-(q_3+q_4)}$, the integration over $q_4$ can be carried out explicitly. While
within the thermal approximation,
    \begin{equation}
        \Psi_\ell(q_3) = - \frac{1}{4} \frac{d\ln f^{(0)}(p_3)}{d\ln p_3} {\cal F}_\ell = \frac{q_3}{4} {\cal F}_\ell \, ,
    \end{equation}
we can define the $D_\ell^{\rm MB}$ function as
    \begin{equation}
        D_\ell^{\rm MB} = \frac{1}{(2\pi)^3 q_1} \int_{0}^{\infty} dq_3 q_3^2 e^{-q_3} \int_{-1}^{1} dy {\cal P}_\ell(y) K(q_1, q_3,y) \, . \label{Dl_MB_red} 
    \end{equation}
Using the short notation $Q_{\pm} = q_1 \pm q_3$ and $P = \sqrt{q_1^2 + q_3^2 - 2q_1q_3 y}$, we have defined
    \begin{equation}
        K(q_1,q_3, y) = K(Q_{+}, Q_{-}, P, y) = \frac{e^{-(Q_{-}+P)/2}}{16P^5} (Q_{-}^2 - P^2)^2 \left[ P^2 (3P^2 - 2P - 4) + Q_{+}^2(P^2 + 6P +12) \right] \, . \label{Kernel}
    \end{equation}
This prescription is very similar to that presented in Refs.~\cite{Oldengott2015, Oldengott2017}; however, here we work with the scaled momenta $q_i$ instead of the comoving momenta $q$. Here, $T = T_\chi (=T_\nu)$ corresponds to the dark fermion (neutrino) bath temperature.

Finally, averaging over the energy density, we obtain the BHEs as in Eq.~\eqref{Boltzmann-eq-relativistic neutrinos}, with ${\cal F}_\ell \to {\cal F}_{\chi,\ell}$, which works well for $T_\chi \lesssim m_\phi/10^2$. The corresponding interaction rate is $\Gamma_{\chi, \ell} = G_{{\rm eff}, \chi}^2 T_\chi^5 \alpha_\ell$, where
    \begin{equation}
        G_{{\rm eff}, \chi} = \frac{\lambda_{\phi\chi}^2}{m_\phi^2} \quad \text{and} \quad \alpha_\ell = \frac{\xi_\chi}{4} \tilde{\alpha}_\ell \, .
    \end{equation}
Here, we have averaged over the energy density $\bar{\rho}_\chi = 3 \xi_\chi/\pi^2$, and
    \begin{equation}
        \tilde{\alpha}_\ell = \frac{1}{6} \int_{0}^\infty dq_1 q_1^2 [A^{\rm MB}(q_1) + B^{\rm MB}_\ell(q_1)-2D^{\rm MB}_\ell(q_1)] \quad \text{and} \quad T_\chi = \left(\frac{T_\chi}{T_\gamma}\right) T_{\gamma,0} (1+z) \, . \label{Alpha_ell-T}
    \end{equation}
Here, $T_{\gamma,0}$ is the photon temperature today, and $z$ is the redshift due to cosmic evolution. Since the HML is determined by neutrinos at $T_{\nu}^{\rm SM}$, which today satisfies $T_{\nu,0}^{\rm SM} = (4/11)^{1/3} (N_{\rm eff}/3)^{1/4}T_{\gamma,0}$ (with $\rho_{\rm DS}=0$), we obtain $\Gamma_\ell = G_{\rm eff}^2 (T_\nu^{\rm SM})^5 \tilde{\alpha}_\ell/4$, with
$\tilde\alpha_\ell$ normalized as in Eq.~(12) of Ref.~\cite{Kreisch2020}. 
Here, we obtain for dark fermions
    \begin{equation}
        \Gamma_{\chi,\ell} = \eta_{\rm sup} G_{{\rm eff},\chi}^2 (T_{\nu,0}^{\rm SM})^5(1+z)^5\tilde{\alpha}_\ell \quad \text{with} \quad \eta_{\rm sup} = \frac{1}{4} \left(\frac{T_\chi}{T_\gamma}\right)^5 \left(\frac{11}{4}\right)^{5/3} \left( \frac{N_\nu}{N_{\rm eff}} \right)^{5/4} \xi_\chi \, .
    \end{equation}
Notice that in Ref.~\cite{Das2025impostor}, the authors used $G_{\rm eff}^2 \equiv G_{{\rm eff},\chi}^2$, this overestimates the interaction rate by a factor of \ipc{$1/\eta_{\rm sup}$ where}
    \begin{equation}
        \eta_{\rm sup} \approx \begin{cases}
            0.126 & \text{for $n_\chi=1$} \\
            5.64 \times 10^{-4} & \text{for $n_\chi=40$}
        \end{cases} \, .
    \end{equation}
    \begin{table}[h!]
    \centering
    \begin{tabular}{|c|c|c|c|c|}
        \hline
        $\ell$ & \makecell{$\tilde{\alpha}_\ell$ with FD statistics \\ and Pauli blocking effects}  & $\tilde{\alpha}_\ell$ with MB approximation & ${\rm D}_\ell$ with Eq.~\eqref{Dl} & ${\rm D}_\ell$ with Eq.~\eqref{Dl_MB_red}\\
        \hline
        \hline
        0 & $0.0$ & $0.0$ & $0.21504$ & $0.21499$\\
        \hline
        1 & $0.0$ & $0.0$ & $0.05374$ & $0.05374$\\
        \hline
        2 & $0.18704$ & $0.19683$ & $0.01988$ & $0.01984$\\
        \hline
        3 & $0.18414$ & $0.19746$ & $0.00874$ & $0.00872$\\
        \hline
        4 & $0.19203$ & $0.20646$ & $0.00420$ & $0.00421$\\
        \hline
        5 & $0.19559$ & $0.21071$ & $0.00216$ & $0.00215$\\
        \hline
        6 & $0.19711$ & $0.21246$ & $0.00113$ & $0.00116$ \\
        \hline
        7 & $0.19785$ & $0.21365$ & $0.00064$ & $0.00065$ \\
        \hline
        8 & $0.19841$ & $0.21428$ & $0.00032$ & $0.00038$\\
        \hline
        9 & $0.19854$ & $0.21465$ & $0.00020$ & $0.00024$ \\
        \hline
        10 & $0.19876$ & $0.21484$ & $0.00015$ & $0.00012$ \\
        \hline
    \end{tabular}
    \caption{$\tilde \alpha_\ell$ and ${\rm D}_\ell $ coefficients; small deviations arise from the VEGAS integration package.}
    \label{tab: Coefficients}
\end{table}

Moreover, the authors of Ref.~\cite{Das2025impostor} use $\alpha_{\ell \geq 2} \approx 0.424$ (from Ref.~[137] therein), which also differs from our direct comparison with $\tilde{\alpha}_\ell$, which saturates at $\tilde{\alpha}_{\ell \geq 10} = 0.215$ (see Table~\ref{tab: Coefficients}).
This discrepancy arises first from the convention used to account for the degrees of freedom: one must take $\alpha_\ell = 2N \tilde{\alpha}_\ell$, where $N$ is a normalization factor typically introduced to match the neutrino energy density to the correct FD distribution function (analogous to our $\xi_\chi$ factor) as $N=7\pi^4/720 <1$, but this is not exact due to the following issues: 
 
\begin{itemize}
    \item [(i)] {\bf Equivalence of the reduction of collision terms.} Using the expressions given in Eqs.~\eqref{A}–\eqref{Dl}, or equivalently those in Eqs.~\eqref{A_MB_red}, \eqref{Bl_MB_red}, and \eqref{Dl_MB_red}, we can rewrite 
        \begin{equation}
            \tilde{\alpha}_\ell = {\rm A} + {\rm B}_\ell - 2{\rm D}_\ell \, .
        \end{equation}
    In particular, using Eqs.~\eqref{A_MB_red} and \eqref{Bl_MB_red}, we obtain the following analytical results:
    \begin{equation}
            A = \frac{1}{6}\int_{0}^{\infty} dq_1 q_1^2 \, A^{\rm MB}(q_1) = \frac{20}{3\pi^3}  \quad \text{and} \quad {\rm B}_\ell = \frac{1}{6} \int_{0}^{\infty} dq_1 q_1^2 \, B_\ell^{\rm MB}(q_1) = \frac{10}{3 \pi^3} \left(2 \delta_{0\ell} -  \delta_{1\ell} + \frac{1}{5} \delta_{2\ell} \right) \, .\label{Alpha_red_1}
        \end{equation}
    We have verified complete agreement with our numerical integration using the expressions given in Eqs.~\eqref{A} and \eqref{Bl}.
    
    The remaining contributions, however, take the form
    \begin{equation}
            {\rm D}_\ell = \frac{1}{6} \int_{0}^{\infty} dq_1 q_1^2 \, D_\ell^{\rm MB}(q_1) \, .
        \end{equation}
    In Table~\ref{tab: Coefficients}, we have computed the ${\rm D}_\ell$ contributions using Eqs.~\eqref{Dl},~\eqref{Dl_MB_red}, and find complete agreement. As $\ell$ increases, the ${\rm D}_\ell$ contribution becomes negligible. Notice that this allows an analytical simplification of expression~\eqref{Alpha_ell-T}, in the large $\ell$ limit, $\tilde{\alpha}_{\ell \geq 10} = 20/(3\pi^3) \approx 0.215$. Finally, the variation between the $\ell=2$ value and the saturated limit with respect to $\alpha_2$ is of order $\sim 9 \%$.
    \item [(ii)] {\bf Impact of the thermal approximation on the weight function.} When deriving the Boltzmann equation for $\nu$SI in the HML within the \emph{thermal approximation}, one uses Eq.~\eqref{Multipoles} with $\Delta_i \to \Delta =0$, which is sometimes called the separable ansatz, followed by averaging over the neutrino energy density. Instead, sometimes the BHEs are written as~\cite{Oldengott2015}
        \begin{equation}
            \frac{\partial \Psi_{\ell}}{\partial \eta} + k \left( \frac{(\ell +1)}{(2 \ell +1)} \Psi_{\ell +1} - \frac{\ell}{(2\ell +1)} \Psi_{\ell -1} \right)
        + \frac{d \ln f_{\nu_i}^{(0)}}{d \ln q} 
        f_{g,\ell} = \frac{a}{f_{\nu_i}^{(0)}} \frac{1}{2(-i)^\ell}\int_{-1}^{1} d\upmu \tilde{C}^{(1)} {\cal P}_\ell(\upmu) \, ,
        \end{equation}
    and then averages over the neutrino energy density, or equivalently, applies the reduction (see Ref.~[89] in~\cite{Perez2026towards});
        \begin{equation}
            {\cal F}_\ell = \frac{\int dq q^3 f_{\nu_i}^{(0)} \Psi_\ell}{\int dq q^3 f_{\nu_i}^{(0)}} = \frac{\int dq_1 q_1^3 f_{\nu_i}^{(0)} \Psi_\ell}{\int dq_1 q_1^3 f_{\nu_i}^{(0)}} \, ,
        \end{equation}
    one can obtain a set of BHEs for massless neutrinos (see~\cite{Oldengott2017}).

    However, this approach is not equivalent to the thermal approximation because after applying the separable ansatz to the first-order collision term, a different weight function appears. For instance, for Eq.~\eqref{Alpha_red_1}:
        \begin{equation}
            A \to \frac{1}{24}\int_{0}^{\infty} dq_1 q_1^3 \, A^{\rm MB}(q_1) = \frac{1}{6 (2\pi)^3}\frac{10}{3}  \int_{0}^{\infty} dq_1 q_1^5 e^{-q_1}  = \frac{25}{3\pi^3} \approx 0.269 \, .
        \end{equation}
    This prescription systematically yields larger $\alpha_\ell$ coefficients than the usual thermal approximation adopted in Ref.~\cite{Kreisch2020} and here. In fact, using $N=7\pi^4/720$, we obtain $\alpha_2 = 0.424$ and
        \begin{equation}
            \alpha_{\ell \geq 10} = 2 \frac{7\pi^4}{720} \frac{25}{3\pi^3} = \frac{35\pi}{216} \approx 0.509 \Rightarrow \left| \frac{\alpha_{\ell\geq 10} - \alpha_2}{\alpha_2}\right| \times 100\% \approx 20 \% \, . 
        \end{equation}
   The results reported in Ref.~\cite{Oldengott2017} use the normalization factor $N$, but their $\alpha_\ell$ coefficients are proportional to $N^2$, leading to smaller values of $\alpha_\ell$, namely $\alpha_2 = 0.40$, $\alpha_3 = 0.43$, $\alpha_4 = 0.46$, $\alpha_5 = 0.47$, and $\alpha_{\ell \geq 6} = 0.48$, rather than the correct result in this approach that is proportional to $N$ (see also Ref.~\cite{Mogensen2026self}).
\end{itemize}

Finally, we compare the conventions for degrees of freedom and symmetry factors used in the recent literature.
In Refs.~\cite{Oldengott2017, Das2025impostor}, as mentioned previously, a factor of $2N$ arises relative to our $\tilde{\alpha}_\ell$ coefficients in Table~\ref{tab: Coefficients} when the same prescription is used. In fact, when adopting the same convention for the number of degrees of freedom in the collision terms, the factor appearing in the comparison would be $16N$, since the Feynman rule for their coupling is $2g$~\cite{Oldengott2015}, rather than $g$. As the authors of~\cite{Oldengott2015, Oldengott2017} define their squared amplitudes by averaging over the initial and final states and including the corresponding symmetry factors for both, they introduce an additional factor of $1/2$ in their convention (see the standard convention in Ref.~[129] of the latest arXiv version of~\cite{Das2025impostor}). Meanwhile, in~\cite{Perez2026towards}, we define the squared amplitude by averaging over the initial spin states, as is customary in collider-physics calculations, and by including the symmetry factors for the final states; we also include a degeneracy factor of ${\rm g}_s=2$, corresponding to the neutrino that scatters off the reference neutrino in constructing the collision term, in order to recover the standard convention. Given that the previous conventions lead to the same squared amplitude in Eq.~\eqref{Amplitude-HML}, and that we omit the factor of $1/4$ in the definition of the $\tilde{\alpha}_\ell$ coefficients to match Eq.~(12) in the FD statistics of Ref.~\cite{Kreisch2020}, the resulting global factor is $16/(4\cdot2)=2$ times the $N$ normalization factor discussed above. Therefore, contrary to the claim made in Ref.~\cite{Barenboim:2026jef}, Fermi--Dirac statistics are, in fact, not significant ($\lesssim 2.4 \%$ when we consider the normalization factor $N$), as can be seen from the first two columns of Table~\ref{tab: Coefficients}.
\end{document}